\documentclass[preprint,aps,eqsecnum,showkeys]{revtex4}
\usepackage{amsfonts}
\usepackage{amsbsy}

\begin{document}



\def\YM/{Yang\discretionary{-}{}{-}Mills}
\def\FT/{Freedman\discretionary{-}{}{-}Townsend}
\def\CS/{Chern\discretionary{-}{}{-}Simons}
\def\CM/{Chapline\discretionary{-}{}{-}Manton}
\def\KV/{Killing vector}

\def\EL/{Euler\discretionary{-}{}{-}Lagrange}
\def\ic/{integrability condition}

\hyphenation{
all
along
anti
ap-pen-dix 
co-tan-gent
equa-tion
equa-tion-s
equiv-a-lent
evo-lu-tion
fields
form
iden-ti-ty
iden-ti-ties
im-por-tant
its
La-grang-ian
La-grang-ian-s
next
nev-er
prod-uct
real
sca-lar
sym-me-try
sym-me-tries
tak-en
tan-gent
term
two
use
use-s
vari-a-tion
vari-a-tion-s
}


\def\eqref#1{(\ref{#1})}
\def\eqrefs#1#2{(\ref{#1}) and~(\ref{#2})}
\def\eqsref#1#2{(\ref{#1}) to~(\ref{#2})}
\def\sysref#1#2{(\ref{#1})--(\ref{#2})}

\def\Eqref#1{Eq.~(\ref{#1})}
\def\Eqrefs#1#2{Eqs.~(\ref{#1}) and~(\ref{#2})}
\def\Eqsref#1#2{Eqs.~(\ref{#1}) to~(\ref{#2})}

\def\secref#1{Sec.~\ref{#1}}
\def\secrefs#1#2{Secs.~\ref{#1} -- \ref{#2}}

\def\appref#1{App.~\ref{#1}}

\def\Ref#1{Ref.~\cite{#1}}
\def\Refs#1{Refs.~\cite{#1}}

\def\Cite#1{${\mathstrut}^{\cite{#1}}$}

\hyphenation{Eq Eqs Sec App Ref Fig}

\def\EQ{\begin{equation}}
\def\EQs{\begin{eqnarray}}
\def\doneEQ{\end{equation}}
\def\doneEQs{\end{eqnarray}}

\def\eqtext#1{\hbox{\rm #1}}

\def\proclaim#1{\medbreak
\noindent{\it {#1}}}
\def\Proclaim#1#2{\medbreak
\noindent{\bf {#1}}{\it {#2}}\par\medbreak}


\def\newline{\hfil\break}
\def\fewquad{\qquad\qquad}
\def\severalquad{\qquad\fewquad}
\def\manyquad{\qquad\severalquad}
\def\manymanyquad{\manyquad\manyquad}

\def\mstrut{\mathstrut}
\def\hp#1{\hphantom{#1}}

\def\ontop#1#2{
\setbox2=\hbox{{$#2$}} \setbox1=\hbox{{$\scriptscriptstyle #1$}} 
\dimen1=0.5\wd2 \advance\dimen1 by 0.5\wd1 \dimen2=1.4\ht2
\ifdim\wd1>\wd2 \raise\dimen2\box1 \kern-\dimen1 \hbox to\dimen1{\box2\hfill}
\else \box2\kern-\dimen1 \raise\dimen2 \hbox to\dimen1{\box1\hfill} \fi }

\def\mixedindices#1#2{{\mstrut}^{\mstrut #1}_{\mstrut #2}}
\def\downindex#1{{\mstrut}^{\mstrut}_{\mstrut #1}}
\def\upindex#1{{\mstrut}_{\mstrut}^{\mstrut #1}}
\def\downupindices#1#2{{\mstrut}_{\mstrut #1}^{\hp{#1}\mstrut #2}}
\def\updownindices#1#2{{\mstrut}^{\mstrut #1}_{\hp{#1}\mstrut #2}}

\def\ind#1#2{{\boldsymbol {#1}}_{#2}}

\def\tensor#1#2#3{{#1}\mixedindices{#2}{#3}}
\def\covector#1#2{{#1}\downindex{#2}}
\def\vector#1#2{{#1}\upindex{#2}}

\def\id#1#2{\delta\downupindices{#1}{#2}}
\def\cross#1#2{\epsilon\,\downupindices{#1}{#2}}
\def\vol#1{\epsilon\,\downindex{#1}}
\def\invvol#1{\epsilon\,\upindex{#1}}
\def\x#1{x\upindex{#1}}

\def\metric#1{g\downindex{#1}}
\def\invmetric#1{g\upindex{#1}}
\def\flatmetric#1{\eta\downindex{#1}}
\def\invflatmetric#1{\eta\upindex{#1}}
\def\detmetric{\sqrt{|g|}}

\def\solder#1#2{\tensor{\sigma}{#1}{#2}}

\def\torsmetric#1#2{{\tilde g}\mixedindices{#1}{#2}}
\def\invtorsmetric#1#2{{\tilde g}\inv\mixedindices{#1}{#2}}

\def\matr#1{\left( \matrix{#1} \right)}

\def\pform#1#2{\tensor{\phi}{#1}{#2}}
\def\pcurl#1#2{\tensor{\Phi}{#1}{#2}}
\def\spform{\phi}
\def\spcurl{\Phi}
\def\pstr#1#2{\tensor{\spstr}{#1}{#2}}
\def\spstr{\Omega_\phi}
\def\pstrY#1#2{Y\mixedindices{#1}{#2}}

\def\A#1#2{\tensor{A}{#1}{#2}}
\def\F#1#2{\tensor{F}{#1}{#2}}
\def\stF#1#2{\tensor{*F}{#1}{#2}}
\def\B#1#2{\tensor{B}{#1}{#2}}
\def\H#1#2{\tensor{H}{#1}{#2}}
\def\stH#1#2{\tensor{*H}{#1}{#2}}

\def\Y#1#2{Y_{\rm A}\mixedindices{#1}{#2}}
\def\invY#1#2{Y\inv_{\rm A}\mixedindices{#1}{#2}}
\def\torsY#1#2{{\tilde Y}{}_{\rm A}\mixedindices{#1}{#2}}
\def\invtorsY#1#2{{\tilde Y}\inv_{\rm A}\mixedindices{#1}{#2}}

\def\torsF#1#2{\tensor{{*\tilde F}}{#1}{#2}}
\def\torsK#1#2{\tensor{{\tilde K}}{#1}{#2}}

\def\chiral{U}
\def\wavemap#1{\varphi\upindex{#1}}

\def\K#1#2{\tensor{K}{#1}{#2}}
\def\stK#1#2{\tensor{*K}{#1}{#2}}

\def\h#1#2{\tensor{h}{#1}{#2}}
\def\e#1#2{\tensor{e}{#1}{#2}}

\def\ymF#1#2{\tensor{F_{\rm A}}{#1}{#2}}
\def\ymstF#1#2{\tensor{*F_{\rm A}}{#1}{#2}}
\def\ftR#1#2{\tensor{R_{\rm K}}{#1}{#2}}
\def\ftstR#1#2{\tensor{*R_{\rm K}}{#1}{#2}}

\def\deformstF#1#2{{*\mkern -3mu {{}^{\mstrut}_{\rm A}\mkern -2mu F}}
\mixedindices{#1}{#2}}

\def\charge#1{\vector{Q}{#1}}
\def\current#1#2{\tensor{J}{#1}{#2}}

\def\R#1#2{\tensor{R_{\conx{}{}}}{#1}{#2}}

\def\T#1#2{T\updownindices{#1}{#2}}

\def\kv#1#2{\tensor{\xi}{#1}{#2}}

\def\conx#1#2{\Gamma\updownindices{#1}{#2}}

\def\v#1#2{v\downupindices{#2}{#1}}

\def\gdeform#1{g_{\rm A}\downindex{#1}}
\def\invgdeform#1{g\inv_{\rm A}\upindex{#1}}
\def\vdeform#1#2{v_{\rm A}\downupindices{#2}{#1}}
\def\voldeform#1#2{\epsilon_{\rm A}\mixedindices{#1}{#2}}
\def\kdeform#1#2{k_{\rm A}\updownindices{#1}{#2}}
\def\gtorsdeform#1#2{{\tilde g}_{\rm A}\mixedindices{#1}{#2}}

\def\Q#1#2{Q\updownindices{#1}{#2}}
\def\p#1#2{p\updownindices{#1}{#2}}
\def\q#1#2{q\updownindices{#1}{#2}}
\def\polarize#1#2{\varepsilon\mixedindices{#1}{#2}}

\def\der#1{\partial\downindex{#1}}
\def\coder#1{\partial\upindex{#1}}

\def\nthder#1#2{\partial\mixedindices{#1}{#2}}

\def\D#1{D\downindex{#1}}
\def\coD#1{D\upindex{#1}}
\def\adD#1{D\mixedindices{*}{#1}}

\def\Parder#1#2{
\mathchoice{\partial{#1} \over\partial{#2}}{\partial{#1}/\partial{#2}}{}{} }
\def\parder#1#2{\partial{#1}/\partial{#2}}

\def\waveop{\square}
\def\Lie#1{{\cal L}{}_{#1}}

\def\div{{\rm div}}
\def\grad{{\rm grad}}

\def\torsder#1{{\tilde\partial}\downindex{#1}}
\def\torscoder#1{{\tilde\partial}\upindex{#1}}

\def\ymD#1{D_{\rm A}\downindex{#1}}
\def\ymcoD#1{D_{\rm A}\upindex{#1}}
\def\ftD#1{D_{\rm K}\downindex{#1}}

\def\E#1#2{\tensor{E}{#1}{#2}}
\def\ymE#1#2{\tensor{E^{\rm YM}}{#1}{#2}}
\def\ftE#1#2{\tensor{E^{\rm FT}}{#1}{#2}}

\def\L#1{\covector{L}{#1}}
\def\sL{L}

\def\ELop#1{{\rm E}_{#1}}

\def\striv{\Theta}
\def\triv#1{\Theta\downindex{#1}}
\def\cotriv#1{{*\Theta}\upindex{#1}}
\def\ibp#1{\Upsilon\upindex{#1}}

\def\solsp{|^{\mstrut}_{E=0}}
\def\nthsolsp#1{|_{\ontop{(#1)}{\scriptstyle E}=0}}

\def\term#1#2#3{\tensor{#1}{#2}{#3}}

\def\X#1{\vector{\sX}{#1}}
\def\Xsub#1#2{\vector{\sX_{#1}}{#2}}

\def\pX#1#2{\tensor{\sX}{#1}{#2}}
\def\pXsub#1#2#3{\tensor{\sX_{#1}}{#2}{#3}}
\def\sX{\zeta}
\def\sXsub#1{\sX_{#1}}

\def\Xrigid{\sX\rigid}

\def\Xvar{\delta_\sX}
\def\Xvarsub#1{\delta_{\sX_{#1}}}
\def\Xrigidvar{\delta_{\sX\rigid}}

\def\ymXvar#1{\delta^{\rm YM}_{\sXsub{#1}}}
\def\ftXvar#1{\delta^{\rm FT}_{\sXsub{#1}}}

\def\var#1{\delta_{#1}}

\def\basis#1#2{{\bf e}\mixedindices{#1}{#2}}
\def\covD#1{{\bf D}_{#1}}
\def\conxform#1{{\boldsymbol \Gamma}_{#1}}
\def\Rform#1{{\bf R}_{#1}}

\def\Aform{{\bf A}}
\def\Fform#1{{\bf F}_{#1}}
\def\Kform{{\bf K}}
\def\Ymap{{\bf Y}_A}
\def\Yinvmap{{\bf Y}_A^{-1}}

\def\Xform{{\boldsymbol \zeta}}

\def\Lform{{\boldsymbol L}}

\def\vform{{\bf v}}
\def\torsd{{\tilde d}}

\def\torsFform#1{{\tilde{\bf F}}{}_{#1}}
\def\torsYmap{{\tilde {\bf Y}}{}_A}

\def\idmap{{\mathbf 1}}

\def\SU#1{SU(#1)}
\def\SO#1{SO(#1)}
\def\U#1{U(#1)}
\def\G{{\cal G}}

\def\vs{{\mathbb V}}
\def\g{{\mathfrak g}}

\def\k#1{\covector{k}{#1}}
\def\invk#1{\vector{k}{#1}}

\def\admap#1{{\rm ad}_{#1}}

\def\ym#1#2{a\updownindices{#1}{#2}}
\def\ft#1#2{b\downupindices{#2}{#1}}
\def\ftcoad#1#2{b\updownindices{#1}{#2}}

\def\liealg#1#2{c\updownindices{#1}{#2}}

\def\semi{ \setbox1=\hbox{{$\times$}} \setbox2=\hbox{{$\scriptstyle |$}}
\box1\kern -1.3\wd2 \raise.2\ht2\box2 }

\def\intprod{ {\scriptstyle\rfloor} }

\def\c#1{\kappa^{\strut}_{\rm #1}}

\def\fieldsp{{\cal S}}
\def\fieldsolsp{{\cal E}}
\def\jetsp{J^{(\infty)}(\fieldsp)}
\def\jetspE{J^{(\infty)}(\fieldsolsp)}
\def\parmsp{{\cal P}}

\def\nthE#1#2#3{\tensor{\ontop{(#1)}{E}}{#2}{#3}}
\def\linE#1#2#3{\ontop{(1)}{E}^{\rm #1}\mixedindices{#2}{#3}}

\def\nthL#1#2{\ontop{(#1)}{L}\downindex{#2}}
\def\nthsL#1{\ontop{(#1)}{\sL}}
\def\nthS#1{\ontop{(#1)}{S}}

\def\nthpstr#1#2#3{\ontop{(#1)}{\spstr}\mixedindices{#2}{#3}}
\def\nthpstrY#1#2#3{\ontop{(#1)}{Y}\mixedindices{#2}{#3}}

\def\nthpX#1#2#3{\ontop{(#1)}{\sX}\mixedindices{#2}{#3}}
\def\nthsX#1{\ontop{(#1)}{\sX}}

\def\nthvar#1#2{\ontop{(#1)}{\delta}{}_{#2}}

\def\nthXvar#1{\ontop{(#1)}{\delta_\sX}}
\def\nthcommXvar#1#2#3{\ontop{(#1)}{[\delta_{\sX_{#2}},\delta_{\sX_{#3}}]}}
\def\nthXvarsub#1#2{\ontop{(#1)}{\delta_{\sX_{#2}}}}
\def\nthvarnthXsub#1#2#3{\ontop{(#1)}{\delta_{\ontop{(#2)}{\scriptstyle{\sX_{#3}}}}}}

\def\nthXrigidvar#1{\ontop{(#1)}{\delta}{}_{\sX\rigid}}
\def\nthXrigidvarsub#1#2{\ontop{(#1)}{\delta}{}_{\sX_{#2}\rigid}}
\def\nthvarnthXrigidsub#1#2#3{\ontop{(#1)}{\delta}_{\ontop{(#2)}{\scriptstyle{\sX_{#3}}}\rigid}}

\def\nthpXsub#1#2#3#4{\ontop{(#1)}{\sX_{#2}}\mixedindices{#3}{#4}}
\def\nthsXsub#1#2{\ontop{(#1)}{\sX_{#2}}}

\def\nthterm#1#2#3#4{\ontop{(#1)}{#2}\mixedindices{#3}{#4}}

\def\nthstriv#1{\ontop{(#1)}{\Theta}}
\def\nthtriv#1#2{\ontop{(#1)}{\Theta}\downindex{#2}}
\def\nthcotriv#1#2{{*\ontop{(#1)}{\Theta}}\upindex{#2}}

\def\nthD#1#2{\ontop{(#1)}{D}\downindex{#2}}

\def\nthT#1#2#3{\ontop{(#1)}{T}\updownindices{#2}{#3}}
\def\nthcurrent#1#2#3{\ontop{(#1)}{J}\mixedindices{#2}{#3}}

\def\nthgdeform#1#2{\ontop{(#1)}{g_{\rm A}}\downindex{#2}}

\def\frac#1#2{{\textstyle {#1 \over #2}}}
\def\sfrac#1#2{{\scriptstyle {#1 \over #2}}}

\def\exchange#1#2{(\sXsub{#1}\leftrightarrow\sXsub{#2})}

\def\rigid{|_{\rm rigid}}

\def\Rnum{\mathbb R}

\def\iso{\simeq}

\def\inv{{}^{-1}}
\def\ad{{}^{\rm T}}
\def\tr{{\rm tr}}
\def\i{{\rm i}}

\def\const{{\rm const.}}
\def\flat{{}_{\rm flat}}
\def\mass{{}_{\rm mass.}}

\def\ie/{i.e.}
\def\eg/{e.g.}


\title{
Gauge theory deformations and novel Yang-Mills Chern-Simons field theories
with torsion }

\author{Stephen C. Anco}
\email{sanco@brocku.ca}
\affiliation{%
Department of Mathematics, Brock University, St Catharines,
Ontario, L2S 3A1, Canada
}

\date{August 6, 2004}

\begin{abstract} 
A basic problem of classical field theory, 
which has attracted growing attention over the past decade,  
is to find and classify 
all nonlinear deformations of linear abelian gauge theories. 
The physical interest in studying deformations is 
to address uniqueness of known nonlinear interactions of gauge fields
and to look systematically for theoretical possibilities for new interactions.
Mathematically, the study of deformations aims 
to understand the rigidity of the nonlinear structure of gauge field theories
and to uncover new types of nonlinear geometrical structures. 

The first part of this paper summarizes and significantly elaborates
a field-theoretic deformation method developed in earlier work.
Some key contributions presented here are, 
firstly, 
that the determining equations for deformation terms 
are shown to have an elegant formulation 
using Lie derivatives in the jet space 
associated with the gauge field variables. 
Secondly, 
the obstructions (integrability conditions)
that must be satisfied by lowest-order deformations terms 
for existence of a deformation to higher orders 
are explicitly identified. 
Most importantly,
a universal geometrical structure 
common to a large class of nonlinear gauge theory examples is uncovered. 
This structure is derived geometrically from the deformed gauge symmetry
and is characterized by 
a covariant derivative operator plus a nonlinear field strength,
related through the curvature of the covariant derivative. 
The scope of these results encompasses
\YM/ theory, \FT/ theory, Einstein gravity theory,
in addition to their many interesting types of novel generalizations
that have been found in the past several years. 

The second part of the paper presents 
a new geometrical type of \YM/ generalization in three dimensions 
motivated from considering torsion in the context of nonlinear sigma models 
with Lie group targets (chiral theories).
The generalization is derived by a deformation analysis of 
linear abelian \YM/ \CS/ gauge theory. 
Torsion is introduced geometrically through a duality with chiral models
obtained from the chiral field form of 
self-dual 2+2 dimensional \YM/ theory under reduction to 2+1 dimensions. 
Field-theoretic and geometric features of 
the resulting nonlinear gauge theories with torsion are discussed. 
\end{abstract}


\maketitle

\section{ Introduction and summary }

Classical gauge field theories are widely studied in many areas of 
mathematics, and their importance in physics almost needs no comment. 
They provide the starting point for formulations of 
physically important theories of quantum fields;
in particular, for the case of electromagnetic and gravitational fields, 
they describe the dynamics of macroscopic quantum states 
in the classical regime where quantum effects are not significant. 

Looked at from the view point of geometry, 
classical gauge theories have a rich nonlinear structure ---
field variables typically are sections of a vector bundle or group bundle 
over spacetime, 
equations of motion for the fields usually involve 
a connection, covariant derivative, and curvature tensor on this bundle,
while gauge symmetries of the fields are described by 
an infinite-dimensional Lie group action or local bundle automorphisms. 
One way to characterize all this structure is by the idea of 
deformations of linear abelian gauge theories. 

A prototype example is non-abelian \YM/ theory. 
The \YM/ field is mathematically defined as a connection 1-form 
in a semisimple Lie group bundle on spacetime, 
in which the \YM/ gauge symmetry is a local automorphism given by
the group action of the bundle. 
If a linearization around a flat connection is considered, 
the theory reduces to a set of linear electromagnetic field equations 
and corresponding spin-one gauge symmetries. 
Non-abelian \YM/ theory describes a nonlinear deformation of 
the structure of this linear abelian (electromagnetic field) theory
\cite{YMth,Des,Wal}. 
Another prototypical example is Einstein gravity theory. 
The gravitational field is a metric tensor on spacetime
to which is naturally associated a metric-connection in the tangent bundle, 
and the gauge symmetries arise from diffeomorphisms 
acting on the spacetime manifold. 
Linearization around a flat metric reduces the theory to 
the linear gravity wave equation 
and its accompanying abelian spin-two gauge symmetry. 
Einstein gravity theory represents a highly nonlinear deformation of 
the structure of this linear abelian (graviton field) theory
\cite{OgiPol,Des,FanFro,Wal}.
In both these examples, 
the gauge covariant field strength in the theory is simply 
the curvature defined geometrically from the field. 

An interesting problem of classical field theory is to find and classify
all nonlinear deformations of a given linear abelian gauge theory
\cite{AMSpaper,Hen1,Sta}. 
Deformations, in general, refer here to 
adding linear and higher power local terms 
to the abelian gauge symmetry,
while also adding quadratic and higher power local terms 
to the linear field equation,
such that there exists a gauge invariant local action functional
for the deformed theory. 
Such a deformation is essentially nonlinear if
the deformed Lagrangian is not equivalent, to within a local divergence,
to the Lagrangian of the linear theory 
under any invertible local nonlinear field redefinitions. 
Deformations in which the abelian gauge symmetry remains undeformed
(for example, by adding gauge invariant Pauli-type interaction terms 
to the Lagrangian) 
are regarded as trivial 
since the abelian gauge invariance (and, hence, geometrical structure) 
is unchanged. 
Importantly, the condition of gauge invariance can be used to formulate
determining equations to solve for all allowed deformation terms 
order-by-order in powers of the fields,
without any need for special assumptions or ansatzes
on possible forms of the deformed field equations, gauge symmetries, 
and gauge group structure. 
There are two main formulations of the deformation determining equations
--- a direct, field-theoretic approach \cite{AMSpaper}
where gauge invariance is expressed elegantly by 
Lie derivative equations for gauge symmetries,
using a jet space formalism in terms of the gauge field variables;
and a powerful BRST approach \cite{Hen1}
based on the field-antifield formalism \cite{BarHen,Sta,StaFulLad1}
in which the condition of a gauge invariant Lagrangian is encoded
by the cohomology of a BRST operator. 

The physical interest in studying deformations is 
to address uniqueness of known nonlinear interactions of gauge fields
and to look systematically for theoretical possibilities for new interactions.
Mathematically, the study of deformations aims 
to understand the rigidity of the nonlinear structure of gauge field theories
and to uncover new types of nonlinear geometrical structures. 
Sparking this subject is the rich interplay between,
on the one hand, 
differential geometry and deformations of graviton theory, 
connections on vector bundles and deformations of electromagnetic theory,
and on the other hand, 
their obvious significance for many developments in mathematics and physics.

Indeed, the last decade has seen a large body of work on 
the study of nonlinear gauge theory deformations, 
which has yielded many interesting new kinds of generalizations of
\YM/ theory and Einstein gravity theory
both as classical theories of spin-one and spin-two fields
and as geometrical theories of Lie-algebra valued 1-forms
and algebra valued vielbeins or metrics.
Rigidity results for these generalizations have been obtained as well
\cite{HenKna,BarBraHen,Hen2,Annalspaper,CQGpaper,JMPpaper}

The first gravity generalizations to be found were motivated by
aspects of classical supergravity theory \cite{classicalSG}
and describe nonlinear multi-graviton theories in four dimensions 
involving the introduction of various algebras 
in which the fields take values \cite{CutWal,Annalspaper}.
Supersymmetric extensions of these generalizations have also been obtained,
along with more novel extensions that involve 
non-commuting classical graviton fields
(and non-anticommuting gravitino fields) \cite{CQGpaper}.
Recently, an exotic type of multi-graviton theory 
with a parity-violating interaction that exists for commuting graviton fields
only in three and five dimensions
has been investigated \cite{BouGua,PhysRevpaper}.
This exotic gravity generalization is most remarkable in that 
its gauge invariance is completely different 
than the familiar diffeomorphism invariance of Einstein gravity,
in contrast to the previous generalizations which all feature 
diffeomorphism invariance extended to an algebra-valued setting. 

\YM/ generalizations were first obtained for 1-form potentials 
in three dimensions \cite{abelianth,nonabelianth}
and come from a \YM/ interaction combined with a \FT/ type interaction
available for 1-forms only when the number of dimensions equals three. 
Recall, \FT/ theory \cite{FTth} 
exists for $d-2$-form potentials in $d\geq 3$ dimensions,
with the dual nonlinear field strength of the potential serving as
a connection 1-form whose curvature vanishes due to the field equations; 
as is well known, because the connection is flat, 
\FT/ theory geometrically describes 
a dual form of principal chiral models. 
Extensions of this type of generalization of \YM/ theory to 
interacting 1-form and 2-form potentials in four dimensions \cite{Dra},
including a \CS/ type mass term, were later obtained \cite{JMPpaper}. 
(It should also be noted that gauge theories containing 2-form potentials 
are of active interest in supergravity contexts \cite{Bra1}.) 
Subsequently a different type of generalization in four dimensions 
was derived in recent work \cite{LMPpaper}
involving a \CM/ type interaction between 1-form and 2-form potentials 
combined with a \YM/ interaction and an extended \FT/ interaction. 
The gauge invariance in all these generalizations unifies the form of
the familiar \YM/ and \FT/ gauge symmetries, 
leading to a nonlinear structure that mixes 
geometrical features of pure \YM/ and pure \FT/ theories 
in an interesting way. 
In particular, the combined \YM/ \CM/ generalization 
exhibits a dual formulation that describes a principal chiral field
with an exotic dilaton coupling to the \YM/ fields
through a generalized Chern-class term \cite{LMPpaper}.

In another direction, 
a novel \YM/ generalization involving a gravity-like interaction of
1-form potentials has been recently derived \cite{Bra2}, 
in which the interaction is constructed using 
conserved currents associated with \KV/s of spacetime. 
Its gauge invariance unites the familiar \YM/ type 
with the vielbein type found in Einstein gravity,
giving a gauge theory of spacetime symmetries (\KV/s) in a direct sense.
However, the geometrical structure and gravity-like aspects of the theory  
remain to be explored more fully. 

A main purpose of this paper will be to present a new geometrical type of
\YM/ generalization in three dimensions motivated from considering
torsion in the context of nonlinear sigma models \cite{torsion}. 
Recall a nonlinear sigma field 
(also known as a wave map) 
is a function on spacetime taking values in a Riemannian target space
\eg/ an $n$-sphere or a compact simple Lie group 
(note the latter defines a chiral field model).
The nonlinear structure of a sigma model is given jointly in terms of
the spacetime metric and the symmetric (Riemannian) metric-connection 
on the target. 
More specifically, 
the field equation is a semilinear geometrical wave equation 
(sometimes called a wave map equation)
consisting of the spacetime wave operator on the sigma field
plus a quadratic interaction given by 
a product of sigma field gradients contracted with 
the target metric-connection and the spacetime metric tensor. 
This interaction readily generalizes to include torsion 
in the antisymmetric part of the connection on the target,
if at the same time a skew tensor is added to the spacetime metric. 

Torsion first arose in this context for 1+1 dimensional sigma models,
motivated by topological Wess-Zumino-Witten terms
which describe the winding number of instanton solutions 
in three dimensional sigma models. 
This is analogous to the relation between 
three dimensional \CS/ terms and four dimensional instantons 
in self-dual \YM/ theory. 
Indeed, torsion sigma models in 2+1 dimensions with Lie group targets
were later found to arise from the chiral field form of 
self-dual \YM/ theory under dimensional reductions 
\cite{Yan,Pol,War}.
As important motivation, 
such reductions yield integrable chiral field equations
that possess many of the same integrability features in 2+1 dimensions
as are known for chiral models in 1+1 dimensions
\cite{integrable}.
These features geometrically stem from tuning the torsion 
in the connection on the Lie group target space 
so as to flatten its generalized Riemannian curvature \cite{torsion}. 
Torsion more generally is of interest in the analytical study
of the Cauchy problem for sigma models in 2+1 dimensions
\cite{AncIse},
since this is the critical dimension in which blow-up may occur
for solutions with smooth initial data of large energy \cite{ShaStr}.

The torsion gauge theories described in this paper
will be geometrical generalizations of the dual form of 
general 2+1 dimensional torsion chiral models. 
In particular, as a main new result, 
torsion will be shown to be consistent with a \YM/ interaction of 
1-form potentials in 2+1 spacetime dimensions 
provided a \CS/ mass term is included in the full theory. 
Because the introduction of a skew tensor 
needed to support the torsion interaction 
breaks Lorentz covariance at each point in spacetime,
the field equations in these new torsion gauge theories 
have a semi-relativistic form resembling that of 
2+1 dimensional torsion chiral models \cite{War}. 

To begin, 
a strengthened geometrical version of the field-theoretic approach to
gauge theory deformations developed in the course of earlier work 
\cite{AMSpaper,Annalspaper,JMPpaper}
will be summarized in \secref{method}. 
The approach given here 
simplifies the steps for finding deformation terms up to second order
and clearly identifies the obstructions (integrability conditions)
that must be satisfied on 1st order deformations 
for existence of a deformation to higher orders. 
In particular, it is shown how the Noether current and commutator 
associated with the 1st order deformed gauge symmetries
explicitly determine the quadratic deformation terms 
in both the field equation and the gauge symmetry. 
This enables finding all obstructions, which is a key contribution. 
As further main results of this comprehensive approach, 
firstly, a simple general quasilinear form for 
the deformation terms is shown to be required by 
preserving the number of initial-data and gauge degrees of freedom
order by order.
Secondly, 
a remarkable universal geometrical structure derived in terms of 
a covariant derivative operator associated with the gauge symmetry, 
along with a nonlinear field strength 
connected to the curvature of this covariant derivative, 
is shown to be common to a large class of deformations. 
The scope of these results is quite general and encompasses
the gauge theory examples discussed earlier. 

In \secref{YMFTCSth}, 
a generalization of nonabelian \YM/ \CS/ theory 
incorporating a \FT/ interaction without torsion is constructed 
by a deformation analysis of linear abelian \YM/ theory
with a \CS/ term for 1-form potentials in three spacetime dimensions.
Geometrical aspects of this generalization are highlighted
and shown to display a duality with a principal chiral field theory
that is coupled in a novel manner to 
nonabelian \CS/ theory with a Proca mass term. 
The torsion generalization of 
three-dimensional nonabelian \YM/ \CS/ gauge theory 
is then derived in \secref{torsionYMth}. 
The derivation applies a deformation analysis to 
linear abelian \YM/ \CS/ theory with the inclusion of torsion 
based on the dual form of torsion chiral models 
for abelian Lie group targets. 
Field-theoretic and geometric features of 
the resulting nonlinear gauge theories with torsion are discussed. 
In \secref{torsionKVth} 
an extension of torsion to the gravity-like generalization of \YM/ theory 
is obtained by a similar derivation. 

Concluding remarks are made in \secref{conclude}. 
Throughout, notation follows that of \Ref{JMPpaper}.

\section{ Deformation method }
\label{method}

For the purpose of a general deformation theory, 
all gauge theories of primary interest in mathematics and physics
can be formulated in common by gauge field variables taken to be 
vector-valued $p$-form potentials in $d$ dimensions, 
with $1\leq p\leq d-2$, 
for some appropriate internal vector space $\vs$. 
Furthermore, these field variables can be chosen such that
the field equations have a zero potential as a solution,
and also such that the gauge symmetries when linearized around this solution
have the form of a set of $\U{1}$ gauge transformations 
involving an arbitrary vector-valued $p-1$-form as the parameter. 

For example, 
the gauge field variable in $d$-dimensional \YM/ theory is a connection 1-form 
represented by a $p=1$ potential $\A{a}{\mu}$
that takes values in the Lie algebra of the \YM/ gauge group
\eg/ $\vs \iso \SU{2}$. 
As discussed in the introduction, 
a linearization about a flat connection $\A{a}{\mu}=0$ yields 
the abelian \YM/ gauge symmetry, given by 
$\Xvar\A{a}{\mu}=\der{\mu}\X{a}$
for a Lie-algebra valued arbitrary function ($0$-form) $\X{a}$
on spacetime. 
\FT/ theory has a similar linearization, involving a $p=d-2$ potential 
$\B{a}{\mu_1\cdots\mu_{d-2}}$. 
As another example, 
$d$-dimensional gravitational theories are accommodated by 
using a vielbein (frame) formulation,
with the gauge field variable 
chosen to be the deviation from a flat vielbein
associated with a Minkowski space solution of 
the gravitational field equations, namely 
$\h{a}{\mu} = \e{a}{\mu}- \solder{a}{\mu}$
where $\e{a}{\mu}$ and $\solder{a}{\mu}$ are respectively 
the curved and flat spacetime vielbeins. 
Thus, 
$\h{a}{\mu}$ is a $p=1$ potential 
with values in a $d$-dimensional internal Minkowski space 
$\vs \iso (\Rnum^d,\flatmetric{ab})$. 
Note the gauge symmetry on the vielbein $\e{a}{\mu}$ 
corresponding to diffeomorphism invariance is simply 
a Lie derivative with respect to an arbitrary vector field $\X{\mu}$
on spacetime. 
From the discussion in the introduction, 
a linearization around flat spacetime $\h{a}{\mu}=0$ 
yields an abelian graviton gauge symmetry that takes the form of 
a linearized Lie derivative 
$\Xvar\h{a}{\mu}= \der{\mu}\X{a}$
where $\X{a} =\X{\mu}\solder{a}{\mu}$ 
is a Minkowski-space valued arbitrary function ($0$-form) on spacetime,
representing the components of 
the vector field $\X{\mu}$ in the flat vielbein. 
This gauge invariance in the linearized theory is of 
the same differential type as in linearized \YM/ theory. 
The only essential difference in the field variables of 
these two linear abelian theories 
is the presence of an auxiliary (extra) gauge freedom coming from 
linearization of the local Lorentz rotation transformations 
on the vielbein $\e{a}{\mu}$,
which yields 
$\delta_\chi\h{a}{\mu}
=\invflatmetric{ac}\solder{b}{\mu}\covector{\chi}{bc}$
where $\covector{\chi}{bc}=-\covector{\chi}{cb}$ is 
an arbitrary antisymmetric matrix function on $M$. 

For generalizations such as multi-graviton theories, 
the vector space $\vs$ is just enlarged \cite{Annalspaper,CQGpaper}
by a tensor product with some kind of internal algebra $\g$;
and for the gravity-like generalization of \YM/ theory, 
$\vs$ is any Lie subalgebra of the isometry group admitted by 
the spacetime metric \cite{Bra2}. 
In addition, supersymmetric gauge theories such as supergravity 
are accommodated by enlarging $\vs$ as a direct sum to include
an appropriate spinor space \cite{CQGpaper}. 

In general it will be convenient to write 
the gauge field variables and their derivatives 
in local coordinates on spacetime 
and assume the underlying spacetime manifold is equipped with 
a fixed metric $\metric{\mu\nu}$ and volume form $\vol{\ind{\mu}{d}}$
for which the coordinate derivative $\der{\alpha}$ is metric-compatible,
$\der{\alpha}\metric{\mu\nu}=0$, $\der{\alpha}\vol{\mu_1\cdots\mu_d}=0$.
As in all the preceding examples, 
for typical gauge theories the geometrical content is 
independent of this background spacetime structure. 

For the sequel, 
a multi-index notation will be used to denote 
collections of $k\geq 1$ spacetime indices 
$\mu_1\cdots\mu_k$ as $\ind{\mu}{k}$; 
index symmetrization and antisymmetrization will be denoted 
by $(\cdots)$ and $[\cdots]$ as usual. 

\subsection{ Linear abelian gauge theory }

So consider a linear abelian gauge theory 
on a $d$-dimensional spacetime manifold $M$
using  vector-valued $p$-form potentials $\pform{a}{\ind{\mu}{p}}$
as field variables whose differential gauge invariance is given by 
$\pform{a}{\ind{\mu}{p}} \rightarrow 
\pform{a}{\ind{\mu}{p}} + \der{[\mu_p}\pX{a}{\ind{\mu}{p-1}]}$
where $\pX{a}{\ind{\mu}{p-1}}$ is vector-valued arbitrary $p-1$-form on $M$. 
The corresponding abelian gauge symmetry for the theory is given by
the infinitesimal field variation 
\EQ\label{abeliangaugesymm}
\nthXvar{0}\pform{a}{\ind{\mu}{p}} = \der{[\mu_p}\pX{a}{\ind{\mu}{p-1}]} .
\doneEQ
The theory is specified by giving 
a quadratic Lagrangian top-form $\nthL{2}{\ind{\mu}{d}}$ 
that is locally constructed from 
the fields $\spform$ and their derivatives $\nthder{k}{}\spform$,
along with a fixed spacetime metric $\metric{\mu\nu}$
and volume form $\vol{\ind{\mu}{d}}$ on $M$,
an inner product $\k{ab}$ fixed on $\vs$,
as well as any extra structure which may be available on $M$ or $\vs$
(for example, a flat vielbein). 
Gauge invariance is expressed by the curl condition 
\EQ\label{abeliangaugeinv}
\nthXvar{0}\nthL{2}{\ind{\mu}{d}} = \der{[\mu_d}\nthtriv{1}{\ind{\mu}{d-1}]}
\doneEQ
for some locally constructed $d-1$-form $\nthtriv{1}{\ind{\mu}{d-1}}$. 
If $\sX$ is of compact support on $M$ then 
the action functional $\nthS{2}[\spform] = \int_M \nthL{2}{\ind{\mu}{d}}$
is gauge invariant since 
$\nthXvar{0}\nthS{2}[\spform] 
= \int_M \der{[\mu_d}\nthtriv{1}{\ind{\mu}{d-1}]} =0$
vanishes by Stokes' theorem. 
Equivalently, 
under the abelian gauge symmetry, 
the scalar Lagrangian 
$*\nthsL{2}= 
\frac{1}{d!} \invvol{\ind{\mu}{d}} \nthL{2}{\ind{\mu}{d}}$
varies into a divergence 
$\nthXvar{0}{*\nthsL{2}} = \frac{1}{d} \der{\mu}\nthcotriv{1}{\mu}$
where 
$\nthcotriv{1}{\mu} = 
\frac{1}{(d-1)!} \invvol{\mu\ind{\mu}{d-1}} \nthtriv{1}{\ind{\mu}{d-1}}$.

The stationary points of the action functional 
yield the linear field equation $\nthE{1}{\ind{\mu}{p}}{a}=0$ 
on $\spform$. 
Gauge invariance of the theory implies this field equation
depends on $\spform$ only through the abelian field strength $p+1$-forms
\EQ\label{linearfieldstr}
\pcurl{a}{\ind{\mu}{p+1}} = \der{[\mu_{p+1}} \pform{a}{\ind{\mu}{p}]}
\doneEQ
and their derivatives.

In typical theories, 
the linear field equation is at most second order in derivatives
and thus has the form 
\EQ\label{linearfieldeq}
\nthE{1}{\ind{\mu}{p}}{a}= 
\tensor{q}{\ind{\mu}{p}\ind{\nu}{p+1}\alpha}{ab} 
\der{\alpha} \pcurl{b}{\ind{\nu}{p+1}}
+ \tensor{p}{\ind{\mu}{p}\ind{\nu}{p+1}}{ab} 
\pcurl{b}{\ind{\nu}{p+1}}
\doneEQ
where the coefficients $q,p$ depend only on the structure available on 
$M$ and $\vs$. 
Correspondingly, 
the Lagrangian $\nthL{2}{\ind{\mu}{d}}$ is at most first order in derivatives,
to within a curl. 

\subsection{ Preliminaries }

A precise mathematical setting for 
writing down deformation terms 
and analyzing the deformation determining equations 
is provided by the field configuration space 
$\fieldsp=\{\pform{a}{\ind{\mu}{p}}(x)\}$, 
defined as the set of all smooth sections of 
the vector bundle of $\vs$-valued $p$-forms on $M$. 
Associated with $\fieldsp$ is the jet space formally defined using 
local coordinates 
\EQ
\jetsp = 
(x,\pform{a}{\ind{\mu}{p}},\der{\nu_1}\pform{a}{\ind{\mu}{p}},
\der{\nu_1}\der{\nu_2}\pform{a}{\ind{\mu}{p}},\ldots)
\doneEQ
where $x$ represents a point in $M$; 
$\pform{a}{\ind{\mu}{p}}$ represents the value of 
the $p$-form field $\spform(x)$ at $x$;
$\nthder{k}{\ind{\nu}{k}}\pform{a}{\ind{\mu}{p}}$ 
represents the values of the $k$th order derivatives of the $p$-form field 
$\nthder{k}{}\spform(x)$ at $x$
(in multi-index notation). 
Note the action of derivatives on these coordinates in $\jetsp$
is given by the total derivative operator 
\EQ
\der{\nu} = 
\Parder{}{\x{\nu}} 
+ \sum_{k\geq 0} \der{\nu}\nthder{k}{\ind{\nu}{k}}\pform{a}{\ind{\mu}{p}}
\Parder{}{\nthder{k}{\ind{\nu}{k}}\pform{a}{\ind{\mu}{p}}}
\doneEQ
where $\nthder{k}{}\spform$ for $k=0$ stands for $\spform$. 
This setting makes clear what will be meant by 
locally constructed deformation terms,
and it allows the use of tools of variational calculus 
\cite{Olv,And,AMSpaper}
that will be relevant for formulating gauge symmetries, field equations, 
and the condition of gauge invariance in a general deformation theory.

Geometrically, a field variation $\delta\pform{a}{\ind{\mu}{p}}$ 
is a vector field on $\fieldsp$. 
Corresponding vector fields on the jet space $\jetsp$
that involve no motion on the spacetime coordinates $\x{\mu}$
and that preserve the derivative relations among the field coordinates
represent field variations that are locally constructed from 
the fields and their derivatives
as well as from any structure available on $M$ and $\vs$. 
Given such a field variation, 
there is a natural Lie derivative operator $\Lie{\delta\spform}$
defined as follows. 
On scalar functions $f$ on $\jetsp$, 
it acts as a total variation 
\EQ
\Lie{\delta\spform} f = \delta f 
= 
\sum_{k\geq 0} \Parder{f}{\nthder{k}{\ind{\nu}{k}}\pform{a}{\ind{\mu}{p}}}
\nthder{k}{\ind{\nu}{k}} \delta\pform{a}{\ind{\mu}{p}} .
\doneEQ
For covector functions $\tensor{f}{\ind{\mu}{p}}{a}$ on $\jetsp$,
the Lie derivative action is given by 
\EQ
\Lie{\delta\spform} \tensor{f}{\ind{\mu}{p}}{a} 
= 
\sum_{k\geq 0} \bigg(
\Parder{ \tensor{f}{\ind{\mu}{p}}{a} }
{ \nthder{k}{\ind{\alpha}{k}}\pform{b}{\ind{\nu}{p}} }
\nthder{k}{\ind{\alpha}{k}} \delta\pform{b}{\ind{\nu}{p}}
+(-1)^k \nthder{k}{\ind{\alpha}{k}}(
\Parder{ \delta\pform{b}{\ind{\nu}{p}} }
{ \nthder{k}{\ind{\alpha}{k}} \pform{\ind{\mu}{p}}{a} } 
\tensor{f}{\ind{\nu}{p}}{b} )
\bigg) .
\doneEQ
This action extends via the Leibniz rule 
to vector functions and, more generally, 
any tensor functions on $\jetsp$. 
The Lie derivative operator on scalar and covector functions
will be central here to the formulation of 
deformation determining equations. 

There are two useful identities that hold for the Lie derivative operator. 
Firstly, 
Lie derivatives with respect to any locally constructed field variations
$\delta_1\spform$ and $\delta_2\spform$ 
satisfy the familiar commutation relation 
\EQ
[\Lie{\delta_1\spform},\Lie{\delta_2\spform}] 
= \Lie{[\delta_1\spform,\delta_2\spform]}
\doneEQ
where $[\delta_1\spform,\delta_2\spform] 
= \delta_{\delta_1\spform} \delta_2\spform -\exchange{1}{2}$
defines the commutator of the field variations. 
Secondly, 
the Jacobi relation is satisfied, 
$[\Lie{\delta_1\spform}, [\Lie{\delta_2\spform},\Lie{\delta_3\spform}] ]
-\eqtext{ cyclic terms }
=0$, 
since for any three locally constructed field variations, 
\EQ
[\delta_1\spform, [\delta_2\spform,\delta_3\spform] ] 
-\eqtext{ cyclic terms }
=0
\doneEQ
holds identically. 
These properties are direct consequences of the representation of
field variations as vector fields on jet space. 

Another important variational operator will be the \EL/ operator 
acting on scalar functions $f$ on $\jetsp$ by 
\EQ
\ELop{\pform{a}{\ind{\mu}{p}}}(f) 
=
\sum_{k\geq 0} (-1)^k \nthder{k}{\ind{\nu}{k}}
\Parder{f}{\nthder{k}{\ind{\nu}{k}}\pform{a}{\ind{\mu}{p}}} .
\doneEQ
This operator takes functions $f$ 
into covector functions 
$\tensor{f}{\ind{\mu}{p}}{a}= \ELop{\pform{a}{\ind{\mu}{p}}}(f)$
and has the property that it annihilates a locally constructed function $f$ 
if and only if on $\jetsp$ 
the dual top-form function is a curl, 
$*\covector{f}{\ind{\mu}{d}} = \der{[\mu_d} \triv{\ind{\mu}{d-1}]}$,
for some locally constructed $d-1$-form function $\triv{\ind{\mu}{d-1}}$. 

The \EL/ and Lie derivative operators are related 
through integration by parts.
In particular, 
for scalar functions $f$ on $\jetsp$, 
repeated integration by parts on a Lie derivative 
with respect to any locally constructed vector field 
$\delta\spform$ on $\jetsp$ 
yields
\EQ
\Lie{\delta\spform} f 
= \delta\pform{a}{\ind{\nu}{p}} \tensor{f}{\ind{\nu}{p}}{a}
+\der{\mu}\ibp{\mu}(f;\delta\spform) ,\quad
\tensor{f}{\ind{\nu}{p}}{a} =\ELop{\pform{a}{\ind{\nu}{p}}}(f) 
\label{ibpid}
\doneEQ
where
\EQ
\ibp{\mu}(f;\delta\spform)
=\sum_{k\geq 0}( \nthder{k}{\ind{\alpha}{k}} \delta\pform{a}{\ind{\nu}{p}} )
\ELop{\nthder{k+1}{\mu\ind{\alpha}{k}} \pform{a}{\ind{\nu}{p}}}(f) .
\label{ibpterm}
\doneEQ

\subsection{ Determining equations for nonlinear deformations }

A deformation of a linear abelian gauge theory for fields $\spform$
consists of adding linear and higher power terms 
to the abelian gauge symmetry \eqref{abeliangaugesymm},
\EQ\label{deformgaugesymm}
\Xvar\pform{a}{\ind{\mu}{p}} 
= 
\nthXvar{0}\pform{a}{\ind{\mu}{p}} 
+ \nthXvar{1}\pform{a}{\ind{\mu}{p}} +\cdots
\doneEQ
while simultaneously adding quadratic and higher power terms 
to the linear field equation \eqref{linearfieldeq},
\EQ\label{deformfieldeq}
\E{\ind{\mu}{p}}{a}
= 
\nthE{1}{\ind{\mu}{p}}{a}
+ \nthE{2}{\ind{\mu}{p}}{a} +\cdots
\doneEQ
such that there exists a top-form Lagrangian 
\EQ\label{deformL}
\L{\ind{\mu}{d}} 
=
\nthL{2}{\ind{\mu}{d}} + \nthL{3}{\ind{\mu}{d}} + \cdots
\doneEQ
required to be gauge invariant to within a curl. 
Here the deformation terms in the field equation and gauge symmetry
are to be locally constructed from powers of 
the fields $\spform$ and their derivatives $\nthder{k}{}\spform$
up to some finite differential order,
with coefficients allowed to depend on 
the spacetime coordinates $\x{\mu}$, 
the spacetime metric $\metric{\mu\nu}$ and volume form $\vol{\ind{\mu}{d}}$,
the internal vector-space inner product $\k{ab}$,
and any other available structure on $M$ or $\vs$. 
As well, the deformation terms in the gauge symmetry are to be linear in
the gauge parameter $\sX$ and its derivatives $\nthder{j}{}\sX$
to a finite differential order. 
Thus, the expressions for 
$\Xvar\pform{a}{\ind{\mu}{p}}$ and $\E{\ind{\mu}{p}}{a}$
will be, respectively, vector and covector functions on $\jetsp$. 
Natural restrictions on the order of derivatives 
appearing in these expressions will be addressed shortly
(see proposition~2.4).

The condition of gauge invariance is stated by 
\EQ
\Xvar\L{\ind{\mu}{d}} 
=
\nthXvar{0}\nthL{2}{\ind{\mu}{d}} 
+\nthXvar{1}\nthL{2}{\ind{\mu}{d}} 
+\nthXvar{0}\nthL{3}{\ind{\mu}{d}} + \cdots
= \der{[\mu_d}\nthtriv{1}{\ind{\mu}{d-1}]}
\label{gaugeinv}
\doneEQ
holding for some $d-1$-form function $\nthtriv{1}{\ind{\mu}{d-1}}$
on $\jetsp$, 
where the Lagrangian \eqref{deformL} 
is related to the field equation \eqref{deformfieldeq} through 
the \EL/ operator,
\EQ
\ELop{\pform{a}{\ind{\mu}{p}}}(*\nthsL{k+1}) 
= \nthE{k}{\ind{\mu}{p}}{a} .
\label{LEeq}
\doneEQ
Necessary and sufficient conditions 
for $\nthE{k}{}{}$ to arise as an \EL/ equation are that 
the Frechet derivative of $\nthE{k}{}{}$ must be a self-adjoint operator
\cite{Olv,AncBlu}.
These conditions can be shown to determine
\EQ
\nthL{k+1}{\ind{\mu}{d}} 
= 
-\frac{1}{k+1} \vol{\ind{\mu}{d}}
\nthE{k}{\ind{\nu}{p}}{b} \pform{b}{\ind{\nu}{p}} 
\doneEQ
to within a curl. 
In terms of the \EL/ operator, 
the condition of gauge invariance is expressed 
through the integration by parts identity \eqref{ibpid} by the equation
\EQ\label{deteq}
\ELop{\pform{a}{\ind{\mu}{p}}}( 
\E{\ind{\nu}{p}}{b} \Xvar\pform{b}{\ind{\nu}{p}} ) =0
\doneEQ
on $\jetsp$.
This constitutes the determining equation
for all allowed deformations of the linear abelian gauge theory
\eqrefs{abeliangaugesymm}{linearfieldeq}. 

The deformation determining equation \eqref{deteq} 
can be reformulated more usefully and geometrically 
as Lie derivative equations,
as shown by results in \cite{AMSpaper}.

{\bf Theorem~2.1}:
Gauge invariance holds iff the Lie derivative of the field equation
with respect to the gauge symmetry vanishes
\EQ\label{liedereq}
\Lie{\Xvar\spform} \E{\ind{\mu}{p}}{a} =0 .
\doneEQ

This equation asserts that the gauge symmetry for each parameter 
is a vector field tangent to the surface in $\jetsp$ 
corresponding to the field equation
(and all its derivatives),
$\jetsp\pmod{\E{}{}=0} \equiv \jetspE \subset \jetsp$.
Due to invariance of the action functional, 
the commutators of these vector fields for all parameters 
have the same tangency property. 

{\bf Theorem~2.2}:
Gauge invariance holds only if the Lie derivative of the field equation
with respect to the gauge symmetry commutators vanishes
\EQ\label{liedercommeq}
\Lie{[\Xvarsub{1}\spform,\Xvarsub{2}\spform]} \E{\ind{\mu}{p}}{a} =0 .
\doneEQ

It is important to remark that here
no conditions are assumed or required on the possibilities 
allowed for the form of the commutators of the deformed gauge symmetries. 
Nevertheless, 
when the gauge symmetries are restricted to the solution space of
the deformed field equations, 
closure of the commutators will be seen to arise order by order, 
stemming from the fact that the abelian gauge symmetries 
generate all of the differential gauge invariance 
present in the solution space of the linear field equations
(\ie/ modulo any auxiliary gauge freedom).
Any deformation therefore will automatically determine
an associated infinitesimal gauge group structure. 
As will be stated in precise form later
(see also \Ref{StaFulLad2}), 
the commutators characterizing this group structure 
may involve local structure functions that depend on the fields $\spform$
(and their derivatives) 
and may fail to close other than when the fields satisfy $\E{}{}=0$.

A related remark is that 
gauge invariance as expressed by the Lie derivative equation \eqref{liedereq}
implies the deformed gauge symmetry $\Xvar\spform$ formally generates 
an infinitesimal symmetry group of the deformed field equation,
since
\EQ
\Xvar\E{\ind{\mu}{p}}{a} =0
\eqtext{ when $\spform$ satisfies $\E{}{}=0$ } .
\label{symmdeteq}
\doneEQ
Geometrically, this equation is precisely the condition 
for the associated vector field $\Xvar\spform$ in $\jetsp$ 
to lie tangent to the solution space surface $\jetspE$.
(This general notion of symmetries 
defined as locally constructed tangent vector fields to $\jetspE$ 
comprises classical point symmetries
as well as generalized or higher order symmetries
of a field equation $\E{}{}=0$; 
see \Refs{Olv,AncBlu,book}.)
However, it is worthwhile to emphasize here that 
the symmetry determining equation \eqref{symmdeteq} is strictly weaker
than the Lie derivative equation \eqref{liedereq},
as not every symmetry of an \EL/ field equation will necessarily generate 
an invariance of its action functional;
for example, symmetries that scale the coordinates of $\jetsp$ 
typically leave invariant the Lagrangian only in certain spacetime dimensions.

There is a related formulation of gauge invariance
from this point of view, which is also useful.
Introduce the gauge parameter space 
$\parmsp =\{\pX{a}{\ind{\mu}{p-1}(x)}\}$
defined as the vector bundle of all smooth $\vs$-valued $p-1$-forms on $M$.
The gauge symmetry \eqref{deformgaugesymm} then can be viewed as 
a linear differential operator
\EQ
\Xvar\pform{a}{\ind{\mu}{p}} = \D{\spform}(\sX)\mixedindices{a}{\ind{\mu}{p}}
\label{Xcovder}
\doneEQ
from $\parmsp$ to $\fieldsp$ that is locally constructed from 
$\spform$, derivatives of $\spform$, 
and any structure available on $M$ and $\vs$.
Note via integration by parts, 
this operator $\D{\spform}$ has a formal adjoint $\adD{\spform}$
that acts as a linear differential operator
from $\fieldsp^*$ into $\parmsp^*$ 
(\ie/ the dual vector bundles of $\fieldsp$ and $\parmsp$).
Now, an application of the \EL/ operator with respect to
$\pX{b}{\ind{\nu}{p-1}}$
in the statement of gauge invariance \eqref{gaugeinv} gives
a Noether divergence identity
\EQ
\adD{\spform}(\E{}{})\mixedindices{\ind{\nu}{p-1}}{b} =0 .
\label{covderXid}
\doneEQ
Conversely, contraction of $\pX{b}{\ind{\nu}{p-1}}$
onto this identity followed by use of the \EL/ relation \eqref{LEeq}
along with repeated integration by parts 
then gives back equation \eqref{gaugeinv}.

{\bf Proposition~2.3}:
Gauge invariance holds iff the field equation satisfies 
the Noether identity \eqref{covderXid}
derived from the gauge symmetry. 

Now to proceed,
an expansion of the Lie derivative equations \eqrefs{liedereq}{liedercommeq}
in powers of the field coordinates in $\jetsp$
gives a hierarchy of determining equations whose solutions yield 
all allowed deformation terms for the field equation and gauge symmetry. 
Before looking at this hierarchy, 
it is important to consider 
the order of derivatives on $\spform$ and $\sX$
allowed in the deformation terms. 

First, a natural requirement is that the number of 
dynamical degrees of freedom of the fields $\spform$
in the linear abelian theory should be preserved order by order
in a nonlinear deformation. 
Otherwise, severe consistency conditions could arise on solutions of 
the field equation, and the deformed theory would not be physically or
mathematically satisfactory.
(These degrees of freedom can be identified as
the number of initial-data functions modulo the gauge symmetry freedom,
provided that the linear field equation is a well-posed system of PDEs
after suitable gauge constraints have been imposed on the fields.)
With this condition, 
for the typical situation in which
the linear field equation is a system of second order PDEs, 
the most general possible form then allowed for deformed field equations
will be a quasilinear second order PDE system, namely, 
highest derivatives are of second order
while the coefficients of these terms depend on 
no higher than first order derivatives, 
\EQ
\nthE{k}{\ind{\mu}{p}}{a} 
= 
\nthterm{k-1}{Q}{\ind{\mu}{p}\ind{\nu}{p}\alpha\beta}{ab}
\der{\alpha}\der{\beta}\pform{b}{\ind{\nu}{p}}
+ \nthterm{k-1}{P}{\ind{\mu}{p}\ind{\nu}{p}\alpha}{ab}
\der{\alpha}\pform{b}{\ind{\nu}{p}}
+ \nthterm{k-1}{R}{\ind{\mu}{p}\ind{\nu}{p}}{ab}
\pform{b}{\ind{\nu}{p}} .
\label{kthEform}
\doneEQ
Here, for later notational ease, 
the lower derivative terms have been written in a factored form 
where 
$\nthterm{k-1}{R}{}{}$ contains no derivatives of $\spform$
and $\nthterm{k-1}{P}{}{}$ contains no higher than 
first derivatives of $\spform$. 
Note that the form of the linear field equation gives
\EQ
\nthterm{0}{Q}{\ind{\mu}{p}\ind{\nu}{p}\nu_{p+1}\alpha}{ab}
= \tensor{q}{\ind{\mu}{p}\ind{\nu}{p+1}\alpha}{ab} ,\quad
\nthterm{0}{P}{\ind{\mu}{p}\ind{\nu}{p}\nu_{p+1}}{ab}
= \tensor{p}{\ind{\mu}{p}\ind{\nu}{p+1}}{ab} ,\quad
\nthterm{0}{R}{\ind{\mu}{p}\ind{\nu}{p}}{ab} 
=0 .
\doneEQ

Next, as a consequence of the quasilinear second-order derivative form 
sought for the deformed field equations,
a bound arises on the order of derivatives allowed 
in the deformed gauge symmetry 
through the Noether divergence identity \eqref{covderXid},
as follows.
The deformed gauge symmetry has the form of a linear differential operator 
on $\sX$,
with coefficients
$\term{U}{a\ind{\nu}{p-1}}{\ind{\mu}{p}b}$ of $\pX{b}{\ind{\nu}{p-1}}$
and $\term{V}{a\ind{\nu}{p-1}\ind{\alpha}{j}}{\ind{\mu}{p}b}$
of $\nthder{j}{\ind{\alpha}{j}}\pX{b}{\ind{\nu}{p-1}}$
for $1\leq j$, up to some highest order $j_{\rm max}$ in derivatives. 
Its associated Noether identity on the deformed field equation 
is then given by
\EQ
0=\adD{\spform}(\E{}{})\mixedindices{\ind{\mu}{p-1}}{a}
= 
\E{\ind{\nu}{p}}{b} \term{U}{b\ind{\mu}{p-1}}{\ind{\nu}{p}a}
+  \sum_{1\leq j\leq j_{\rm max}} (-1)^j \nthder{j}{\ind{\alpha}{j}}( 
\E{\ind{\nu}{p}}{b} \term{V}{b\ind{\mu}{p-1}\ind{\alpha}{j}}{\ind{\nu}{p}a} ) .
\doneEQ
When this identity is expanded in powers of 
the field coordinates in $\jetsp$, 
taking into account the form of the abelian gauge symmetry, 
it gives divergence equations of the form 
\EQ
\der{\mu_p} \nthE{k}{\ind{\mu}{p}}{a} 
= 
\sum_{1\leq i\leq k-1} \bigg(
\nthE{i}{\ind{\nu}{p}}{b} \nthterm{k-i}{U}{b\ind{\mu}{p-1}}{\ind{\nu}{p}a}
+ \sum_{1\leq j\leq j_{\rm max}} (-1)^j \nthder{j}{\ind{\alpha}{j}}( 
\nthE{i}{\ind{\nu}{p}}{b} 
\nthterm{k-i}{V}{b\ind{\mu}{p-1}\ind{\alpha}{j}}{\ind{\nu}{p}a} )
\bigg)
\label{diveqs}
\doneEQ
for each $k\geq 2$.
In general, as $k$ increases, 
these equations imply an unbounded escalation in the order of derivatives
on $\spform$ in the deformation terms $\nthE{k}{}{}$
compared with $\nthE{1}{}{}$, 
unless $\nthterm{k}{V}{}{}$ vanishes for $j\geq 2$. 
Moreover, for a balance to hold in the order of derivatives appearing
on both sides of the divergence equations
under the condition that $\nthE{k}{}{}$ 
is quasilinear second order in derivatives of $\spform$, 
$\nthterm{k}{V}{}{}$ for $j=1$ should contain 
no second or higher order derivatives of $\spform$
while $\nthterm{k}{U}{}{}$ should contain 
second derivatives of $\spform$ at most linearly. 

Therefore, given a general quasilinear second-order derivative form 
for the deformed field equation, 
the most general possible form allowed for the deformed gauge symmetry 
will be a first-order linear differential operator on $\sX$
\EQ
\nthXvar{k}\pform{a}{\ind{\mu}{p}} 
=
\nthterm{k}{U}{a\ind{\nu}{p-1}}{\ind{\mu}{p}b} \pX{b}{\ind{\nu}{p-1}}
+ \nthterm{k}{V}{a\ind{\nu}{p-1}\alpha}{\ind{\mu}{p}b} 
\der{\alpha}\pX{b}{\ind{\nu}{p-1}}
\label{kthUVform}
\doneEQ
whose coefficients $\nthterm{k}{V}{}{}$ and $\nthterm{k}{U}{}{}$
respectively have no dependence on higher than 
first and second order derivatives of $\spform$,
while $\nthterm{k}{U}{}{}$ is at most linear 
in second derivatives of $\spform$.
Note, from the form of the abelian gauge symmetry, 
the coefficients at lowest order are given by 
\EQ
\nthterm{0}{U}{a\ind{\nu}{p-1}}{\ind{\mu}{p}b} 
=0 ,\quad
\nthterm{0}{V}{a\ind{\nu}{p-1}\alpha}{\ind{\mu}{p}b} 
= \id{b}{a} \id{\ind{\mu}{p}}{\ind{\nu}{p-1}\alpha} .
\label{0thUV}
\doneEQ

However, the divergence equations \eqref{diveqs} show that, 
with $j_{\rm max}=1$, 
any dependence on first derivatives of $\spform$ in $\nthterm{k}{V}{}{}$ 
or second derivatives of $\spform$ in $\nthterm{k}{U}{}{}$ 
will lead to an escalating total number of derivatives in $\nthE{k}{}{}$
as $k$ increases \cite{Wal,Annalspaper},
in which case the deformed field equation will be non-polynomial 
in first (lowest order) derivatives of $\spform$.
In particular, 
count one derivative as a first derivative appearing linearly,
two derivatives as a first derivative appearing quadratically
or a second derivative appearing linearly,
and so on. 
Now, if $\nthE{k}{}{}$ is to be only polynomial in first derivatives
(and at most linear in second derivatives),
then clearly the number of derivatives 
on both sides of 
\EQ
\der{\alpha}( \E{\ind{\nu}{p}}{b} 
\term{V}{b\ind{\mu}{p-1}\alpha}{\ind{\nu}{p}a} )
=
\E{\ind{\nu}{p}}{b} \term{U}{b\ind{\mu}{p-1}}{\ind{\nu}{p}a}
\label{noetherid}
\doneEQ
will balance only if 
$\nthterm{k}{V}{}{}$ contains no derivatives of $\spform$
and $\nthterm{k}{U}{}{}$ contains first order derivatives of $\spform$
at most linearly and no second order derivatives of $\spform$,
for every order $k\geq 1$. 
Furthermore, at each successive order $k$, 
this balance in the number of derivatives will imply $\nthE{k}{}{}$
should only contain the same number of derivatives 
as contained in $\nthE{1}{}{}$. 

{\bf Proposition~2.4}:
With a polynomial restriction on lowest-order derivatives of $\spform$ 
in the quasilinear form \eqref{kthEform} 
sought for the deformed field equation,
the most general compatible form for the deformed gauge symmetry
will be linear in first derivatives of both $\spform$ and $\sX$. 
The deformed field equation, in turn, must be only of 
a semilinear second-order derivative form,
with no derivatives of $\spform$ allowed to appear in the coefficient of
the second derivative (highest order) term
and at most quadratic dependence on derivatives of $\spform$ allowed for
the first derivative (lowest order) term. 

Next, the Noether divergence identity \eqref{noetherid}
leads to an interesting relation 
connecting the gauge symmetry and the field equation.
Associated with the deformed gauge symmetry is a rigid symmetry 
defined by the field variation
\EQ
\Xrigidvar\pform{a}{\ind{\mu}{p}}
= \term{U}{a\ind{\nu}{p-1}}{\ind{\mu}{p}b} 
\pX{b}{\ind{\nu}{p-1}}\rigid ,\quad
\sX\rigid =\const 
\doneEQ
Due to gauge invariance, 
this field variation leaves the deformed Lagrangian invariant to within a curl,
\EQ
\Xrigidvar\L{\ind{\mu}{d}} = 
\der{[\mu_d}\triv{\ind{\mu}{d-1}]} , 
\doneEQ
and hence it generates a conserved current of the deformed field equation 
by Noether's theorem \cite{Olv,AncBlu}.
The current arises from the identity
\EQs
\Lie{\Xrigidvar\spform}(*\sL) &&
= \Xrigidvar{*\sL} = \frac{1}{d} \der{\mu}\cotriv{\mu}
\nonumber\\&&
= \E{\ind{\mu}{p}}{a} \Xrigidvar\pform{a}{\ind{\mu}{p}} 
+ \der{\mu} \ibp{\mu}
\doneEQs
where $\ibp{\mu}$ is the integration by parts term \eqref{ibpterm}
relating the Lie derivative and \EL/ operators. 
This yields
\EQ
\der{\mu}\current{\mu}{}(\sX\rigid)
= \E{\ind{\mu}{p}}{a} \Xrigidvar\pform{a}{\ind{\mu}{p}} ,\quad
\current{\mu}{}(\sX\rigid)= \frac{1}{d} \cotriv{\mu} -\ibp{\mu}
= \current{\mu\ind{\nu}{p-1}}{b} \pX{b}{\ind{\nu}{p-1}} , 
\label{conscurr}
\doneEQ
from which the Noether current $\current{}{}$ is seen to be conserved
on the solution space, $\E{}{}=0$.
Now, a comparison of the conservation equation \eqref{conscurr}
with the divergence identity \eqref{noetherid} reveals the relation
$\tensor{J'}{\mu}{}(\sX\rigid) 
= \tensor{J'}{\mu\ind{\nu}{p-1}}{b} \pX{b}{\ind{\nu}{p-1}}
= \E{\ind{\alpha}{p}}{a} \term{V}{a\ind{\nu}{p-1}\mu}{\ind{\alpha}{p}b}
\pX{b}{\ind{\nu}{p-1}}\rigid$
where the factorized current $\tensor{J'}{}{}$ is equivalent to 
$\current{}{}$ 
to within a locally constructed, identically divergence-free expression.

{\bf Proposition~2.5}: 
The deformed field equation can be written in terms of 
a rigid Noether current
\EQ
\E{\ind{\alpha}{p}}{a} 
= \term{V^{-1}}{\ind{\alpha}{p}b}{a\ind{\nu}{p-1}\mu}
\tensor{J'}{\mu\ind{\nu}{p-1}}{b}
\doneEQ
which is associated with the deformed gauge symmetry
\EQ
\der{\mu} \tensor{J'}{\mu\ind{\nu}{p-1}}{b} 
= \E{\ind{\alpha}{p}}{a} \term{U}{a\ind{\nu}{p-1}}{\ind{\alpha}{p}b}
\doneEQ
where $\term{V^{-1}}{}{}$ denotes the formal left-inverse of $\term{V}{}{}$
defined in powers of the field coordinates through equation \eqref{0thUV}.

Finally, 
deformations that are related by a locally constructed invertible
change of field variables
\EQ
\pform{a}{\ind{\mu}{p}} \rightarrow \tensor{\spform'}{a}{\ind{\mu}{p}}
= 
\pform{a}{\ind{\mu}{p}} 
+ \ontop{(2)}{\spform'}\mixedindices{a}{\ind{\mu}{p}} 
+\cdots
\label{fieldredef}
\doneEQ
or of gauge parameters
\EQ
\pX{a}{\ind{\mu}{p-1}} \rightarrow \tensor{\sX'}{a}{\ind{\mu}{p-1}}
= 
\pX{a}{\ind{\mu}{p-1}} 
+ \ontop{(1)}{\sX'}\mixedindices{a}{\ind{\mu}{p-1}} 
+\cdots
\label{parmredef}
\doneEQ
will be regarded as equivalent. 
In order to stay within the general form given for deformation terms
\eqrefs{kthEform}{kthUVform}, 
these transformation terms 
$\ontop{(k)}{\spform'}\mixedindices{a}{\ind{\mu}{p}}$
and $\ontop{(k-1)}{\sX'}\mixedindices{a}{\ind{\mu}{p-1}}$
for $k\geq 2$ must contain no derivatives of $\spform$, 
and $\ontop{(k-1)}{\sX'}\mixedindices{a}{\ind{\mu}{p-1}}$ must also 
depend linearly on $\sX$ (and contain no derivatives of $\sX$). 

\subsection{ Solving the determining equations }

Steps will now be outlined for how to solve 
for the deformation terms in an efficient, explicit manner
from the hierarchy of determining equations given by theorems~2.1 and~2.2. 
In this hierarchy
\EQs
0th: && 
\Lie{\nthXvar{0}\spform} \nthE{1}{}{} =0 ,
\label{0thdeteq}
\\&&
\Lie{\nthXvarsub{0}{1}\nthXvarsub{1}{2}\spform} \nthE{1}{}{} 
-\exchange{1}{2} =0 ,
\label{0thcommdeteq}
\\
1st: &&
\Lie{\nthXvar{0}\spform} \nthE{2}{}{} 
+ \Lie{\nthXvar{1}\spform} \nthE{1}{}{} =0 ,
\label{1stdeteq}
\\&&
\Lie{\nthXvarsub{0}{1}\nthXvarsub{1}{2}\spform} \nthE{2}{}{} 
+ \Lie{\nthXvarsub{0}{1}\nthXvarsub{2}{2}\spform} \nthE{1}{}{} 
-\exchange{1}{2} 
+ \Lie{[\nthXvarsub{1}{1},\nthXvarsub{1}{2}]\spform} \nthE{1}{}{} 
=0 ,
\label{1stcommdeteq}
\\
2nd: && \cdots
\nonumber\\ &&\nonumber\\
\vdots &&
\nonumber
\doneEQs
observe that the unknowns $\nthXvar{k}\spform$ and $\nthE{k+1}{}{}$
for $k\geq 1$ are partly decoupled since they first enter the equations 
at $(k-1)$th and $k$th orders, respectively. 
Also, notice the bottom equation \eqref{0thdeteq} holds 
as a gauge-invariance identity 
on $\nthXvar{0}\spform$ and $\nthE{1}{}{}$
from the given linear abelian theory. 

At each order $k\geq 1$, 
an important feature of the hierarchy is that 
the unknowns $\nthXvar{k}\spform$ and $\nthE{k+1}{}{}$
occur only through variations 
$\nthXvarsub{0}{2}(\nthXvarsub{k}{1}\spform)$
and $\nthXvar{0}\nthE{k+1}{}{}$
involving the undeformed gauge symmetry $\nthXvar{0}\spform$. 
Such variational expressions can be annihilated in two different ways,
leading to \ic/s in solving for 
the quadratic and higher power deformation terms. 
One \ic/ is that 
an antisymmetrized variation with respect to the undeformed gauge symmetry
vanishes, since by commutativity,
\EQ
\nthXvarsub{0}{2}( \nthXvarsub{0}{1}\nthE{k+1}{}{} )
-\exchange{1}{2}
= [\nthXvarsub{0}{2},\nthXvarsub{0}{1}]\nthE{k+1}{}{} 
=0 ,
\doneEQ
and similarly
\EQ
\nthXvarsub{0}{3}( \nthXvarsub{0}{2}(\nthXvarsub{k}{1}\spform) )
-\eqtext{ cyclic terms }
= [\nthXvarsub{0}{3},\nthXvarsub{0}{2}] \nthXvarsub{k}{1}\spform 
-\exchange{1}{2} + \exchange{1}{3}
=0 .
\doneEQ
A second \ic/ arises because the undeformed gauge symmetry itself vanishes
if its parameter $\sX$ is taken to be rigid, \ie/ constant, 
so that $\nthXrigidvar{0}\spform=0$. 
Hence, by rigidity, 
\EQ
\nthXrigidvar{0}\nthE{k+1}{}{} =0
\doneEQ
and 
\EQ
\nthXrigidvarsub{0}{2}(\nthXrigidvarsub{k}{1}\spform) =0 .
\doneEQ
These \ic/s will be referred to as 
``commutativity-type'' and ``rigidity-type''. 
As a main result, 
because of the decoupled structure of the hierarchy,
the commutativity-type \ic/ will turn out to be trivial,
as at each order it will hold 
due to the Jacobi identity for the deformed gauge symmetry
and the closure of the associated gauge group, 
discussed earlier. 
In contrast, 
the rigidity-type \ic/ will be central to the deformation analysis.

Connected with the identity \eqref{0thdeteq} at the bottom of the hierarchy,
the property that the abelian gauge symmetry generates all of 
the differential gauge invariance in the linear field equation
leads to a very useful result. 

{\bf Lemma~2.6}:
For any locally constructed field variation
$\nthvar{k}{}\spform(\sX)$ depending linearly on 
at least one arbitrary gauge parameter $\sX$, 
the following three conditions are equivalent
(provided auxiliary gauge freedom is excluded):
(i) It has the form of an undeformed gauge symmetry
\EQ
\nthvar{k}{}\pform{a}{\ind{\mu}{p}}(\sX) 
= \der{[\mu_p} \ontop{(k)}{\sX'}\mixedindices{a}{\ind{\mu}{p-1}]}(\sX)
\doneEQ
whose parameter $\ontop{(k)}{\sX'}$ depends linearly on $\sX$
(and its derivatives). 
(ii) It is identically curl-free 
\EQ
\nthvar{k}{}\pcurl{a}{\ind{\mu}{p+1}} 
= \der{[\mu_{p+1}} \nthvar{k}{}\pform{a}{\ind{\mu}{p}]}(\sX) 
=0 .
\doneEQ
(iii) It identically satisfies the undeformed field equation
\EQ
\nthvar{k}{}\nthE{1}{\ind{\mu}{p}}{a} 
= \nthE{1}{\ind{\mu}{p}}{a}( \nthvar{k}{}\spform(\sX) )
=0 .
\doneEQ
Moreover, the same equivalences hold 
whenever $\spform$ satisfies $\nthE{1}{}{}(\spform)=0$. 

The proof $(i)\Rightarrow (ii)\Rightarrow (iii)$ is immediate,
since $\nthE{1}{}{}$ is gauge invariant under $\nthXvar{0}\spform$.
For the converse, $(iii)\Rightarrow (ii)$ is a consequence of 
$\nthXvar{0}\spform$ exhausting all the gauge invariance 
admitted by $\nthE{1}{}{}$
(under the assumption no auxiliary gauge freedom is present), 
while  $(ii)\Rightarrow (i)$ holds because on jet space
any closed $p$-form function is exact \cite{And}. 
These arguments are not difficult to extend to the solution space 
$\nthE{1}{}{}=0$ by means of the jet-space tensor techniques
from \Ref{Annalspaper}. 

This lemma will be a key tool in the analysis to follow.
In examples, the condition of no auxiliary gauge freedom 
is met by linear abelian \YM/ and \FT/ theories, 
including abelian \CS/ and torsion generalizations, 
and their extension to other $p$-form fields. 
The main examples of a linear abelian theory having auxiliary gauge freedom
are graviton theories. 
The simplest situation occurs when this auxiliary gauge freedom
does not enter into solutions of the determining equations 
for the deformed field equation and deformed gauge symmetry, 
order by order. 
For solutions of that type the deformation terms in the gauge symmetry
will satisfy the equivalences stated in the lemma.
This is what happens for the deformation of linear abelian graviton theory
corresponding to the Einstein gravitational theory 
and its multi-graviton (algebra-valued) generalizations, 
in $d\geq 3$ dimensions \cite{Annalspaper,Hen2}.
In contrast, 
the exotic parity-violating multi-graviton theories in $d=3,5$ dimensions
involve a more complicated situation where the auxiliary gauge freedom
from the linear theory is itself deformed nontrivially 
in the course of solving the determining equations \cite{PhysRevpaper}. 

So, as the simplest case, 
only those deformations obtained by postulating the lemma 
to hold independently of any auxiliary gauge freedom 
in solving the determining equations 
will be considered hereafter. 
Note this assumption is trivially met whenever 
the abelian gauge symmetry exhausts the gauge freedom present 
in the linear field equation. 

To begin the analysis, 
the first step is to solve 
the $0$th order Lie derivative commutator equation \eqref{0thcommdeteq}
for the linear deformation terms in the gauge symmetry
\EQ
\nthXvar{1}\pform{a}{\ind{\mu}{p}} 
=
\nthterm{1}{U}{a\ind{\nu}{p-1}}{\ind{\mu}{p}b} \pX{b}{\ind{\nu}{p-1}}
+ \nthterm{1}{V}{a\ind{\nu}{p-1}\alpha}{\ind{\mu}{p}b} 
\der{\alpha}\pX{b}{\ind{\nu}{p-1}} .
\doneEQ
Through lemma~2.6, 
equation \eqref{0thcommdeteq} is equivalent to 
\EQ
0 = 
\der{[\nu} \nthXvarsub{0}{2}\nthXvarsub{1}{1}\pform{a}{\ind{\mu}{p}]}
-\exchange{1}{2} ,
\label{0thclosureeq}
\doneEQ
which expands out to give 
\EQs
0 = &&
\der{[\nu|} \bigg(
\sum_{0\leq k\leq 2} 
\pXsub{1}{b}{\ind{\nu}{p-1}}
\nthder{k+1}{\ind{\alpha}{k}\beta_p} \pXsub{2}{c}{\ind{\beta}{p-1}}
\parder{ \nthterm{1}{U}{a\ind{\nu}{p-1}}{|\ind{\mu}{p}]b} }
{ \nthder{k}{\ind{\alpha}{k}}\pform{c}{\ind{\beta}{p}} }
\nonumber
\\&&\qquad
+ \sum_{0\leq k\leq 1} 
\der{\alpha}\pXsub{1}{b}{\ind{\nu}{p-1}}
\nthder{k+1}{\ind{\alpha}{k}\beta_p} \pXsub{2}{c}{\ind{\beta}{p-1}}
\parder{ \nthterm{1}{V}{a\ind{\nu}{p-1}\alpha}{|\ind{\mu}{p}]b} }
{ \nthder{k}{\ind{\alpha}{k}}\pform{c}{\ind{\beta}{p}} }
\bigg)
-\exchange{1}{2} .
\label{1stUVdeteq}
\doneEQs
Substitution into this equation 
using an explicit general linear form for 
$\nthterm{1}{V}{}{}$ and $\nthterm{1}{U}{}{}$
in terms of $\spform$ and derivatives of $\spform$
yields determining equations on their coefficients.
These equations are readily solved by using 
the jet-space (algebraic) tensor techniques
shown in \Ref{Annalspaper}.

The solution for $\nthterm{1}{V}{}{}$ and $\nthterm{1}{U}{}{}$
determines the lowest order part of 
the infinitesimal gauge group structure 
$\nthcommXvar{0}{1}{2} = \nthXvarsub{0}{3}$
where the parameter $\sXsub{3}$ is given in terms of 
$\sXsub{1}$ and $\sXsub{2}$ to lowest order by the closure relation 
\EQ
\nthXvarsub{0}{1}\nthXvarsub{1}{2}\pform{a}{\ind{\mu}{p}}
-\exchange{1}{2}
= \der{[\mu_p}\nthpXsub{0}{3}{a}{\ind{\mu}{p-1}]}
\label{closed0th}
\doneEQ
which follows from equation \eqref{0thclosureeq}. 

Closure of the gauge group at 1st order is derived as the next step
from the following analysis of the determining equations at $1$st order
in the hierarchy. 
Consider the Lie derivative commutator equation \eqref{1stcommdeteq}
minus the Lie derivative equation \eqref{1stdeteq}
applied to the gauge group commutator parameter $\sX=\nthsXsub{0}{3}$, 
\EQ
0= 
\Lie{\nthvar{1}{[\sXsub{1},\sXsub{2}]}\spform} \nthE{1}{\ind{\mu}{p}}{a}
= 
\nthvar{1}{[\sXsub{1},\sXsub{2}]} \nthE{1}{\ind{\mu}{p}}{a}
+ \nthE{1}{}{} \eqtext{ terms }
\label{1stcommeq}
\doneEQ
where 
\EQ
\var{[\sXsub{1},\sXsub{2}]} \pform{a}{\ind{\mu}{p}}
= [\Xvarsub{1},\Xvarsub{2}]\pform{a}{\ind{\mu}{p}}
- \Xvarsub{3}\pform{a}{\ind{\mu}{p}}
\doneEQ
denotes the field variation representing the deviation from closure
of the deformed gauge symmetry group. 
When $\spform$ is taken to satisfy the linear field equation,
the $\nthE{1}{}{}$ terms drop out of equation \eqref{1stcommeq},
yielding 
\EQ
0 = 
( \nthvar{1}{[\sXsub{1},\sXsub{2}]} \nthE{1}{\ind{\mu}{p}}{a} 
)\nthsolsp{1} .
\doneEQ
Lemma~2.6 implies 
\EQ
\nthvar{1}{[\sXsub{1},\sXsub{2}]}\pform{a}{\ind{\mu}{p}}
= \der{[\mu_p}\nthpXsub{1}{3}{a}{\ind{\mu}{p-1}]}
\label{closed1st}
\doneEQ
for some locally constructed parameter $\nthsXsub{1}{3}$,
which determines the deformation of the gauge group structure at 1st order. 
Hence a closure result is obtained. 

{\bf Theorem~2.7}:
The deformed gauge symmetry generates a gauge group that closes
up to 1st order on the solution space deformed field equation
\EQ
\var{[\sXsub{1},\sXsub{2}]}\spform \solsp
= \nthvar{0}{[\sXsub{1},\sXsub{2}]}\spform
+ \nthvar{1}{[\sXsub{1},\sXsub{2}]}\spform\nthsolsp{1} +\cdots
=0 \eqtext{ up to 1st order } .
\label{closureeq}
\doneEQ

Closure can be shown to extend in a similar way to higher orders,
leading to an infinitesimal gauge group structure
\EQ
[\Xvarsub{1},\Xvarsub{2}]\solsp 
= \Xvarsub{3}\solsp
\doneEQ
with 
\EQ
\sXsub{3} = \nthsXsub{0}{3} + \nthsXsub{1}{3} + \cdots
\doneEQ
depending bilinearly on $\sXsub{1},\sXsub{2}$ (and their derivatives).

This theorem gives rise to an obstruction for existence of 
quadratic terms in the deformed gauge symmetry. 
Take the parameter to be rigid, $\Xrigid=\const$, 
so then the curl of the gauge group closure equation \eqref{closureeq}
restricted to rigid symmetries yields
\EQs
0 = &&
\der{[\nu} ( 
[\nthXrigidvarsub{1}{1},\nthXrigidvarsub{1}{2}] \pform{a}{\ind{\mu}{p}]}
- \nthvarnthXrigidsub{1}{0}{3} \pform{a}{\ind{\mu}{p}]} )\nthsolsp{1} 
\nonumber\\
= &&
\der{[\nu|} \bigg( 
\pXsub{2}{b}{\ind{\nu}{p-1}}\rigid 
\pXsub{1}{d}{\ind{\rho}{p-1}}\rigid
\sum_{0\leq k\leq 2} 
( \nthder{k}{\ind{\alpha}{k}}
\nthterm{1}{U}{c\ind{\rho}{p-1}}{\ind{\beta}{p}d} )
\parder{ \nthterm{1}{U}{a\ind{\nu}{p-1}}{|\ind{\mu}{p}]b} }
{ \nthder{k}{\ind{\alpha}{k}}\pform{c}{\ind{\beta}{p}} }
-\exchange{1}{2}
\nonumber\\&&\fewquad
- \nthpXsub{0}{3}{b}{\ind{\nu}{p-1}}\rigid
\nthterm{1}{U}{a\ind{\nu}{p-1}}{|\ind{\mu}{p}]b}
\bigg)\nthsolsp{1} .
\label{1stUrigidcommeq}
\doneEQs
This equation is a rigidity-type \ic/ on the linear deformation terms,
namely, further necessary determining equations 
on the coefficients of $\spform$ and derivatives of $\spform$ 
in $\nthterm{1}{U}{}{}$. 

An additional rigidity-type \ic/ on the linear deformation terms 
arises from the 1st order Lie derivative equation \eqref{1stdeteq}
with $\sX$ taken to be constant,
$0=\Lie{\nthXrigidvar{1}\spform} \nthE{1}{\ind{\mu}{p}}{a}$. 
Factorization of the parameter $\pX{c}{\ind{\alpha}{p-1}}\rigid$ 
out of this equation gives
\EQ
0=
\sum_{0\leq k\leq 2}
( \nthder{k}{\ind{\alpha}{k}} 
\nthterm{1}{U}{b\ind{\rho}{p-1}}{\ind{\nu}{p}c} )
\parder{ \nthE{1}{\ind{\mu}{p}}{a} }
{ \nthder{k}{\ind{\alpha}{k}}\pform{b}{\ind{\nu}{p}} }
+ (-1)^k \nthder{k}{\ind{\alpha}{k}} (
\nthE{1}{\ind{\nu}{p}}{b}
\parder{ \nthterm{1}{U}{b\ind{\alpha}{p-1}}{\ind{\nu}{p}c} }
{ \nthder{k}{\ind{\alpha}{k}}\pform{a}{\ind{\mu}{p}} } 
) .
\label{1stUrigideq}
\doneEQ
This imposes more determining equations
on the coefficients of $\spform$ and derivatives of $\spform$ 
in $\nthterm{1}{U}{}{}$. 

The preceding equations are readily solved by 
the same algebraic techniques as discussed for 
the commutator equation \eqref{1stUVdeteq}. 
There is a useful remark to make here. 
If a classification of locally constructed linear symmetries 
depending on a rigid parameter $\sX\rigid$ is already known 
for the linear field equation, 
these symmetries then comprise all possible 
linear deformation terms for $\nthterm{1}{U}{}{}$,
since existence of a rigid symmetry
$\nthXrigidvar{1}\pform{a}{\ind{\mu}{p}}
= \term{U}{a\ind{\nu}{p-1}}{\ind{\mu}{p}b} 
\pX{b}{\ind{\nu}{p-1}}\rigid$
is necessary to satisfy the \ic/s \eqrefs{1stUrigidcommeq}{1stUrigideq}.
Moreover, the Noether currents associated with these symmetries
comprise the corresponding possible quadratic deformation terms for 
$\nthE{2}{}{}$, 
since the conservation law for a Noether current is 
the same as the gauge invariance identity 
$\der{\mu}( \nthE{2}{\mu\ind{\nu}{p-1}}{b} \pX{b}{\ind{\nu}{p-1}}\rigid ) 
= \nthE{1}{\ind{\alpha}{p}}{a} 
\nthterm{1}{U}{a\ind{\nu}{p-1}}{\ind{\alpha}{p}b} 
\pX{b}{\ind{\nu}{p-1}}\rigid$
from proposition~2.5,
and hence 
$\nthE{2}{\mu\ind{\nu}{p-1}}{b} \pX{b}{\ind{\nu}{p-1}}\rigid$
necessarily agrees with the Noether current 
to within an identically divergence-free term. 
Thus, any symmetry of a rigid form 
that may be available in the linear abelian theory 
can be used as a starting point 
for solving the deformation determining equations. 
This observation is often referred to as the 
``method of gauging a rigid symmetry'' 
or the ``Noether coupling method''
and it has been used extensively in supergravity contexts \cite{Des,SG}.
Of course, it is unable to provide 
a complete classification or a uniqueness result for deformations. 

Once the linear deformation terms have been determined, 
the next steps are to solve for 
the quadratic deformation terms 
in the gauge symmetry and in the field equation, respectively, 
from the 1st order closure equation \eqref{closed1st}
\EQ
0 = 
\der{[\nu} ( 
\nthXvarsub{0}{1}\nthXvarsub{2}{2}] \pform{a}{\ind{\mu}{p}]}
-\exchange{1}{2}
+ [\nthXvarsub{1}{1},\nthXvarsub{1}{2}] \pform{a}{\ind{\mu}{p}]}
- \nthvarnthXsub{1}{0}{3} \pform{a}{\ind{\mu}{p}]} )\nthsolsp{1} ,
\label{2ndXdeteq}
\doneEQ
and from the 1st order Lie derivative equation \eqref{1stdeteq}
\EQ
0= \nthXvar{0} \nthE{2}{\ind{\mu}{p}}{a}
+ \Lie{\nthXvar{1}\spform} \nthE{1}{\ind{\mu}{p}}{a} ,
\label{2ndEdeteq}
\doneEQ
with the gauge parameters no longer being rigid. 
Through the closure equation \eqref{closed0th} at lowest order, 
the commutativity-type \ic/ on equation \eqref{2ndEdeteq}
reduces to 
\EQ
0= 
\nthXvarsub{0}{3} \nthE{1}{\ind{\mu}{p}}{a} 
+ \Lie{\nthXvarsub{1}{1}\spform}\nthXvarsub{0}{2} \nthE{1}{\ind{\mu}{p}}{a} 
-\exchange{1}{2}
\doneEQ
which is satisfied due to abelian gauge invariance. 
Then, an analysis of the commutativity-type \ic/ 
on equation \eqref{2ndXdeteq} shows it is satisfied 
due to the lowest order part of the Jacobi identity applied to 
the gauge symmetries. 
Hence no obstructions arise for solving 
the $1$st order Lie derivative equations \eqrefs{2ndEdeteq}{2ndXdeteq}. 

To continue, 
after the expressions determined for 
$\nthterm{1}{U}{}{}$ and $\nthterm{1}{V}{}{}$ 
from prior equations are substituted,
the determining equations \eqrefs{2ndEdeteq}{2ndXdeteq}
expand out to have the form
\EQs
&&
\der{[\nu|} \bigg( \sum_{0\leq k\leq 2}
\pXsub{2}{b}{\ind{\nu}{p-1}}
\nthder{k+1}{\ind{\alpha}{k}\beta_p} \pXsub{1}{c}{\ind{\beta}{p-1}}
\parder{ \nthterm{2}{U}{a\ind{\nu}{p-1}}{|\ind{\mu}{p}]b} }
{ \nthder{k}{\ind{\alpha}{k}}\pform{c}{\ind{\beta}{p}} }
\nonumber\\&&\qquad
+ \sum_{0\leq k\leq 1}
\der{\alpha}\pXsub{2}{b}{\ind{\nu}{p-1}} 
\nthder{k+1}{\ind{\alpha}{k}\beta_p} \pXsub{1}{c}{\ind{\beta}{p-1}}
\parder{ \nthterm{2}{V}{a\ind{\nu}{p-1}\alpha}{|\ind{\mu}{p}]b} }
{ \nthder{k}{\ind{\alpha}{k}}\pform{c}{\ind{\beta}{p}} }
\bigg) 
-\exchange{1}{2}
\nonumber\\
&&
= \der{[\nu|} \bigg( \bigg( 
\sum_{0\leq k\leq 2}
\pXsub{1}{b}{\ind{\nu}{p-1}}
\nthder{k}{\ind{\alpha}{k}}( 
\nthterm{1}{U}{c\ind{\gamma}{p-1}}{\ind{\beta}{p}d} 
\pXsub{2}{d}{\ind{\gamma}{p-1}}
+ \nthterm{1}{V}{c\ind{\gamma}{p-1}\beta}{\ind{\beta}{p}d} 
\der{\beta}\pXsub{2}{d}{\ind{\gamma}{p-1}} )
\parder{ \nthterm{1}{U}{a\ind{\nu}{p-1}}{|\ind{\mu}{p}]b} }
{ \nthder{k}{\ind{\alpha}{k}}\pform{c}{\ind{\beta}{p}} }
\nonumber\\&&\qquad
+ \sum_{0\leq k\leq 1}
\der{\alpha}\pXsub{1}{b}{\ind{\nu}{p-1}} 
\nthder{k}{\ind{\alpha}{k}}( 
\nthterm{1}{U}{c\ind{\gamma}{p-1}}{\ind{\beta}{p}d} 
\pXsub{2}{d}{\ind{\gamma}{p-1}}
+ \nthterm{1}{V}{c\ind{\gamma}{p-1}\beta}{\ind{\beta}{p}d} 
\der{\beta}\pXsub{2}{d}{\ind{\gamma}{p-1}} )
\parder{ \nthterm{1}{V}{a\ind{\nu}{p-1}\alpha}{|\ind{\mu}{p}]b} }
{ \nthder{k}{\ind{\alpha}{k}}\pform{c}{\ind{\beta}{p}} } \bigg)
\nonumber\\&&\qquad
-\exchange{1}{2}
- \nthterm{1}{U}{a\ind{\nu}{p-1}}{|\ind{\mu}{p}]b} 
\nthpXsub{0}{3}{b}{\ind{\nu}{p-1}}
- \nthterm{1}{V}{a\ind{\nu}{p-1}\alpha}{|\ind{\mu}{p}]b} 
\der{\alpha}\nthpXsub{0}{3}{b}{\ind{\nu}{p-1}}
\bigg)
\label{2ndUVdeteq}
\doneEQs
and
\EQs
&&
\sum_{0\leq k\leq 2}
\nthder{k+1}{\ind{\alpha}{k}\beta_p} \pX{c}{\ind{\beta}{p-1}}
\parder{ \nthE{2}{\ind{\mu}{p}}{a} }
{ \nthder{k}{\ind{\alpha}{k}}\pform{c}{\ind{\beta}{p}} }
\nonumber\\&&\qquad
= 
-\sum_{0\leq k\leq 2}
\nthder{k}{\ind{\alpha}{k}}( 
\nthterm{1}{U}{c\ind{\gamma}{p-1}}{\ind{\beta}{p}d} 
\pX{d}{\ind{\gamma}{p-1}}
+ \nthterm{1}{V}{c\ind{\gamma}{p-1}\beta}{\ind{\beta}{p}d} 
\der{\beta}\pX{d}{\ind{\gamma}{p-1}} )
\parder{ \nthE{1}{\ind{\mu}{p}}{a} }
{ \nthder{k}{\ind{\alpha}{k}}\pform{c}{\ind{\beta}{p}} }
\nonumber\\&&\qquad\quad
- \sum_{0\leq k\leq 1} (-1)^k \nthder{k}{\ind{\alpha}{k}}( 
\der{\beta}\pX{c}{\ind{\beta}{p-1}} \nthE{1}{\ind{\nu}{p}}{b} 
\parder{ \nthterm{1}{V}{b\ind{\beta}{p-1}\beta}{\ind{\nu}{p}c} }
{ \nthder{k}{\ind{\alpha}{k}}\pform{a}{\ind{\mu}{p}} } )
\nonumber\\&&\qquad\quad
- \sum_{0\leq k\leq 2} (-1)^k \nthder{k}{\ind{\alpha}{k}}( 
\pX{c}{\ind{\beta}{p-1}} \nthE{1}{\ind{\nu}{p}}{b} 
\parder{ \nthterm{1}{U}{b\ind{\beta}{p-1}}{\ind{\nu}{p}c} }
{ \nthder{k}{\ind{\alpha}{k}}\pform{a}{\ind{\mu}{p}} } )
\label{2ndQPRdeteq}
\doneEQs
where
\EQ
\nthE{2}{\ind{\mu}{p}}{a}
=
\nthterm{1}{Q}{\ind{\mu}{p}\ind{\nu}{p}\alpha\beta}{ab}
\der{\alpha}\der{\beta}\pform{b}{\ind{\nu}{p}}
+ \nthterm{1}{P}{\ind{\mu}{p}\ind{\nu}{p}\alpha}{ab}
\der{\alpha}\pform{b}{\ind{\nu}{p}}
+ \nthterm{1}{R}{\ind{\mu}{p}\ind{\nu}{p}}{ab}
\pform{b}{\ind{\nu}{p}} .
\doneEQ
Despite their complicated appearance, 
it is computationally straightforward to solve these equations 
for $\nthterm{2}{U}{}{}$ and $\nthterm{2}{V}{}{}$ 
by use of the jet-space (algebraic) tensor techniques in \Ref{Annalspaper}.

\subsection{ Geometrical structure of lowest-order deformations }

Strikingly, a simple explicit solution for the quadratic deformation terms
can be constructed out of the linear deformations terms 
$\nthterm{1}{U}{}{}$ and $\nthterm{1}{V}{}{}$
under a mild restriction on the form sought 
for the deformed field equation and gauge symmetry. 
The result relies on the solution of 
the $0$th order Lie derivative commutator equation. 

{\bf Theorem~2.8}:
Restrict deformations so the field equation 
and the gauge symmetry contain no more derivatives of $\spform$ and $\sX$,
in total, than the number in the linear field equation \eqref{linearfieldeq}
and the abelian gauge symmetry \eqref{abeliangaugesymm}
(namely, in notation \eqrefs{kthEform}{kthUVform}, 
$Q$, $R$, and $V$ have no dependence on derivatives of $\spform$,
while $P$ and $U$ have at most linear dependence on derivatives of $\spform$,
with higher order derivatives excluded). 
Then, up to a nonlinear change of field variable
and of gauge parameter, 
the linear deformation terms have the form
\EQ
\nthterm{1}{V}{a\ind{\nu}{p-1}\alpha}{\ind{\mu}{p}b} 
=0 ,\quad
\nthterm{1}{U}{a\ind{\nu}{p-1}}{\ind{\mu}{p}b} 
= \tensor{u_0}{\ind{\nu}{p-1}a}{bc} \pform{c}{\ind{\mu}{p}}
+ \tensor{u_1}{\ind{\nu}{p-1}\ind{\alpha}{p+1}a}{\ind{\mu}{p}bc} 
\pcurl{c}{\ind{\alpha}{p+1}}
\label{sol1stUV}
\doneEQ
for some coefficients such that 
$u_0$ is constant, skew in its lower internal indices for $p=1$,
and vanishes for $p>1$;
the quadratic deformation terms are explicitly given by 
\EQs
\nthterm{2}{V}{a\ind{\nu}{p-1}\alpha}{\ind{\mu}{p}b} 
&&
=0 , 
\label{sol2ndV}\\
\nthterm{2}{U}{a\ind{\nu}{p-1}}{\ind{\mu}{p}b} 
&&
= 
\nthpstr{2}{c}{\ind{\alpha}{p+1}}
\parder{ \nthterm{1}{U}{a\ind{\nu}{p-1}}{\ind{\mu}{p}b} }
{ \pcurl{c}{\ind{\alpha}{p+1}} }
\label{sol2ndU}\\
&&
= 
-\tensor{u_1}{\ind{\nu}{p-1}\beta\ind{\alpha}{p}a}{\ind{\mu}{p}bc} 
( \frac{1}{2}\tensor{u_0}{\ind{\gamma}{p-1} c}{ij} 
\pform{j}{\ind{\alpha}{p}}
+ \tensor{u_1}{\ind{\gamma}{p-1}\ind{\lambda}{p+1} c}{\ind{\alpha}{p}ij} 
\pcurl{j}{\ind{\lambda}{p+1}} )
\pform{i}{\beta\ind{\gamma}{p-1}} , 
\nonumber
\\
\nthE{2}{\ind{\mu}{p}}{a} 
&&
= 
( \der{\alpha}\nthpstr{2}{b}{\ind{\nu}{p+1}}
+\pform{c}{\alpha\ind{\beta}{p-1}} 
\der{[\nu_{p+1}}\nthterm{1}{U}{b\ind{\beta}{p-1}}{\ind{\nu}{p}]c} )
\parder{ \nthE{1}{\ind{\mu}{p}}{a} }
{ \der{\alpha}\pcurl{b}{\ind{\nu}{p+1}} }
\\&&\qquad
+\nthpstr{2}{b}{\ind{\nu}{p+1}}
\parder{ \nthE{1}{\ind{\mu}{p}}{a} }
{ \pcurl{b}{\ind{\nu}{p+1}} }
+ \nthE{1}{\ind{\nu}{p}}{b} \pform{c}{\alpha\ind{\beta}{p-1}} 
\parder{ \nthterm{1}{U}{b\ind{\beta}{p-1}}{\ind{\nu}{p}c} }
{ \pcurl{a}{\alpha\ind{\mu}{p}} }
\nonumber\\&&
= -\tensor{q}{\ind{\mu}{p}\ind{\nu}{p+1}\alpha}{ab} 
\Big( 
\der{\alpha}(
( \frac{1}{2}\tensor{u_0}{\ind{\gamma}{p-1} b}{ij} 
\pform{j}{\ind{\nu}{p}}
+ \tensor{u_1}{\ind{\gamma}{p-1}\ind{\lambda}{p+1} b}{\ind{\nu}{p}ij} 
\pcurl{j}{\ind{\lambda}{p+1}} ) \pform{i}{\nu_{p+1}\ind{\gamma}{p-1}} )
\nonumber\\&&\qquad
+ \pform{c}{\alpha\ind{\beta}{p-1}}( 
\tensor{u_0}{\ind{\beta}{p-1} b}{ce} \pcurl{e}{\ind{\nu}{p+1}}
+\der{[\nu_{p+1}}( 
\tensor{u_1}{\ind{\beta}{p-1}\ind{\gamma}{p+1} b}{\ind{\nu}{p}]ce} 
\pcurl{e}{\ind{\gamma}{p+1}} ))
\Big) 
\nonumber\\&&\qquad
-\tensor{p}{\ind{\mu}{p}\ind{\nu}{p+1}}{ab} 
 ( \frac{1}{2}\tensor{u_0}{\ind{\gamma}{p-1} b}{ij} 
\pform{j}{\ind{\nu}{p}}
+ \tensor{u_1}{\ind{\gamma}{p-1}\ind{\lambda}{p+1} b}{\ind{\nu}{p}ij} 
\pcurl{j}{\ind{\lambda}{p+1}} ) \pform{i}{\nu_{p+1}\ind{\gamma}{p-1}}
\nonumber\\&&\qquad
+\tensor{u_1}{\ind{\gamma}{p-1}\beta\ind{\mu}{p} b}{\ind{\nu}{p} ea}
( \tensor{q}{\ind{\nu}{p}\ind{\lambda}{p+1}\alpha}{bc}
\der{\alpha}\pcurl{c}{\ind{\lambda}{p+1}}
+ \tensor{p}{\ind{\nu}{p}\ind{\lambda}{p+1}}{bc}
\pcurl{c}{\ind{\lambda}{p+1}} )
\pform{e}{\beta\ind{\gamma}{p-1}} 
\label{sol2ndE}
\\&&
= \nthterm{1}{Q}{\ind{\mu}{p}\ind{\nu}{p}\alpha\beta}{ab}
\der{\alpha}\der{\beta}\pform{b}{\ind{\nu}{p}}
+ \nthterm{1}{P}{\ind{\mu}{p}\ind{\nu}{p}\alpha}{ab}
\der{\alpha}\pform{b}{\ind{\nu}{p}}
+ \nthterm{1}{R}{\ind{\mu}{p}\ind{\nu}{p}}{ab}
\pform{b}{\ind{\nu}{p}} , 
\nonumber
\doneEQs
where
\EQ
\nthpstr{2}{c}{\ind{\alpha}{p+1}}
= 
-( \frac{1}{2}\tensor{u_0}{\ind{\gamma}{p-1} c}{ij} 
\pform{j}{[\ind{\alpha}{p}|}
+ \tensor{u_1}{\ind{\gamma}{p-1}\ind{\lambda}{p+1} c}{[\ind{\alpha}{p}|ij} 
\pcurl{j}{\ind{\lambda}{p+1}} )
\pform{i}{|\alpha_{p+1}]\ind{\gamma}{p-1}} .
\doneEQ
Note the coefficients $u_0$ and $u_1$
are also subject to additional algebraic equations that come from 
the \ic/s \eqrefs{1stUrigidcommeq}{1stUrigideq}.

{\bf Remark}: As a result of proposition~2.4,
the restriction assumed on derivatives is necessary 
if deformations to all orders are required both 
to preserve the number of dynamical degrees of freedom 
of $\spform$ from the linear theory
and to contain a bounded total number of derivatives of $\spform$. 

In outline, 
the proof of the theorem rests on 
using the determining equation \eqref{1stUVdeteq}
to show, firstly, that the solution for $\nthterm{1}{V}{}{}$
vanishes after a suitable change of field variable \eqref{fieldredef}
and gauge parameter \eqref{parmredef} are made;
secondly, the solution for $\nthterm{1}{U}{}{}$ establishes
that the curl of this expression is gauge invariant
with respect to the $\nthXvar{0}\spform$
and that the exterior product of 
$\nthterm{1}{U}{}{}$ and $\nthXvar{0}\spform$ 
is an exact variation, 
\EQ
\nthXvar{0} \der{[\alpha} \nthterm{1}{U}{a\ind{\nu}{p-1}}{\ind{\mu}{p}]b}
=0 ,\qquad
\nthterm{1}{U}{a\ind{\nu}{p-1}}{[\ind{\mu}{p}| b}
\der{|\alpha]} \pX{b}{\ind{\nu}{p-1}} 
= -\nthXvar{0} \nthpstr{2}{a}{\ind{\mu}{p}\alpha} .
\label{1stUVvareq}
\doneEQ
This analysis is a straightforward application of 
the jet-space tensor techniques from \Ref{Annalspaper}.
Then, from expressions \eqref{1stUVvareq},
the Lie derivative equation \eqref{2ndQPRdeteq} is easily solved 
for $\nthE{2}{}{}$,
as is the closure equation \eqref{2ndUVdeteq}
for $\nthXvar{2}\spform$ in terms of the commutator of $\nthXvar{1}\spform$.
Note the expression obtained for $\nthE{2}{}{}$ is essentially 
a construction of the Noether current of 
the rigid symmetry $\nthvar{1}{\rm rigid}\spform = \nthterm{1}{U}{}{}$,
due to the Noether gauge-invariance identity stated in proposition~2.5.
Uniqueness of both $\nthE{2}{}{}$ and $\nthXvar{2}\spform$
will be established later (see theorem~2.10).

These results \sysref{sol1stUV}{sol2ndE}
express the following geometrical structure
arising from the linear and quadratic deformation terms:
First observe, since $\nthterm{1}{V}{}{}=\nthterm{2}{V}{}{}=0$,
the deformed gauge symmetry takes the form of a covariant derivative
on the gauge parameter,
\EQ
\Xvar\pform{a}{\ind{\mu}{p}} 
= \D{\spform}(\sX)\mixedindices{a}{\ind{\mu}{p}}
= \der{[\mu_p} \pX{a}{\ind{\mu}{p-1}]}
+ ( \tensor{u_0}{\ind{\nu}{p-1}a}{bc} \pform{c}{\ind{\mu}{p}}
+ \tensor{u_1}{\ind{\nu}{p-1}\ind{\alpha}{p+1}a}{\ind{\mu}{p}bc} 
\pcurl{c}{\ind{\alpha}{p+1}} )
\pX{b}{\ind{\nu}{p-1}}
+\cdots
\doneEQ
(to linear order)
where $\D{\spform}$ is viewed as an exterior derivative operator 
from the gauge parameter vector bundle $\parmsp$ into $\fieldsp$
(\ie/ $\D{\spform}: \Lambda_{p-1}(M)\otimes\vs \rightarrow 
\Lambda_{p}(M)\otimes\vs$). 
This derivative operator has a natural extension to 
the vector bundle $\fieldsp$
(\ie/ $\D{\spform}: \Lambda_{p}(M)\otimes\vs \rightarrow 
\Lambda_{p+1}(M)\otimes\vs$), 
in particular 
acting on $\Xvar\spform$ itself by 
\EQs
\D{\spform}(\Xvar\spform)\mixedindices{a}{\ind{\mu}{p+1}}
&& 
= \der{[\mu_{p+1}} \Xvar\pform{a}{\ind{\mu}{p}]}
+ ( \tensor{u_0}{\ind{\nu}{p-1}a}{bc} \pform{c}{[\ind{\mu}{p}|}
+ \tensor{u_1}{\ind{\nu}{p-1}\ind{\alpha}{p+1}a}{[\ind{\mu}{p}|bc} 
\pcurl{c}{\ind{\alpha}{p+1}} )
\Xvar\pform{b}{|\mu_{p+1}]\ind{\nu}{p-1}}
+\cdots
\nonumber\\&&
= \der{[\mu_{p+1}|} ( \tensor{u_0}{\ind{\nu}{p-1}a}{bc} \pform{c}{|\ind{\mu}{p}]}
+ \tensor{u_1}{\ind{\nu}{p-1}\ind{\alpha}{p+1}a}{|\ind{\mu}{p}]bc} 
\pcurl{c}{\ind{\alpha}{p+1}} )
\pX{b}{\ind{\nu}{p-1}}
\nonumber\\&&\qquad
+2(1-1/p) ( \tensor{u_0}{\nu\ind{\nu}{p-2}a}{bc} \pform{c}{[\ind{\mu}{p}|}
+ \tensor{u_1}{\nu\ind{\nu}{p-2}\ind{\alpha}{p+1}a}{[\ind{\mu}{p}|bc} 
\pcurl{c}{\ind{\alpha}{p+1}} )
(\der{|\mu_{p+1}]} \pX{b}{\nu\ind{\nu}{p-1}} 
\nonumber\\&&\qquad
+\der{\nu} \pX{b}{|\mu_{p+1}]\ind{\nu}{p-2}} )
+\cdots
\doneEQs
(to linear order).
Next, the relation \eqref{1stUVvareq} on $\nthterm{1}{U}{}{}$
gives rise to a nonlinear deformation of the abelian field strength $\spcurl$
as defined by
\EQ
\pstr{a}{\ind{\mu}{p+1}}= 
\pcurl{a}{\ind{\mu}{p+1}} + \nthpstr{2}{a}{\ind{\mu}{p+1}} +\cdots ,\quad
\pstrY{a\ind{\nu}{p+1}}{\ind{\mu}{p+1}b}
\Xvar\pstr{b}{\ind{\nu}{p+1}} 
= \D{\spform}(\Xvar\spform)\mixedindices{a}{\ind{\mu}{p+1}}
\doneEQ
for some locally constructed linear map 
\EQ
\pstrY{a\ind{\nu}{p+1}}{\ind{\mu}{p+1}b}
= \id{b}{a}\id{\ind{\mu}{p+1}}{\ind{\nu}{p+1}}
+ \nthpstrY{1}{a\ind{\nu}{p+1}}{\ind{\mu}{p+1}b}
+\cdots .
\doneEQ
Notice this expands out to 
$\der{[\mu_{p+1}} \nthXvar{0}\pform{a}{\ind{\mu}{p}]}
= \nthXvar{0}\pcurl{a}{\ind{\mu}{p+1}}=0$ at lowest order,
which holds as an identity;
then at 1st order, the cancellation of terms 
$\der{[\mu_{p+1}} \nthXvar{1}\pform{a}{\ind{\mu}{p}]}
= \nthXvar{1}\pcurl{a}{\ind{\mu}{p+1}}$
leaves 
$\nthD{1}{\spform}(\nthXvar{0}\spform) \mixedindices{a}{\ind{\mu}{p+1}}
= \nthXvar{0}\nthpstr{2}{a}{\ind{\mu}{p+1}}$,
which reduces to equation \eqref{1stUVvareq}.
The linear map $\pstrY{}{}$ does not enter until next order. 
Thus, the nonlinear field strength $p+1$-form is given by 
\EQ
\pstr{a}{\ind{\mu}{p+1}} 
= 
\der{[\mu_{p+1}} \pform{a}{\ind{\mu}{p}]}
+ ( \tensor{u_0}{\ind{\nu}{p-1}a}{bc} \pform{c}{[\ind{\mu}{p}|}
+ \tensor{u_1}{\ind{\nu}{p-1}\ind{\alpha}{p+1}a}{[\ind{\mu}{p}|bc} 
\pcurl{c}{\ind{\alpha}{p+1}} )
\pform{b}{|\mu_{p+1}]\ind{\nu}{p-1}}
+\cdots 
\doneEQ
(to quadratic order).
As a result, 
the geometrical content of deformations in theorem~2.8 
is seen to be encoded in
the covariant derivative operator $\D{\spform}$ 
and the nonlinear field strength $\spstr$
emerging from the solution for the linear terms 
$\nthterm{1}{U}{}{}$, $\nthterm{1}{V}{}{}$.

{\bf Corollary~2.9}:
In geometrical terms, 
the form of $\nthterm{2}{U}{}{}$ is given by
a replacement of the abelian field strength $\spcurl$ in $\nthterm{1}{U}{}{}$
by the quadratic part of the nonlinear field strength $\spstr$.
Similarly, the form of $\nthE{2}{}{}$ is given by
a simultaneous replacement of 
the quadratic field strength $\nthpstr{2}{}{}$ for $\spcurl$
and the linear part of the covariant derivative $\nthD{1}{\spform}$
for the coordinate derivative operator,
up to additional terms that vanish on the solution space $\nthE{1}{}{}=0$.

The scope of this theorem encompasses
\YM/ and \FT/ theories, 
Einstein gravity theory and its multi-graviton generalizations,
as well as the gravity-like generalization of \YM/ theory,
in any dimension $d\geq 3$.
Moreover, in $d=3$ dimensions, 
\YM/ \CS/ theory is included, as is combined \YM/ \FT/ theory,
in addition to the torsion generalizations of \YM/ theory 
presented in \secrefs{YMFTCSth}{torsionKVth}. 
The results in the theorem thus expose a hitherto unrecognized 
universal nonlinear structure present 
among a wide class of important examples of nonlinear gauge theories. 

The only significant example excluded by the theorem is 
the exotic parity-violating multi-graviton theories in $d=3,5$ dimensions,
as they involve a more complicated type of deformation 
in which the form of the gauge symmetry is heavily coupled to 
a deformed auxiliary gauge freedom on the vielbein. 
In contrast, 
the auxiliary gauge freedom in ordinary gravitational theories
is consistent with the form of the deformations given in the theorem.
Basically, the expression for $\nthE{2}{}{}$
can be shown to preserve the number of degrees of gauge freedom
from the linearized local Lorentz transformations 
present on the (linearized vielbein) field variable in $\nthE{1}{}{}$. 
A full discussion will be left to a forthcoming paper
addressed to more general gravity-like nonlinear $p$-form theories
for $p\geq 1$ in $d$ dimensions. 

\subsection{ Remarks on integrability }

The trivial nature of the commutativity-type \ic/s 
up to 1st order in the hierarchy of determining equations
extends to higher orders.
However, the rigidity-type \ic/s remain nontrivial at least up to next order,
as follows. 
The Lie derivative equation at 2nd order gives 
\EQ
0=
\Lie{\nthXvar{1}\rigid\spform} \nthE{2}{\ind{\mu}{p}}{a}
+ \Lie{\nthXvar{2}\rigid\spform} \nthE{1}{\ind{\mu}{p}}{a}
\label{2ndrigiddeteq}
\doneEQ
which constitutes an \ic/ on the linear deformation terms,
since the quadratic deformation terms 
$\nthXvar{2}\spform$ and $\nthE{2}{}{}$ 
are determined from $\nthXvar{1}\spform$ by theorem~2.8. 
Use of jet-space tensor techniques then reduces the \ic/ 
to determining equations 
on the coefficients of $\spform$ and derivatives of $\spform$ 
in $\nthterm{1}{U}{}{}$. 
These equations are nontrivial, as seen in examples. 

In \FT/ theory and its \YM/ generalizations, 
the gauge groups close only on the solution space
and their structure emerges purely from 
the integrability equation \eqref{2ndrigiddeteq} at 2nd order.
In contrast, for \YM/ theory and algebra-valued gravity theory,
the nature of the closure and integrability equations is reversed ---
the gauge groups in those theories close off the solution space
and their structure is determined entirely by the 1st order \ic/
coming from the closure equation \eqref{closed1st},
while the integrability equation \eqref{2ndrigiddeteq} at 2nd order 
gives no extra structure or algebraic constraints. 
Finally, the exotic parity-violating multi-graviton theories
share some features of all these other examples,
with both the closure equation and integrability equation 
contributing to the structure of the gauge group. 

\subsection{ Uniqueness (rigidity) of deformations }

A strong uniqueness result for nonlinear deformation terms 
can be established by an induction argument 
in powers of the fields $\spform$
under the same restriction on derivatives in the form sought for
deformations as assumed in theorem~2.8.
To begin, 
consider two solutions of the determining equations to all orders, 
and let $\Delta\nthE{k}{}{}$ and $\Delta\nthXvar{k}\spform$
denote the difference of the $k$th power deformation terms 
in the solutions. 
Suppose these solutions agree for $k<n$ up to some power $n>1$, so 
$\Delta\nthE{k}{}{}=0$ and $\Delta\nthXvar{k}\spform=0$,
$0\leq k\leq n-1$,
with the lowest power terms automatically
fixed by the form of the abelian gauge symmetry and linear field equation.
Subtraction of the Lie derivative equations satisfied for each solution 
then yields
\EQ
0= \nthXvar{0}( \Delta\nthE{n}{}{} )
\doneEQ
to lowest nonvanishing order. 
From this equation it follows that $\Delta\nthE{n}{}{}$ is
a gauge-invariant expression with respect to the abelian gauge symmetry
and hence has the form of a homogeneous polynomial of degree $n>1$
in the abelian field strength $\spcurl$. 
The restriction assumed on derivatives of $\spform$ requires
$\Delta\nthE{n}{}{}$ to contain at most two derivatives in total,
hence $\Delta\nthE{n}{}{}$ must vanish if $n>2$. 
In the case $n=2$, a simple analysis shows that 
the Frechet derivative of 
any homogeneous quadratic polynomial in $\spcurl$ 
fails to be self-adjoint, 
which implies such polynomials cannot have the form of an \EL/ equation
\cite{Olv,AncBlu}.
Thus $\Delta\nthE{n}{}{}$ must also vanish if $n=2$. 

Now, 
by subtraction of the Lie derivative commutator equations for each solution, 
$\Delta\nthXvar{n}\spform$ satisfies 
\EQ
0=\nthXvarsub{0}{2}( \Delta\nthXvarsub{n}{1}\nthE{1}{}{} )\nthsolsp{1} 
-\exchange{1}{2} 
\doneEQ
to lowest nonvanishing order. 
An analysis of this equation using lemma~2.6 shows that 
the curl of 
$\nthXvarsub{0}{2}( \Delta\nthXvarsub{n}{1}\spform )$ is zero,
after all change of variable freedom \eqrefs{fieldredef}{parmredef}
in $\spform$ and $\sX$ is exhausted. 
Hence $\Delta\nthXvar{n}\spform$ must have the form of 
a homogeneous polynomial of degree $n>1$ 
in the abelian field strength $\spcurl$. 
But, with the restriction assumed on derivatives of $\spform$, 
$\Delta\nthXvar{n}\spform$ is required to contain 
at most one derivative in total, 
and thus it must vanish if $n\geq 2$. 

Induction on the power $n$ 
therefore establishes the following uniqueness (rigidity) result. 

{\bf Theorem~2.10}:
Up to a nonlinear change of field variable and of gauge parameter, 
deformations that contain, in total, 
no more derivatives of $\spform$ and $\sX$ 
than the number in the linear abelian theory 
are uniquely determined by their lowest-order terms $\nthXvar{1}\spform$.

\section{ \YM/ \CS/ gauge theory with \FT/ interaction }
\label{YMFTCSth}

The starting point here will be the linear abelian theory of 
a 1-form potential $\A{a}{\mu}$ 
with an associated internal vector space structure $\vs$ 
of dimension $n\geq 1$,
on a $d=3$ dimensional spacetime. 
For simplicity, the spacetime is taken to be flat,
with a metric $\metric{\mu\nu}$ and a volume form $\vol{\mu\nu\alpha}$
given in Minkowski coordinates.
The standard Lagrangian for this theory is expressed in terms of 
the abelian field strength 2-form $\F{a}{\mu\nu}=\der{[\mu}\A{a}{\nu]}$
by
\EQ
\nthsL{2} = \frac{1}{2} \invmetric{\mu\nu} \k{ab} \stF{a}{\mu} \stF{b}{\nu}
\label{quadrL}
\doneEQ
where $\stF{a}{\mu} =\cross{\mu}{\alpha\beta} \F{a}{\alpha\beta}$
is the dual field strength 1-form. 
The Lagrangian is gauge invariant under the abelian gauge symmetry 
\EQ
\nthXvar{0}\A{a}{\mu} = \der{\mu}\X{a}
\label{zerothXvar}
\doneEQ
for an arbitrary function $\X{a}$ on $M$ with values in $\vs$. 
The linear field equation obtained via 
the Euler operator applied to $\nthsL{2}$ 
is second order in derivatives and also is gauge invariant,
as given by the curl of the dual field strength
\EQ
\nthE{1}{\mu}{a} 
= \k{ab} \invvol{\mu\alpha\beta} \der{\alpha} \stF{b}{\beta} .
\doneEQ
Note this is the source-free Maxwell equation on $\A{a}{\mu}$
in three dimensions. 
An abelian \CS/ term is straightforwardly introduced through the Lagrangian 
\EQ
\nthsL{2}_{\rm CS} 
= \frac{1}{2} m \invmetric{\mu\nu} \k{ab} \A{a}{\mu} \stF{b}{\nu} 
\doneEQ
where $m$ is the \CS/ coupling constant. 
Under the abelian gauge symmetry $\nthXvar{0}\A{a}{\mu}$,
the total Lagrangian $\nthsL{2} + \nthsL{2}_{\rm CS}$
remains invariant to within a divergence
and thus contributes a gauge-invariant first derivative term 
\EQ
\linE{m}{\mu}{a} = m \k{ab} \invmetric{\mu\nu} \stF{b}{\nu}
\doneEQ
to the linear field equation $\nthE{1}{\mu}{a} + \linE{m}{\mu}{a} =0$.
Thus, the potential $\A{a}{\mu}$ satisfies the field equation
\EQ
\cross{\mu}{\nu\alpha } \der{\nu} \stF{a}{\alpha} 
= -m \stF{a}{\mu}
\label{linfieldeq}
\doneEQ
along with the identity 
\EQ
\coder{\mu} \stF{a}{\mu} =0 .
\label{linfieldstrid}
\doneEQ

The physical content of solutions $\A{a}{\mu}$ describes 
a set of free spin-one fields of mass $m$,
as seen in a gauge invariant manner 
from the curl of the field equation \eqref{linfieldeq},
\EQ
\waveop\stF{a}{\mu} = m^2 \stF{a}{\mu} .
\doneEQ
Alternatively, in Lorentz gauge $\coder{\mu}\A{a}{\mu}=0$,
when $m=0$ the field equation reduces to 
a set of massless linear vector wave equations 
\EQ
\waveop\A{a}{\mu} =0
\doneEQ
modulo the residual gauge freedom 
$\A{a}{\mu} \rightarrow \A{a}{\mu} +\der{\mu}\X{a}$
such that $\waveop\X{a} =0$. 
When $m\neq 0$, 
since both sides of the field equation \eqref{linfieldeq} are a curl, 
this implies 
\EQ
\stF{a}{\mu} +m\A{a}{\mu} =\der{\mu}\psi ,\quad
\waveop\psi =0 , 
\doneEQ
to within a gradient term $\psi$ that satisfies the scalar wave equation
as a consequence of the Lorentz gauge on $\A{a}{\mu}$
and the divergence free nature of $\stF{a}{\mu}$. 
This $\psi$ term can be removed by the residual gauge freedom
in $\A{a}{\mu}$, so then 
\EQ
\stF{a}{\mu} = -m\A{a}{\mu} ,
\label{linmCSeq}
\doneEQ
and thus the field equation \eqref{linfieldeq} becomes 
a set of massive linear vector wave equations
\EQ
\waveop\A{a}{\mu} = m^2 \A{a}{\mu} .
\doneEQ
It is interesting to note the relation \eqref{linmCSeq} 
between the dual field strength and the potential in Lorentz gauge
is equivalent to a pure \CS/ theory with an added mass term
given by the Proca Lagrangian 
$\nthsL{2}_{\rm m} =
\frac{1}{2} m^2 \invmetric{\mu\nu} \k{ab} \A{a}{\mu}\A{b}{\nu}$.
In both the massless and massive cases, 
two of the three independent components of each field $\A{a}{\mu}$ 
are able to be gauged away, leaving a single dynamical degree of freedom
for solutions of the field equation. 

Nonlinear deformations of this linear abelian theory 
fall into two distinct classes, 
differing in how much of the spacetime structure they use.
The first class consists of deformations that only rely on
the spacetime metric and volume form
along with exterior derivatives on the field and gauge parameter. 
Such deformations preserve the manifest Lorentz covariance of 
the linear abelian theory. 
The second class of deformations make essential use of \KV/s of
the spacetime metric and thus break Lorentz covariance. 
These non-covariant deformations will be considered in \secref{torsionKVth}.

Covariant deformations in the massless case were first studied systematically 
in \Ref{abelianth}, 
using the basic field-theoretic approach of \secref{method},
with deformation terms restricted to 
a quasilinear first-order derivative form for the gauge symmetry 
and a semilinear second-order derivative form for the field equation. 
These deformations to lowest order in powers of the fields 
comprise just the \YM/ type
\EQ
\nthXvar{1}\A{a}{\mu} = \c{YM} \ym{a}{bc} \A{b}{\mu} \X{c} ,\quad
\nthsL{3} = \frac{1}{2} \c{YM} \ym{}{abc} \invvol{\mu\nu\alpha} 
\stF{a}{\mu} \A{b}{\nu} \A{c}{\alpha} , 
\doneEQ
and the \FT/ type 
\EQ
\nthXvar{1}\A{a}{\mu} = \c{FT} \ftcoad{a}{bc} \stF{b}{\mu} \X{c} ,\quad
\nthsL{3} = \frac{1}{2} \c{FT} \ftcoad{}{abc} \invvol{\mu\nu\alpha} 
\stF{a}{\mu} \stF{b}{\nu} \A{c}{\alpha} , 
\doneEQ
where the coefficients 
$\ym{}{abc} =\k{ad}\ym{d}{bc}$ and $\ftcoad{}{abc}=\k{ad}\ftcoad{d}{bc}$ 
are constant internal tensors on $\vs$ satisfying the algebraic relations
\EQs
&&
\ym{}{a(bc)} =\ym{}{(ab)c} =0 ,\quad
\ym{d}{[bc} \ym{e}{a]d} =0 ,
\label{ymalg}\\
&&
\ftcoad{}{a(bc)}=0 .
\label{ftalg}
\doneEQs
An additional algebraic relation 
\EQ
\ftcoad{d}{e[a} \ftcoad{}{bc]d} =0
\label{ftalg'}
\doneEQ
arises from an integrability condition at next order. 
Here $\c{YM},\c{FT}$ denote coupling constants. 

Strong classification results have been obtained
for lowest-order deformations of 
massless linear abelian $p$-form gauge theories in $d\geq 3$ dimensions
\cite{HenKna,BarBraHen},
relaxing the restrictions made on derivatives in the deformation terms 
in \Ref{abelianth}.
In particular, the results establish the uniqueness of these two types of 
covariant deformations in $d=3$ dimensions.

\subsection{ Deformations with \CS/ term }

The deformation analysis is readily extended to the massive case,
because the addition of a \CS/ term preserves the abelian gauge invariance. 
In outline, 
the $0$th order closure equation on the deformed gauge symmetry is unchanged,
and so the same linear deformation terms $\nthXvar{1}\A{a}{\mu}$ are obtained. 
As the key step, next the deformed gauge symmetry is found to remain compatible
with the abelian \CS/ term in the linear field equation 
when the integrability condition from the $1$st order determining equation 
on the deformed field equation is considered. 
Thus the same second derivative part for the quadratic deformation terms 
$\nthE{2}{\mu}{a}$ is obtained, while only the lower derivative part is changed
due to the \CS/ term.
This yields
\EQ
\nthsL{3}_{\rm CS} = 
\frac{1}{6} m  \c{YM} \ym{}{abc} \invvol{\mu\nu\alpha}
\A{a}{\mu} \A{b}{\nu} \A{c}{\alpha} .
\doneEQ
As a result, lowest-order covariant deformations in the massive case  
comprise the same types as in the massless case, 
apart from the presence of the \CS/ term. 
However, an algebraic integrability condition occurs at the next order. 

Additional algebraic integrability conditions emerge as well
when the \YM/ and \FT/ types of deformations are combined. 

{\bf Theorem~3.1}:
At lowest order, 
the most general allowed covariant deformation of 
the linear abelian gauge theory \sysref{quadrL}{linfieldstrid}
restricted to 
a quasilinear first-order derivative form for $\nthXvar{1}\A{a}{\mu}$
and semilinear second-order derivative form for $\nthE{1}{\mu}{a}$
is a combination of the pure \YM/ and \FT/ types. 
The combined type involves an algebraic compatibility condition
\EQ
4\ym{}{e[c|[a} \ftcoad{e}{b]|d]} + \c{}\ym{e}{cd} \ftcoad{}{abe}
-2 m \c{} \ftcoad{e}{[b|c} \ftcoad{}{e|a]d} =0
\label{combinedalg}
\doneEQ
relating the constant coefficient tensors,
where $\c{}=\c{FT}/\c{YM}$ is the ratio of coupling constants. 

It is straightforward to complete these deformations to all higher orders,
as carried out for the massless case in \Ref{nonabelianth}.

The algebraic structure \eqref{ymalg} states that 
the tensor $\ym{a}{bc}$
represents the structure constants of a Lie algebra on $\vs$
for which the symmetric tensor $\k{ab}$ is an invariant metric.
In the case where $\k{ab}$ is positive definite, 
this is well known to imply 
$(\vs,\ym{a}{bc})_{\rm YM}$ is a semisimple Lie algebra 
and $\k{ab}$ is its Cartan-Killing metric \cite{liealgebra} 
(up to a scaling factor).
In contrast, from the algebraic structure \eqrefs{ftalg}{ftalg'},
the tensor $\ftcoad{a}{bc}$ is the co-adjoint representation of 
the structure constants $\ft{e}{cd} = \invk{ed}\k{da} \ftcoad{a}{bc}$
of a Lie algebra on $(\vs,\k{ab})$,
where the adjoint is defined with respect to the vector space metric $\k{ab}$.
Note $\k{ab}$ is not required to be an invariant metric, 
so this Lie algebra $(\vs,\ft{a}{bc})_{\rm FT}$ 
is allowed to be non-semisimple
(for instance it can be solvable or nilpotent \cite{liealgebra}).

For the combined \YM/ \FT/ type of covariant deformation,
the algebraic relation \eqref{combinedalg} is satisfied by 
the identification of the two Lie algebras
\EQ
\ft{a}{bc} = \ym{a}{bc} \eqtext{ iff } \c{}=1/m 
\doneEQ
in which case $\ym{a}{bc}$ and $\ft{a}{bc}$ are necessarily 
the structure constants of a semisimple Lie algebra on $\vs$.
Indeed, the lowest-dimensional example 
satisfying all the algebra structure 
for the combined deformation in the massive case 
is readily found to be 
$\c{FT}= m\inv \c{YM}$ 
and
$(\vs,\ym{a}{bc})_{\rm YM}=(\vs,\ft{a}{bc})_{\rm FT} \iso \SU{2}$,
\ie/ $\vs$ is a three dimensional vector space
on which the Lie algebra structure constants are given by 
$\liealg{a}{bc}= \invk{ad}\liealg{}{bcd}$
where $\liealg{}{bcd}$ and $\k{ab}$ 
are a compatible volume form and positive-definite metric.
However, this Lie algebra structure is incompatible for the massless case. 
In particular, for the example of 
a semisimple Lie algebra of lowest dimension, 
$(\vs,\ym{a}{bc})_{\rm YM}\iso \SU{2}$,
the compatibility relation \eqref{combinedalg} implies
$(\vs,\ft{a}{bc})_{\rm FT}$
is the Lie algebra of
translations and a dilation in the Euclidean plane,
as shown in \Ref{nonabelianth}. 
This Lie algebra is degenerate, in contrast with $\SU{2}$.
More generally, a similar degeneracy occurs for 
$(\vs,\ft{a}{bc})_{\rm FT}$
when higher dimensional semisimple Lie algebras are considered
for $(\vs,\ym{a}{bc})_{\rm YM}$.
A full investigation of the solutions of 
the algebraic compatibility relation \eqref{combinedalg}
will be given elsewhere. 

The algebraic obstruction in the massless case to combining
\YM/ and \FT/ types of deformations with no \CS/ term
reflects a basic difference in their geometrical structures.
Consider a general semisimple Lie algebra 
$(\vs,\liealg{a}{bc})=\g$
in which the potential $\A{a}{\mu}$ is assigned to take values,
putting $\ym{a}{bc}=\ft{a}{bc}=\liealg{a}{bc}$
for the \YM/ and \FT/ Lie algebra structure tensors. 
In \YM/ theory, recall
the gauge symmetry on $\A{a}{\mu}$ geometrically is a covariant derivative
\EQ
\ymXvar{}\A{a}{\mu} = \der{\mu}\X{a} +\c{YM}\ym{a}{bc} \A{b}{\mu} \X{c} 
= \ymD{\mu}\X{a}
\doneEQ
for which the connection 1-form is $\c{YM}\ym{a}{bc}\A{b}{\mu}$.
The associated curvature $[\ymD{},\ymD{}]$ of this connection 
yields the \YM/ field strength 2-form
\EQ
\ymF{a}{\mu\nu} = 
\der{[\mu}\A{a}{\nu]} + \frac{1}{2}\c{YM} \ym{a}{bc} \A{b}{\mu} 
\label{ymfieldstr}
\doneEQ
which satisfies the Bianchi divergence identity
\EQ
\ymcoD{\mu} \ymstF{a}{\mu} =0
\doneEQ
due to $[\ymD{},[\ymD{},\ymD{}]] =0$.
The field equation in three dimensions is given by 
the covariant curl of the dual field strength 
\EQ
\ymE{a}{\mu} = \cross{\mu}{\nu\alpha} \ymD{\nu} \ymstF{a}{\alpha} =0
\doneEQ
arising from the Lagrangian 
\EQ
\sL_{\rm YM} = 
\frac{1}{2} \invmetric{\mu\nu} \k{ab} \ymstF{a}{\mu} \ymstF{b}{\nu} .
\doneEQ
The gauge group generated by $\ymXvar{}$ 
is an infinite-dimensional representation of $\g$, 
\EQ
[\ymXvar{1},\ymXvar{2}] = \ymXvar{3} , \quad
\sXsub{3} = [\sXsub{1},\sXsub{2}]
\doneEQ
where $[\sXsub{1},\sXsub{2}]$ 
denotes the Lie algebra commutator, \ie/
$\Xsub{3}{a}=\ym{a}{bc} \Xsub{1}{b} \Xsub{2}{c}$.

By comparison, in \FT/ theory 
the main geometrical role is played by the dual field strength
\EQ
\K{a}{\mu} = \invY{a \nu}{\mu b} \stF{b}{\nu}
\label{ftK}
\doneEQ
as defined using the inverse of
the symmetric linear map $\Y{}{}$ on Lie-algebra valued 1-forms
\EQ
\Y{a \nu}{\mu b} = 
\id{a}{b} \id{\nu}{\mu} 
-\c{FT} \ftcoad{a}{bc} \cross{\mu}{\nu\sigma} \A{c}{\alpha}
\label{Ylinmap}
\doneEQ
(note the inverse exists as a well-defined power series 
whenever $\A{}{}$ is restricted such that $\det \Y{}{}\neq 0$). 
This field strength appears as a connection 1-form $\ftcoad{a}{bc} \K{b}{\mu}$
introduced through the gauge symmetry
\EQ
\ftXvar{}\A{a}{\mu} = \der{\mu}\X{a} +\c{FT}\ftcoad{a}{bc} \K{b}{\mu} \X{c} 
= \ftD{\mu}\X{a}
\doneEQ
which defines a covariant derivative $\ftD{}$. 
Geometrically, the potential $\A{a}{\mu}$ is related to 
the connection by a covariant curl
\EQ
\cross{\mu}{\nu\alpha} \ftD{\nu} \A{a}{\alpha} = \K{a}{\mu} .
\label{ftAK}
\doneEQ
The field equation is given by the dual of the associated curvature 2-form
\EQ
\ftE{a}{\mu} = 
\cross{\mu}{\nu\alpha}( 
\der{\nu}\K{a}{\alpha} + \frac{1}{2}\c{FT} \ft{a}{bc} \K{b}{\nu} \K{c}{\alpha} )
= \ftstR{a}{\mu} =0
\label{ftfieldeq}
\doneEQ
and arises from the Lagrangian
\EQ
\sL_{\rm FT} = \frac{1}{2} \invmetric{\mu\nu} \K{a}{\mu} \stF{b}{\nu}
= \frac{1}{2} \Y{\mu\nu}{ab} \K{a}{\mu} \K{b}{\nu} 
\label{ftL}
\doneEQ
with $\Y{\mu\nu}{ab} = \k{ac} \invmetric{\mu\alpha} \Y{c \nu}{\alpha b}$
denoting the quadratic form corresponding to the linear map \eqref{Ylinmap}.
This Lagrangian is gauge invariant to within a divergence. 
The gauge symmetry generates a gauge group 
that is abelian on the solution space
\EQ
[\ftXvar{1},\ftXvar{2}]|_{\ftE{}{}=0} =0 .
\doneEQ
Due to the abelian Bianchi identity $\coder{\mu} \stF{a}{\mu}=0$,
the field strength satisfies a divergence identity 
\EQ
\coder{\mu} \K{a}{\mu} =0
\doneEQ
for solutions. 
Moreover, the curvature $\ftR{}{}$ vanishes on solutions,
so $\K{}{}$ is a flat connection. 
Hence
(if $M$ is assumed to be a simply connected manifold,
on which $(\vs,\ft{a}{bc})$ is viewed as a trivial bundle $\vs\times M$),
the connection 1-form is given by a chiral field strength
\EQ
\K{a}{\mu} = (\chiral\inv\der{\mu} \chiral)\upindex{a}
\label{ftchiralK}
\doneEQ
where $\chiral$ is a map from spacetime $M$ into 
the simply-connected semisimple Lie group $\G$ whose Lie algebra is $\g$,
with the tangent space of the group manifold at the identity element 
identified with the Lie algebra. 
As is customary, this map will be taken to be 
a principal matrix representation of the Lie group, 
so thus $\chiral$ is a matrix-valued function. 
The divergence identity on $\K{}{}$ now implies $\chiral$ satisfies 
the massless chiral field equation
\EQ
\coder{\mu}( \chiral\inv \der{\mu} \chiral) =0
\eqtext{ or equivalently }
\waveop\chiral = \coder{\mu}\chiral \chiral\inv \der{\mu}\chiral .
\label{chiralfieldeq}
\doneEQ
In physical terms, solutions $\chiral$ describe a nonlinearly coupled set of 
massless spin-zero (matrix) fields. 

One interesting feature common to both theories
is a geometrical relation linking the gauge symmetry, field strength,
and action of the gauge group through the covariant derivative operator,
given by 
\EQs
&&
\ymD{[\mu} \ymXvar{} \A{a}{\nu]} 
= \ym{a}{bc} \ymF{b}{\mu\nu} \X{c} 
= \ymXvar{} \ymF{a}{\mu\nu} ,
\\
&&
\ftD{[\mu} \ftXvar{} \A{a}{\nu]} 
= \ftcoad{a}{bc} \ftR{b}{\mu\nu} \X{c} 
= \cross{\mu\nu}{\alpha} \Y{a \beta}{\alpha b} \ftXvar{} \K{b}{\beta} ,
\doneEQs
in accordance with the deformation theorem~2.8 in \secref{method}.

The combined deformation with nonabelian \CS/ term 
mixes separate nonlinear aspects of \YM/ and \FT/ theories
in an interesting way. 
To proceed with the construction of the deformation to all orders,
first put $\c{FT}=m\inv \c{YM}$
and replace $\stF{}{}$ in the \FT/ field strength \eqref{ftK}
by the \YM/ field strength $\ymstF{}{}$,
\EQ
\K{a}{\mu} = \invY{a \nu}{\mu b} \ymstF{b}{\nu}
\doneEQ
where $\Y{}{}$ is unchanged. 
In terms of this nonpolynomial field strength,
the combined \YM/ and \FT/ gauge symmetry in the deformation is given by 
\EQ
\Xvar\A{a}{\mu} =
 \der{\mu}\X{a} +\c{YM}\ym{a}{bc} (\A{b}{\mu} +m\inv\K{b}{\mu})\X{c} .
\label{ftymXvar}
\doneEQ
The deformed Lagrangian is obtained by directly combining 
the \CS/ Lagrangian 
and the \FT/ Lagrangian for the deformed field strength 
\EQ
\sL = 
\frac{1}{2} \K{a}{\mu} \K{b}{\nu} \Y{\mu\nu}{ab}
+ \frac{1}{2} m \c{YM} \ym{}{abc} \invvol{\alpha\mu\nu} \A{a}{\alpha}
( \der{\mu} \A{b}{\nu} + \frac{1}{3} \ym{b}{cd} \A{c}{\mu} \A{d}{\nu} ) .
\doneEQ
Substitution of the field strength $\ymF{}{}=\Y{}{}\K{}{}$
into the \CS/ term 
yields the equivalent expression
\EQ
\sL = 
\frac{1}{2} \invmetric{\mu\nu} \k{ab} \K{b}{\nu} ( m\A{a}{\mu} +\K{a}{\mu} )
-\frac{1}{2} \c{YM} m\inv \ym{}{abc} \invvol{\mu\nu\alpha} 
( ( m\A{a}{\mu} +\K{a}{\mu} ) \K{b}{\nu} 
+m^2 \frac{1}{6} \A{a}{\mu} \A{b}{\nu} ) \A{c}{\alpha} .
\doneEQ
This Lagrangian is invariant to within a divergence 
under the gauge symmetry \eqref{ftymXvar},
\EQ
\Xvar\sL =
\der{\mu}( 
\frac{1}{2} \invvol{\mu\alpha\nu} \ym{}{abc} 
( \A{a}{\nu} ( m\A{a}{\alpha} +\K{a}{\alpha} ) 
+m\inv \K{a}{\nu} \K{b}{\alpha} ) \X{c}
+ m\invmetric{\mu\nu} \k{ab} \ymstF{a}{\nu} \X{b} 
) .
\doneEQ
The field equation obtained from the \EL/ operator applied to $\sL$ 
is given by a \CS/ term 
plus a \FT/ curvature deformed with an extra \YM/ coupling 
\EQ
\E{\mu}{a} =\invmetric{\mu\nu} \k{ab}(
m\stF{b}{\nu} +\cross{\nu}{\alpha\beta} (
\der{\alpha}\K{b}{\beta} 
+\c{YM} \frac{1}{2} \ym{b}{cd} \K{c}{\alpha} 
( m\inv \K{d}{\beta} +2 \A{d}{\beta} )
)) =0 .
\doneEQ
Through the relation between the \YM/ and \FT/ field strengths,
the field equation simplifies elegantly to
\EQ
\cross{\nu}{\alpha\beta} (
\der{\alpha}\K{b}{\beta} 
+\c{YM} m\inv \frac{1}{2} \ym{a}{cd} \K{c}{\alpha} \K{d}{\beta} )
+ m\K{a}{\mu}
=0
\doneEQ
which involves only the \FT/ field strength itself. 
The Bianchi identity on $\ymstF{}{}$ leads to 
a covariant divergence identity on this field strength
\EQ
\invmetric{\mu\nu}( 
\ftD{\mu} \K{a}{\nu} - \ftcoad{a}{bc} \E{b}{\mu} \A{c}{\nu} )
=0 .
\doneEQ
Since the Lie algebra is semisimple,
this identity reduces to an ordinary divergence
\EQ
\coder{\mu} \K{a}{\mu} =0
\doneEQ
holding on solutions of the field equation.

\subsection{ Geometrical structure }

There is a highly interesting geometrical structure behind 
this nonlinear generalization of \YM/ \CS/ gauge theory, 
blending features of 
pure \YM/ theory and pure \FT/ theory
involving their connections 
and associated covariant derivatives and curvatures. 
It will be useful to employ an index-free notation:
$\Aform = \A{a}{\mu} d\x{\mu} \basis{}{a}$ 
and $\Kform = \K{a}{\mu} d\x{\mu} \basis{}{a}$ 
are $\g$-valued 1-forms,
where $\basis{}{a}$ denotes a fixed basis for $\g$;
likewise $\Xform = \X{a} \basis{}{a}$ is $\g$-valued function.
Hereafter, $\c{YM}=m\c{FT}=1$ for simplicity.

First, the potential $\Aform$ is related to the field strength $\Kform$
by a deformed \YM/ curvature of $\covD{A}$
\EQ
-{*\Kform} = \covD{\sfrac{1}{m}K} \Aform +\frac{1}{2} [\Aform,\Aform]
\doneEQ
in which the \FT/ covariant derivative $\covD{\sfrac{1}{m}K}$
replaces an ordinary exterior derivative. 
The field equation relates the curvature of $\covD{\sfrac{1}{m}K}$
to $\Kform$ itself,
\EQ
*\Rform{\sfrac{1}{m}K} = -\Kform .
\label{ymftcurveq}
\doneEQ
The Bianchi identity associated with this curvature 
$0= \covD{\sfrac{1}{m}K} \Rform{\sfrac{1}{m}K} 
= \covD{\sfrac{1}{m}K} {*\Kform}$
reduces to an ordinary (abelian) zero curl equation 
\EQ
d{*\Kform} =0
\label{ymftcurleq}
\doneEQ
due to the algebraic identity $[\Kform,{*\Kform}]=0$.
Surprisingly, the curvature and curl equations \eqrefs{ymftcurveq}{ymftcurleq}
together constitute a massive nonabelian \CS/ Proca theory 
for the potential 
$\Aform\mass = m\inv \Kform$ in Lorentz gauge,
\EQ
\div \Aform\mass =0 ,\quad
*\Fform{A\rm mass} = -m\Aform\mass , 
\doneEQ
whose Lagrangian is 
\EQ
\Lform_{\rm CS} +\Lform_{\rm m} =
\frac{1}{2} \tr( \Aform\mass \wedge d\Aform\mass 
+\frac{1}{3} \Aform\mass\wedge [\Aform\mass,\Aform\mass]
+m^2 \Aform\mass \wedge *\Aform\mass )
\doneEQ
where $\tr$ denotes 
the inner product with respect to the Cartan-Killing metric on $\g$. 
The physical content of this theory thus describes
a massive spin-one nonabelian vector field $\Aform\mass$.

Second, the gauge symmetry defines a covariant derivative operator
\EQ
\Xvar\Aform = 
\covD{A} \Xform +m\inv [\Kform,\Xform] =\covD{K}\Xform +[\Aform,\Xform]
=\covD{A+\sfrac{1}{m}K} \Xform
\doneEQ
based on a combined connection 1-form 
$\admap{A+\sfrac{1}{m}K}=[\Aform+m\inv \Kform,\cdot]$.
Through the field equation, the curvature of this connection is zero
\EQ
\Rform{A+\sfrac{1}{m}K} =0 
\label{zerocurvcombined}
\doneEQ
on solutions $\Aform$.
Since in addition this connection transforms in the expected way
under the gauge group on solutions, 
\EQ
\Xvar( \Aform +m\inv \Kform ) = \covD{A+\sfrac{1}{m}K} \Xform ,
\doneEQ 
Lorentz gauge can be imposed 
\EQ
\div( \Aform +m\inv \Kform ) =0
\label{divgaugecombined}
\doneEQ
(which is achieved by a suitable finite gauge transformation,
analogously to \YM/ theory).
Then these equations \eqrefs{zerocurvcombined}{divgaugecombined}
are identical with the field strength equations in pure \FT/ theory
for $\Kform\flat = \Aform +m\inv \Kform$, 
\EQ
\Rform{K\flat} =0 ,\quad
\div \Kform\flat =0 .
\doneEQ
As a result, $\Kform\flat$ is a flat connection
and hence is given by 
\EQ
\Kform\flat = \chiral\inv d\chiral
\doneEQ
in terms of a chiral field $\chiral$ introduced as in pure \FT/ theory,
namely $\chiral$ is a function from spacetime 
into the simply-connected Lie group of matrices whose Lie algebra is $\g$.
Since $\chiral$ satisfies the massless spin-zero chiral equation,
the potential $\Aform$ has the decomposition
\EQ
\Aform = \chiral\inv d\chiral -\Aform\mass
\doneEQ
given by a massive spin-one part $\Aform\mass$ satisfying Lorentz gauge
and a massless spin-zero part $\chiral\inv d\chiral$ 
representing the residual gauge freedom left in Lorentz gauge.
Interestingly, if Lorentz gauge is relaxed on $\Aform$,
then $\chiral$ becomes coupled as a pure gauge field to $\Aform\mass$,
as will be seen in \secref{torsionYMth}
when the torsion generalization of this theory is discussed. 

Thus, the nonlinear generalization of \YM/ \CS/ gauge theory 
possesses a striking geometrical duality with 
pure massless chiral field theory combined with massive \CS/ gauge theory. 

Lastly, in conformance with the deformation theorem~2.8,
the gauge symmetry and field strength are linked by the geometrical relation
\EQ
\covD{A+\sfrac{1}{m}K} \Xvar\Aform 
= [\Rform{A+ \sfrac{1}{m}K},\Xform] = \Ymap(\Xvar\Kform)
\doneEQ
where $\Ymap = \idmap - \frac{1}{m}*[\Aform,\cdot]$ 
is the linear map on $\g$-valued 1-forms.

\section{ Torsion generalization of \YM/ \CS/ gauge theory }
\label{torsionYMth}

A natural setting for torsion is provided  
through the duality of \FT/ gauge theory and chiral field theory 
for a general semisimple Lie group 
$\G$ with Lie algebra $\g=(\vs,\liealg{a}{bc})$.
To express this duality at the Lagrangian level,
it is convenient to pass to a 1$st$ order formalism 
for the \FT/ Lagrangian
\EQ
\sL^{1st}_{\rm FT} =
\invmetric{\mu\nu}\k{ab}( 
\A{a}{\mu} \ftstR{b}{\nu} -\frac{1}{2} 
\K{a}{\mu}\K{b}{\nu} )
\doneEQ
where the Lie-algebra valued potential $\A{a}{\mu}$ 
and field strength $\K{a}{\mu}$
are treated as independent field variables. 
The respective field equations yield the curl equation \eqref{ftAK}
relating $\A{a}{\mu}$ to $\K{a}{\mu}$,
and the zero curvature equation \eqref{ftfieldeq} on $\K{a}{\mu}$.
Note that a substitution of $\K{a}{\mu}$ into $\sL^{1st}_{\rm FT}$
through the nonpolynomial relation \eqref{ftK} in terms of $\A{a}{\mu}$
recovers the \FT/ Lagrangian \eqref{ftL}.
A dual formulation is obtained if instead $\A{a}{\mu}$ is eliminated
by substituting a zero-curvature expression \eqref{ftchiralK} 
for $\K{a}{\mu}$ into $\sL^{1st}_{\rm FT}$ 
in terms of a chiral field.
To do this, it is useful geometrically 
to view the Lie group $\G$ in the well-known way \cite{liegrouptarget}
as a Riemannian symmetric space whose $\G$-invariant metric is 
the Cartan-Killing tensor $\k{AB}$ on the Lie group
and to write the chiral field as a map $\wavemap{A}$ 
from spacetime into $\G$, 
using local coordinates (denoted by upper case latin indices)
for the group manifold. 
Consider a left-invariant orthonormal frame $\e{A}{a}$ on this space,
satisfying the commutator relation
\EQ
[\e{}{a}, \e{}{b}]^B = 
2\e{A}{[a} \der{A} \e{B}{b]} = \c{FT} \e{B}{c} \liealg{c}{ab}
\label{framecomm}
\doneEQ
and the orthonormality relation
\EQ
\e{A}{a} \e{B}{b} \k{AB} = \k{ab}
\doneEQ
where $\liealg{c}{ab}$ is the structure tensor of the Lie algebra
and $\k{ab}=-\liealg{c}{ad} \liealg{d}{bc}$ is the Cartan-Killing metric.
By means of a corresponding coframe $\e{a}{A}$, 
which is a left-invariant Lie-algebra valued 1-form on $\G$,
the zero-curvature \FT/ field strength is given by 
\EQ
\K{a}{\mu} = \e{a}{A}(\wavemap{}) \der{\mu} \wavemap{A} .
\label{ftsigmaK}
\doneEQ
Here $\e{a}{A}(\wavemap{})$ denotes the coframe evaluated 
at the point in the group manifold given by the chiral field $\wavemap{A}$.
Substitution of $\K{a}{\mu}$ into $\sL^{1st}_{\rm FT}$ then gives
the chiral theory Lagrangian
\EQ
\sL_{\rm chiral} 
= -\frac{1}{2} \invmetric{\mu\nu} \k{AB}(\wavemap{})
\der{\mu}\wavemap{A} \der{\nu}\wavemap{B}
= -\frac{1}{2} \invmetric{\mu\nu} \k{ab} 
\e{a}{A}(\wavemap{})\der{\mu}\wavemap{A} 
\e{b}{B}(\wavemap{})\der{\nu}\wavemap{B}
\doneEQ
with the standard field equation
\EQ
\E{A}{\rm chiral} = 
\invmetric{\mu\nu}( 
\der{\mu}\der{\nu} \wavemap{A} 
+\conx{A}{BC}(\wavemap{}) \der{\mu}\wavemap{B} \der{\nu}\wavemap{C} )
=0
\doneEQ
where
\EQ
\conx{A}{BC} = \invmetric{AD}( \der{(B}\k{C)D} -\frac{1}{2}\der{D}\k{BC} )
\doneEQ
is the Christoffel symbol of the Lie group Cartan-Killing metric $\k{AB}$.
Geometrically, $\conx{A}{BC}$ describes a symmetric Riemannian connection 
determined by this metric. 
Following the approach in \Ref{AncIse},
torsion will now be added to the theory 
by a geometrical term in the Lagrangian of the form
\EQ
\sL_{\rm tors.} =
-\frac{1}{2} \v{\mu\nu}{} \p{}{AB}(\wavemap{})
\der{\mu}\wavemap{A} \der{\nu}\wavemap{B}
\label{sigmatorsL}
\doneEQ
which involves introducing 
a constant skew tensor $\v{\mu\nu}{}$ on spacetime
and a left-invariant 2-form $\p{}{AB}$ on the Lie group. 
This contributes a torsion term to the chiral field equation
\EQ
\E{A}{\rm tors.} =
\v{\mu\nu}{} \Q{A}{BC}(\wavemap{})
\der{\mu}\wavemap{B} \der{\nu}\wavemap{C}
\doneEQ
where
\EQ
\Q{A}{BC} = -\frac{3}{2} \invk{AD} \der{[D} \p{}{BC]}
\doneEQ
which serves geometrically as torsion 
added into the connection on the Lie group. 
The field equation with torsion is thus given by 
\EQ
\E{A}{\rm chiral} +\E{A}{\rm tors.} =
( \invmetric{\mu\nu} + \v{\mu\nu}{} )
( \der{\mu}\der{\nu} \wavemap{A} 
+( \conx{A}{BC}(\wavemap{}) + \Q{A}{BC}(\wavemap{}) ) 
\der{\mu}\wavemap{B} \der{\nu}\wavemap{C} )
=0 .
\doneEQ

Because the torsion potential $\p{}{BC}$ is $\G$-invariant,
it has a well defined pull back to the Lie algebra,
so 
\EQ
\p{}{ab} = \e{A}{a} \e{B}{b} \p{}{AB}
\doneEQ
is a constant tensor
as will be necessary for torsion to be compatible with duality. 
Then, through the relation 
\EQ
\liealg{c}{ab} = -2\e{A}{a} \e{B}{b} \der{[A} \e{c}{B]}
\doneEQ
derived from the commutator \eqref{framecomm},
the torsion in the connection leads to a corresponding constant tensor
\EQ
\Q{a}{bc} = \frac{3}{2}\invk{ae} \p{}{d[e} \liealg{d}{bc]}
= \e{a}{A} \e{B}{b} \e{C}{c} \Q{A}{BC}
\label{tors}
\doneEQ
in terms of the metric $\k{ab}$ and torsion potential $\p{}{ab}$ on $\vs$.
This allows the torsion Lagrangian \eqref{sigmatorsL}
to be expressed entirely in terms of the field strength \eqref{ftsigmaK} 
by 
\EQ
\sL^{1st}_{\rm tors.} = 
-\frac{1}{2} \v{\mu\nu}{} \p{}{ab} \K{a}{\mu} \K{b}{\nu} .
\doneEQ

A dual formulation of the torsion chiral theory Lagrangian 
$\sL_{\rm chiral}+\sL_{\rm tors.}$
is now obtained simply by adding $\sL^{1st}_{\rm tors.}$ to 
the 1$st$ order form of the \FT/ Lagrangian $\sL^{1st}_{\rm tors.}$.
This yields
\EQ
\sL^{1st}_{\rm FT} + \sL^{1st}_{\rm tors.}
= \invmetric{\mu\nu} \k{ab} \A{a}{\mu} \ftstR{b}{\nu}
-\frac{1}{2} \torsmetric{\mu\nu}{ab} \K{a}{\mu} \K{b}{\nu}
\doneEQ
where it is convenient to introduce
\EQ
\torsmetric{\mu\nu}{ab} =
\invmetric{\mu\nu} \k{ab} +\v{\mu\nu}{} \p{}{ab}
\label{gmetric}
\doneEQ
which defines a symmetric inner product on
Lie-algebra valued 1-forms.
The effect of torsion in this Lagrangian is to generalize 
the field equation of $\K{a}{\mu}$ so it becomes
\EQ
\K{a}{\mu} = \invtorsY{a \nu}{\mu b} \stF{b}{\nu}
\label{fttorsK}
\doneEQ
using the inverse of the linear map $\torsY{}{}$
covariantly constructed in terms of the potential $\A{a}{\mu}$,
where
\EQ
\torsY{\mu\nu}{ab} = 
\torsmetric{\mu\nu}{ab} 
-\invvol{\mu\nu\alpha} \c{FT}\liealg{}{abc} \A{c}{\alpha}
= \torsY{c \mu}{\alpha b} \invmetric{\mu\alpha} \k{ac}
\doneEQ
is the corresponding quadratic form on Lie-algebra valued 1-forms. 
A simple algebraic analysis shows that the inner product \eqref{gmetric}
is nondegenerate, so the linear map 
\EQ
\torsY{a \nu}{\mu b} = 
\id{b}{a}\id{\mu}{\nu} +\p{a}{b} \v{\nu}{\mu} 
- \cross{\mu}{\nu\alpha} \c{FT}\liealg{a}{bc} \A{c}{\alpha}
\label{torsYlinmap}
\doneEQ
is invertible by a power series in $\A{}{}$
(under the restriction $\det \torsY{}{}\neq 0$),
with $\p{a}{b} = \invk{ac} \p{}{cb}$ 
and $\v{\nu}{\mu} = \metric{\mu\alpha}\v{\alpha\nu}{}$.
Substitution of the field strength \eqref{fttorsK}
back into $\sL^{1st}_{\rm FT} + \sL^{1st}_{\rm tors.}$
then yields a torsion generalization of \FT/ gauge theory 
\EQ
\sL^{\rm tors.}_{\rm FT} 
= \torsmetric{\mu\nu}{ab} \K{a}{\mu} \K{b}{\nu}
= \invtorsmetric{\mu\nu}{ab} \stF{a}{\mu} \stF{b}{\nu} .
\doneEQ
In this generalization,
the Lagrangian remains gauge-invariant to within a divergence, 
such that the gauge symmetry has the same form given by 
the covariant derivative of the gauge parameter $\X{a}$
using $\K{a}{\mu}$ as a Lie-algebra valued connection 1-form. 
This is an immediate consequence of the gauge invariance of
$\sL^{1st}_{\rm FT}$ and $\sL^{1st}_{\rm tors.}$ under 
$\delta^{1st}_{\sX}\K{a}{\mu} =0$, 
$\delta^{1st}_{\sX}\A{a}{\mu} =\ftD{\mu}\X{a}$.
Moreover, from the form of $\sL^{1st}_{\rm FT}$,
the field equation coming from $\sL^{\rm tors.}_{\rm FT}$
retains the form of a zero-curvature equation on $\K{a}{\mu}$.
The only place where torsion enters explicitly is 
in the divergence equation on $\K{a}{\mu}$
\EQ
\coder{\mu} \K{a}{\mu} + \Q{a}{bc} \v{\mu\nu}{} \K{b}{\mu} \K{c}{\nu}
= \ftR{b}{\mu\nu} 
( \invvol{\mu\nu\alpha} \c{FT} \liealg{a}{bc} \A{c}{\alpha} 
-\v{\mu\nu}{} \p{a}{b} )
\doneEQ
derived through the identity 
$0= \coder{\mu} \stF{a}{\mu} 
= \torsmetric{a \nu}{\mu b} \coder{\mu} \K{b}{\nu}$.
On solutions of the field equation,
this divergence equation reduces precisely to the dual form of 
the torsion chiral field equation \eqref{chiralfieldeq}.

Finally, it is worth noting that the algebraic formulation \eqref{tors}
of torsion in the Lie algebra gives rise to a slight restriction on
the dimension of the Lie group \cite{AncIse}.

{\bf Proposition~4.1}:
For semisimple Lie groups $\G$, 
if the dimension of $\G$ is three, 
then every torsion potential has the form $\p{}{bc} =\p{}{d} \liealg{d}{bc}$
for some vector $\p{}{d}$ in the Lie algebra
and hence the torsion vanishes by the Jacobi identity. 
This is not the case if $\G$ has dimension larger than three.
In particular, 
every such Lie group possesses 
an abelian subalgebra of dimension at least two,
allowing a non-zero torsion tensor $\Q{a}{bc}$ to exist 
with a potential given by the bi-vector form 
$\p{}{ab} = 2\p{}{[a} \q{}{b]}$, 
where $\p{}{a},\q{}{a}$ are any two (linearly independent) commuting vectors
in the Lie algebra. 

\subsection{ Linear abelian theory with torsion }

The main goal now will be to derive a \YM/ generalization of
the \FT/ gauge theory with torsion,
in the setting of a deformation analysis as follows. 
To begin, consider a linearization of the torsion \FT/ theory 
around the zero potential $\A{a}{\mu}=0$
by turning off the \FT/ coupling $\c{FT}=0$.
This gives a novel abelian gauge theory with torsion 
for the potential $\A{a}{\mu}$.
The gauge-invariant quadratic Lagrangian takes the form 
\EQ
\nthsL{2} = \invtorsmetric{\mu\nu}{ab} \stF{a}{\mu} \stF{b}{\nu}
\label{torsquadrL}
\doneEQ
where $\invtorsmetric{\mu\nu}{ab}$ denotes the inverse of
the nondegenerate symmetric inner product \eqref{gmetric}
with respect to the spacetime metric $\metric{\mu\nu}$
and the internal vector space metric $\k{ab}$,
such that
$\invtorsmetric{\mu\alpha}{ac} \metric{\alpha\beta} \invk{cd}
\invtorsmetric{\beta\nu}{db} = \invmetric{\mu\nu} \k{ab}$.
There is no change to the form of 
the abelian gauge symmetry, 
$\nthXvar{0}\A{a}{\mu} = \der{\mu}\X{a}$. 
The linear field equation of this theory 
\EQ
\nthE{1}{\mu}{a} = -\cross{\alpha}{\nu\mu} \der{\nu} \torsF{\alpha}{a}
\doneEQ
is given by the curl of the gauge-invariant torsion field strength
\EQ
\torsF{\mu}{a} 
= \invtorsmetric{\mu\nu}{ab} \stF{b}{\nu}
= \invtorsmetric{\mu\nu}{ab} \cross{\nu}{\alpha\beta} 
\der{\alpha} \A{b}{\beta} .
\label{torsF}
\doneEQ
It is interesting to include an abelian \CS/ term in the theory 
by adding the Lagrangian $\nthsL{2}_{\rm CS}$. 
This contributes the same linear \CS/ term to the field equation
as in the case without torsion, which now gives
\EQ
\linE{CS}{\mu}{a} 
= m\invmetric{\mu\nu} \k{ab} \stF{b}{\nu}
= m\torsmetric{\mu b}{a \nu} \torsF{\nu}{b}
= m\torsF{\mu}{a} +m\v{\mu}{\nu} \p{b}{a} \torsF{\nu}{b}
\doneEQ
after inverting the field strength relation \eqref{torsF}.
Thus, $\A{a}{\mu}$ satisfies the gauge-invariant field equation
$\linE{}{\mu}{a} +\linE{CS}{\mu}{a}=0$. 

To understand the physical meaning of 
torsion in this theory,
it is useful to look at the equations satisfied by 
the gauge-invariant field strength $\torsF{\mu}{a}$. 
First write the skew tensor 
\EQ
\v{\mu\nu}{} = \invvol{\mu\nu\alpha} \v{}{\alpha}
\doneEQ
by introducing the dual vector $\v{}{\alpha}$ on spacetime. 
Note this vector represents an extra background structure
introduced on spacetime, which breaks Lorentz covariance. 
Then the field equation $\linE{}{\mu}{a} +\linE{CS}{\mu}{a}=0$
has the explicit non-covariant form
\EQ
\cross{\alpha}{\nu\mu} \der{\nu} \torsF{\alpha}{a}
= m \torsF{\mu}{a} 
-m\p{b}{a} \cross{\alpha}{\nu\mu} \v{}{\nu} \torsF{\alpha}{b} .
\label{torslinfieldeq}
\doneEQ
Next the abelian Bianchi identity 
$0= \coder{\mu} \stF{a}{\mu} 
= \torsmetric{b \mu}{\nu a} \der{\mu} \torsF{\nu}{b}$
gives the divergence equation
\EQ
\der{\mu} \torsF{\mu}{a} = -m\p{b}{a} \v{}{\mu} \torsF{\mu}{b}
\doneEQ
after simplifications of the curl term through the field equation. 
These curl and divergence equations on $\torsF{\mu}{a}$ 
have a more elegant form if expressed via the derivative operator 
\EQ
\torsder{\mu} = \der{\mu} +m \p{b}{a} \v{}{\mu}
\label{torsder}
\doneEQ
based on a connection $m \p{b}{a} \v{}{\mu}$ 
for Lie-algebra valued differential forms on $M$. 
Because $\v{}{\mu}$ and $\p{b}{a}$ are constant on $M$
and have vanishing exterior product, this connection is flat.
Thus, the field strength $\torsF{\mu}{a}$ satisfies 
\EQ
\torsder{\mu} \torsF{\mu}{a} =0 ,\quad
\cross{\alpha}{\beta\mu} \torsder{\beta} \torsF{\alpha}{a} = -m\torsF{\mu}{a} ,
\doneEQ
which differ from the standard field strength equations only by
the torsion parts in the definition of $\torsF{\mu}{a}$ and $\torsder{\mu}$. 

Consequently, as in the case without torsion,
a further curl applied to the curl equation on the field strength
yields a vector wave equation
\EQ
\widetilde\waveop \torsF{\mu}{a} = m^2 \torsF{\mu}{a} , 
\doneEQ
and hence the torsion field strength physically describes
a set of free spin-one fields of mass $m$, 
each with a single dynamical degree of freedom. 
In the massless case $m=0$, 
the torsion does not explicitly appear in this wave equation, 
so the spin-one vector fields $\torsF{\mu}{a}$ have the same propagation
as in the torsionless covariant theory, 
$\waveop\torsF{\mu}{a}=0$. 
But in the massive case $m\neq 0$, 
the corresponding propagation is altered by 
a novel non-covariant ``twist'' effect
due to the connection term $m\p{b}{a}\v{}{\mu}$ 
in the derivative operator $\torsder{\mu}$ 
as result of the torsion. 
More precisely, for plane wave solutions 
$\torsF{\mu}{a} = \polarize{\mu}{a} \exp(\i \k{\alpha}\x{\alpha})$,
the polarization $\polarize{\mu}{a}=\const$ 
is found to be internally aligned 
relative to the eigenvectors of $\p{b}{a}$,
while the mass shell condition on the propagation vector 
$\k{\alpha}=\const$
is found to undergo a shift proportional to $\v{}{\mu}$
determined by the eigenvalues of $\p{b}{a}$.

The field-theoretic content of torsion is best seen
by passing to the dual form of the free Lagrangian 
$\nthsL{2}+\nthsL{2}_{\rm CS}$. 
Observe the terms in the linear field equation \eqref{torslinfieldeq}
have the form of a curl, 
$\linE{}{\mu}{a} +\linE{CS}{\mu}{a} = 
\invvol{\mu\nu\alpha} \k{ab}\der{\nu}( \torsF{b}{\alpha} +m\A{b}{\alpha} )$.
So, for solutions $\A{a}{\mu}$,
\EQ
\torsF{a}{\mu} +m\A{a}{\mu} =\der{\mu}\wavemap{a}
\doneEQ
is a gradient. 
Here, the $\vs$-valued scalar field 
$\wavemap{a}=\e{a}{A}\wavemap{A}$ 
geometrically represents an abelian chiral field
in a left-invariant frame $\e{a}{A}$ on the abelian Lie group
whose Lie algebra is identified with the vector space $\vs$.
Elimination of the torsion field strength $\torsF{a}{\mu}$
from $\nthsL{2}+\nthsL{2}_{\rm CS}$ leads to the equivalent dual Lagrangian
\EQ
\nthsL{2}_{\rm dual}
= -\frac{1}{2} \invtorsmetric{\mu\nu}{ab}
( \der{\mu}\wavemap{a} -m\A{a}{\mu} )( \der{\nu}\wavemap{a} -m\A{a}{\nu} )
-\frac{1}{2} \k{ab} m \invvol{\mu\nu\alpha} \A{a}{\mu}\der{\nu} \A{b}{\alpha}
\doneEQ
which describes an abelian chiral theory for $\wavemap{a}$
coupled as a pure gauge field to an abelian \CS/ theory for $\A{a}{\mu}$,
containing torsion. 

\subsection{ Nonlinear deformation }

It is now of obvious interest to consider the compatibility of torsion
with the \YM/ and \FT/ types of covariant deformations of 
this linear abelian gauge theory for the potential $\A{a}{\mu}$.
Because the abelian gauge symmetry $\nthXvar{0}\A{a}{\mu}$ is
independent of torsion,
the closure equations up to $1$st order for determining 
the linear deformation terms in the gauge symmetry
continue to have solutions of the \YM/ and \FT/ types 
\EQ
\nthXvar{1}\A{a}{\mu} = 
\c{YM} \ym{a}{bc} \A{b}{\mu} \X{c} 
+ \c{FT} \ftcoad{a}{bc} \torsF{b}{\mu} \X{c}
\label{combinedlinXvar}
\doneEQ
where the constant coefficient tensors 
$\ym{}{abc} =\k{ad}\ym{d}{bc}$ and $\ft{}{abc}=\k{ad}\ftcoad{d}{bc}$ 
satisfy the algebraic relations \eqrefs{ymalg}{ftalg},
and $\c{YM},\c{FT}$ are independent coupling constants.
Note these deformation terms are constructed covariantly 
from the potential $\A{a}{\mu}$ 
and the torsion field strength $\torsF{\mu}{a}$.
A straightforward analysis of the determining equations for
the quadratic deformation terms in the field equation yields
\EQ
\nthsL{3} = \frac{1}{2} \c{YM} \ym{a}{bc} \cross{\mu}{\nu\alpha} 
\torsF{\mu}{a} \A{b}{\nu} \A{c}{\alpha}
+ \frac{1}{2} \c{FT} \ft{ab}{c} \cross{\mu\nu}{\alpha} 
\torsF{\mu}{a} \torsF{\nu}{b} \A{c}{\alpha}
+\frac{1}{6} m \invvol{\mu\nu\alpha} \c{YM} \ym{}{abc} 
\A{a}{\mu} \A{b}{\nu} \A{c}{\alpha}
\label{combinedcubicL}
\doneEQ
together with the algebraic relations \eqrefs{ftalg'}{combinedalg}
in addition to 
\EQ
\p{}{d[a}( \ym{d}{b]c} -m\c{} \ft{d}{b]c}) =0
\label{torsalg}
\doneEQ
where $\c{}=\c{FT}/\c{YM}$.
Notice that, up to this order in the deformation analysis,
only the algebraic relation \eqref{torsalg} involves torsion. 
The algebraic structure 
\eqref{ymalg}, \eqref{ftalg}, \eqref{ftalg'}, 
\eqref{combinedalg}, \eqref{torsalg}
comprises all integrability conditions at 1st and 2nd orders. 
This establishes the following main result.

{\bf Theorem~4.2}:
The algebraic condition \eqref{torsalg} 
is the only obstruction to combining \YM/ and \FT/ types of deformation
of the linear abelian theory with torsion
\sysref{torsquadrL}{torslinfieldeq}, including a \CS/ mass. 

As in the case with no torsion, 
the other algebraic relations require $\ym{a}{bc}$ to be 
the structure constants of a semisimple Lie algebra on $\vs$
whose Cartan-Killing metric is $\k{ab} = -\ym{d}{ac} \ym{c}{bd}$.
In the massless case, the torsion obstruction states
$\p{}{d[a}\ym{d}{b]c} =0$.
Contraction with $\ym{bc}{e}$ leads to 
$\p{}{ae} = \frac{1}{2} \ym{}{bae} \p{b}{}$ 
by the Jacobi identity,
where $\p{b}{} = \p{}{ad} \ym{adb}{}$.
However, contraction with $\ym{ab}{e}$ gives
$0=\ym{ab}{e} \ym{d}{bc} \p{}{ad} = \frac{1}{2} \p{b}{} \ym{}{bce}$,
which implies the vector $\p{b}{}$
lies in the center of the Lie algebra $(\vs,\ym{a}{bc})_{\rm YM}$.
Hence $\p{b}{}$ vanishes since the Lie algebra is semisimple,
and so $\p{}{ae}=0$. 
This algebraic analysis therefore gives a strong rigidity result. 

{\bf Corollary~4.3}:
At lowest order, 
the only deformation of covariant form 
\sysref{combinedlinXvar}{combinedcubicL}
that is compatible with torsion in the linear abelian theory 
in the massless case (without \CS/ term)
consists of the pure \FT/ type. 

In contrast, 
the torsion obstruction in the massive case has a quite different nature.
Recall, the algebraic relations \eqrefs{ftalg}{ftalg'} without torsion
hold if $\ft{a}{bc}$ is identified with $\ym{a}{bc}$
as the structure constants of a semisimple Lie algebra on $\vs$
and if $\c{}$ equals $1/m$. 
This algebraic structure also trivially satisfies 
the obstruction \eqref{torsalg}, 
and thus the combined \YM/ and \FT/ type of covariant deformation 
at lowest order is compatible with torsion in the linear abelian theory,
when the \CS/ mass is nonzero. 
A construction of this deformation to all orders amounts to 
turning on the \YM/ \CS/ interaction in the torsion \FT/ gauge theory.
Alternatively, the construction can be carried out more directly 
by adding torsion to the combined \FT/ \YM/ gauge theory 
with \CS/ term presented earlier, 
as will be now shown. 

As with the pure \FT/ gauge theory, 
the introduction of torsion is accomplished by generalizing 
the symmetric linear map $\Y{a \nu}{\mu b}$ to $\torsY{a \nu}{\mu b}$,
which enters into constructing the nonlinear torsion field strength 
appearing in the gauge symmetry, field equation, and Lagrangian. 
This generalization is a nonlinear analog of replacing the metrics 
$\invmetric{\mu\nu}$ and $\k{ab}$
in the Lagrangian of the linear abelian gauge theory \eqref{quadrL}
by the symmetric inner product $\torsmetric{\mu\nu}{ab}$
to arrive at the Lagrangian for the torsion theory \eqref{torsquadrL}. 

Thus, consider a general semisimple Lie algebra 
$(\vs, \liealg{a}{bc})=\g$
and put $\c{FT}=m\inv \c{YM}$.
Now introduce the torsion field strength 
\EQ
\torsK{a}{\mu} = \invtorsY{a \nu}{\mu b} \ymstF{b}{\nu}
\doneEQ
where $\ymstF{}{}$ is the \YM/ dual field strength \eqref{ymfieldstr}
and $\torsY{}{}$ is the linear map \eqref{torsYlinmap}
with inverse $\invtorsY{}{}$.
The complete Lagrangian with torsion is given by 
combining the \CS/ Lagrangian and the \FT/ torsion Lagrangian 
with the field strength replaced by $\torsK{a}{\mu}$, 
\EQs
\sL_{\rm tors.}
&&
= \frac{1}{2} \torsY{\mu\nu}{ab} \torsK{a}{\mu} \torsK{b}{\nu}
+ \frac{1}{2} m \c{YM} \ym{}{abc} 
\A{a}{\alpha} ( \der{\mu} \A{b}{\nu} 
+\frac{1}{3}\ym{b}{cd} \A{c}{\mu} \A{d}{\nu} )
\label{torsYMFTCSL}
\\&&
= 
\frac{1}{2} \torsmetric{\mu\nu}{ab} 
\torsK{b}{\nu} ( m\A{a}{\mu} +\torsK{a}{\mu} )
-\frac{1}{2} \c{YM} m\inv \ym{}{abc} \invvol{\mu\nu\alpha} 
( ( m\A{a}{\mu} +\torsK{a}{\mu} ) \torsK{b}{\nu} 
+m^2 \frac{1}{6} \A{a}{\mu} \A{b}{\nu} ) \A{c}{\alpha} .
\nonumber
\doneEQs
Similarly, the gauge symmetry on $\A{a}{\mu}$
is obtained by replacing $\torsK{a}{\mu}$
for the field strength in $\Xvar\A{a}{\mu}$ for \FT/ torsion theory.
Under this gauge symmetry
\EQ
\Xvar\A{a}{\mu} =
 \der{\mu}\X{a} +\c{YM}\ym{a}{bc} (\A{b}{\mu} +m\inv\torsK{b}{\mu})\X{c} , 
\doneEQ
the Lagrangian \eqref{torsYMFTCSL} is invariant to within a divergence.
In the field equation obtained from this Lagrangian,
\EQ
\E{\mu}{a} =\invmetric{\mu\nu} \k{ab}(
m\stF{b}{\nu} +\cross{\nu}{\alpha\beta} (
\der{\alpha}\torsK{b}{\beta} 
+\c{YM} \frac{1}{2} \torsK{c}{\alpha} ( m\inv \torsK{d}{\beta} +2\A{d}{\beta} )
) ,
\doneEQ
substitution of $\ymstF{}{}= \torsY{}{}\torsK{}{}$ 
leads to the elegant simplification
\EQ
\cross{\nu}{\alpha\beta} (
\torsder{\alpha}\torsK{b}{\beta} 
+\c{YM} \frac{1}{2} m\inv \torsK{c}{\alpha} \torsK{d}{\beta} )
=-m\torsK{a}{\mu}
\doneEQ
where $\torsder{\mu}$ is the derivative operator \eqref{torsder}
given by the flat connection 
$m\p{a}{b} \v{}{\mu}$ for Lie-algebra valued differential forms 
as introduced in the linear abelian theory. 
Likewise, the field strength divergence equation coming from 
the \YM/ Bianchi identity $0=\ymcoD{\mu}\ymstF{a}{\mu}$
simplifies to 
\EQ
\torscoder{\mu} \torsK{a}{\mu} 
+ \Q{a}{bc} \v{\mu\nu}{} \torsK{b}{\mu} \torsK{c}{\nu}
=0
\doneEQ
on solutions of the field equation,
where $\Q{a}{bc} = \frac{3}{2} \invk{ae} \p{}{d[e} \liealg{d}{bc]}$
is the torsion tensor on $\vs$.

Some field-theoretic properties of this novel torsion generalization of 
\YM/ \CS/ theory are worth noting. 
The gauge group generated by the gauge symmetry on solutions
is an infinite-dimensional representation of $\g$, 
\EQ
[\Xvarsub{1},\Xvarsub{2}] = \Xvarsub{3} ,\quad
\Xsub{3}{a} = \liealg{a}{bc} \Xsub{1}{b} \Xsub{2}{c} .
\doneEQ
Remarkably, the field strength for solutions is gauge invariant 
under this gauge group,
\EQ
\Xvar\torsK{a}{\mu} =0 .
\doneEQ
A natural subgroup of the gauge group consists of rigid symmetries
\EQ
\Xrigidvar \A{a}{\mu} =
 \c{YM} \liealg{a}{bc}( \A{a}{\mu} +m\inv \torsK{a}{\mu} ) \X{c}\rigid
\doneEQ
where $\X{c}\rigid=\const$
(note the dimension of this subgroup is equal to 
the internal number of fields,
\ie/ the dimension of $\g$).
Application of Noether's theorem to these symmetries
yields a set of conserved currents
\EQ
\current{a}{\mu} = 
\liealg{a}{bc} \cross{\mu}{\nu\alpha} \torsK{b}{\nu} \torsK{c}{\alpha} 
+ m\torsmetric{a \nu}{\mu b}\torsK{b}{\nu}
\doneEQ
which are gauge invariant $\Xvar\current{a}{\mu} = 0$
and divergence free $\coder{\mu}\current{a}{\mu} = 0$
on solutions $\A{a}{\mu}$.
Given a two dimensional hypersurface $\Sigma$ in spacetime $M$,
the flux of these currents defines internal charges 
carried by the field $\A{a}{\mu}$,
$\charge{a} =\int_\Sigma \current{a}{\mu} d\Sigma^\mu$
where $\vol{\mu\nu\alpha} d\Sigma^\mu$ 
is the volume element pulled back to  $\Sigma$.
For solutions $\A{a}{\mu}$, these charges are evaluated by a line integral 
around the boundary at infinity on $\Sigma$ by Stokes' theorem,
$\charge{a} = \int_{\partial\Sigma} \K{a}{\mu} dS^\mu$
where $dS^\mu$ is the line element on $\partial\Sigma$.
When there is no current flow across $\partial\Sigma$,
the charges are conserved and gauge invariant. 

Under the action of diffeomorphisms on spacetime,
the Lagrangian gives rise to a stress-energy tensor
\EQ
\T{}{\mu\nu} = 
\k{ab}( \torsK{a}{\mu} \torsK{b}{\nu} 
-\frac{1}{2} \metric{\mu\nu} \invmetric{\alpha\beta} 
\torsK{a}{\alpha} \torsK{b}{\beta} )
-\p{}{ab} \frac{1}{2} \cross{\mu}{\alpha\beta} \v{}{\nu}
\torsK{a}{\alpha} \torsK{b}{\beta} 
\doneEQ
which is conserved $\coder{\mu}\T{}{\mu\nu}=0$ 
and gauge-invariant $\Xvar\T{}{\mu\nu}=0$
for solutions of the field equation.
Notice, this stress-energy tensor is non-symmetric, due to the torsion term. 
As a result, 
contraction of $\T{}{\mu\nu}$ with 
a \KV/ $\kv{\nu}{}$ of the spacetime metric 
$\Lie{\kv{}{}} \invmetric{\mu\nu}= -2\coder{(\mu}\kv{\nu)}{}=0$
yields a conserved current $\coder{\mu}( \T{}{\mu\nu}\kv{\nu}{} )=0$
if $\kv{\nu}{}$ is either a translation, 
or a rotation/boost in a hyperplane orthogonal to the vector $\v{\nu}{}$,
\ie/ $\Lie{\v{}{}} \kv{\mu}{} = \v{}{\nu}\coder{[\mu} \kv{\nu]}{}=0$.
If a two dimensional hypersurface $\Sigma$ in spacetime $M$ 
is considered as before, these conserved currents then define 
gauge-invariant fluxes of energy-momentum and angular momentum
carried by the field $\A{a}{\mu}$ on $\Sigma$,
$\charge{}{}_{\kv{}{}} = \int_\Sigma \T{}{\mu\nu}\kv{\nu}{} d\Sigma^\mu$.

\subsection{ Geometrical structure}

The nonlinear structure of this theory has several striking aspects that merge
important geometrical features of pure \FT/ theory with torsion 
and combined \YM/ \FT/ \CS/ theory without torsion. 
Here it will be useful to return to an index-free notation,
$\Aform = \A{a}{\mu} d\x{\mu} \basis{}{a}$ 
and $\Kform = \K{a}{\mu} d\x{\mu} \basis{}{a}$.
Write $\vform=\v{}{\mu}d\x{\mu}$ as a covariantly constant 1-form on $M$
and $\p{}{}(\cdot)=\p{a}{b} \basis{}{a} \basis{b}{}$ 
as a constant linear map on $\vs$.
Regard the torsion tensor $\Q{a}{bc}$ on $\vs$ 
as a bilinear skew-symmetric map 
$\Q{}{}(\cdot)$ from $\vs\times\vs$ into $\vs$.
To proceed, firstly, 
the field equation has the geometrical content that 
the Lie-algebra valued 1-form 
$\Aform +m\inv \tilde\Kform = \Aform\flat$
is a flat connection
\EQ
\Rform{A\flat} = d\Aform\flat +\frac{1}{2}[\Aform\flat,\Aform\flat] =0 .
\doneEQ
This connection therefore has a zero-curvature form given by 
\EQ
\Aform\flat=\chiral\inv d\chiral
\doneEQ
where $\chiral$ is a chiral field,
namely a function from spacetime into 
the simply-connected Lie group of matrices whose Lie algebra is $\g$.
Secondly, 
the field strength part $m\inv\tilde\Kform$ of this flat connection 
satisfies the curl and divergence equations
\EQs
&&
m\inv \torsd\tilde\Kform +\frac{1}{2} m^{-2}[\tilde\Kform,\tilde\Kform]
= -\tilde\Kform ,
\label{torsKcurveq}
\\&&
m\inv \torsd*\tilde\Kform 
+m^{-2} \vform\wedge \Q{}{}(\tilde\Kform \wedge \tilde\Kform) ,
\label{torsKdiveq}
\doneEQs
involving the exterior derivative $\torsd =d -m\vform \p{}{}(\cdot)$
on Lie-algebra valued differential forms on $M$.
Note this derivative obeys $\torsd{}^2=0$
and reduces to $\torsd=d$ when acting on Lie-algebra valued functions on $M$.
The respective field strength equations \eqrefs{torsKcurveq}{torsKdiveq}
geometrically comprise a nonabelian \CS/ theory with a Proca mass term 
for the 1-form potential $m\inv \tilde\Kform = \Aform\mass$
in a nonabelian Lorentz-torsion gauge
\EQ
*\torsFform{A\mass} +m \Aform\mass =0 ,\quad
\widetilde\div \Aform\mass 
+*\vform \cdot \Q{}{}( \Aform\mass\wedge \Aform\mass ) 
=0
\doneEQ
with $\torsFform{A\mass} 
= \torsd\Aform\mass + \frac{1}{2}[\Aform\mass,\Aform\mass]$.
Solutions $\Aform\mass$ describe 
a set of coupled spin-one vector fields of mass $m$
whose propagation on spacetime involves a ``twist'' effect
due to the nature of $\torsd$ 
discussed in the linear abelian theory. 
Moreover, if the same nonabelian Lorentz-torsion gauge condition
is imposed on $\Aform\flat$, 
\EQ
\widetilde\div \Aform\flat 
+*\vform \cdot \Q{}{}( \Aform\flat\wedge \Aform\flat ) =0 ,
\label{torsAflatgauge}
\doneEQ
then $\chiral$ is found to satisfy 
a ``twisted'' massless chiral field equation
\EQ
0= \widetilde\div( \chiral\inv \widetilde\grad\chiral) 
\eqtext{ or equivalently }
\widetilde\waveop\chiral 
= \widetilde\grad\chiral \cdot( \chiral\inv \widetilde\grad\chiral ) .
\doneEQ
In particular, solutions $\chiral$ describe
a set of coupled massless spin-zero matrix fields
exhibiting ``twist'' in their propagation on spacetime. 
Note the gauge condition \eqref{torsAflatgauge}
can be achieved through 
a suitable finite gauge transformation on $\Aform\flat$,
since the gauge symmetry acts as
\EQ
\Xvar\Aform\flat = d\Xform +[\Aform\flat,\Xform] ,\quad
\Xvar\Aform\mass = 0
\doneEQ
on solutions of the field equation. 
Therefore, the 1-form potentials $\Aform\flat$ and $\Aform\mass$
give a geometrical decomposition of $\Aform =\Aform\flat -\Aform\mass$
into massive and massless parts. 

Without the torsion-Lorentz gauge on $\Aform\flat$, 
the chiral field $\chiral$ exhibits a pure gauge coupling to
the massive \YM/ field $\Aform\mass$. 
The precise nature of this coupling is best seen through 
the dual formulation of the torsion Lagrangian $\sL_{\rm tors.}$
obtained by elimination of $\tilde\Kform$ in $1$st order form,
which yields
\EQ
-\frac{1}{m} \sL^{\rm tors.}_{\rm dual} 
= 
m \torsmetric{}{}( \chiral\inv d\chiral-\Aform, \chiral\inv d\chiral-\Aform )
+ {*\tr}( -\Aform \wedge( \covD{U}(-\Aform) +\frac{1}{3}[-\Aform,-\Aform] ))
\doneEQ
in terms of $-\Aform = \Aform\mass -\chiral\inv d\chiral$, 
where $\covD{U}= d+(\chiral\inv d\chiral) = \chiral\inv d(\chiral\cdot)$
is a zero-curvature derivative operator, $(\covD{U})^2=0$. 

Lastly, $\Aform,\Aform\flat,\Aform\mass$ 
are also linked geometrically by the action of the gauge group
\EQ
\covD{A\flat} \Xvar\Aform 
= [\Rform{A\flat},\Xform] = \torsYmap(\Xvar\Aform\mass)
\doneEQ
where $\torsYmap = \idmap-*(\vform\wedge\p{}{}(\cdot)) - *[\Aform,\cdot]$,
as holds from the deformation theorem~2.8 in \secref{method}.

\section{ Gravity-like \YM/ \CS/ theory with torsion }
\label{torsionKVth}

Gravity-like interactions of a 1-form potential $\A{a}{\mu}$
arise in a natural fashion from consideration of 
non-covariant deformations involving stress-energy currents
for linearized \YM/ gauge theory.
Recall, in the linear abelian theory \sysref{quadrL}{linfieldstrid}, 
such currents are produced from the stress-energy tensor
\EQ
\nthT{2}{}{\mu\nu} = 
\k{ab}( \stF{a}{\mu} \stF{b}{\nu} 
-\sfrac{1}{2}\metric{\mu\nu} \invmetric{\alpha\beta} 
\stF{a}{\alpha}\stF{b}{\beta} )
\doneEQ
when it is contracted with any \KV/ $\kv{\mu}{}$ of 
the flat spacetime metric $\metric{\mu\nu}$.
The current 
$\nthcurrent{2}{}{\mu}(\kv{}{}) =\nthT{2}{}{\mu\nu} \kv{\nu}{}$
is conserved 
$\coder{\mu}\nthcurrent{2}{}{\mu}(\kv{}{})=0$
and gauge invariant 
$\nthXvar{0}\nthcurrent{2}{}{\mu}(\kv{}{}) =0$
under the abelian gauge symmetry \eqref{zerothXvar}
on solutions $\A{a}{\mu}$ of the linear field equation \eqref{linfieldeq}.
As a starting point to write down 
a pure gravity-like deformation of the linear abelian theory, 
consider a set of commuting \KV/s 
\EQ
0=\Lie{\kv{}{a}} \invmetric{\mu\nu} =-2\coder{(\mu} \kv{\nu)}{a} ,\quad
0=\Lie{\kv{}{a}}\kv{\mu}{b} = 2\kv{\nu}{[b} \der{\nu}\kv{\mu}{a]}
= [\kv{}{b},\kv{}{a}]^\mu
\doneEQ
equal in number to the dimension of the internal vector space $\vs$,
\ie/ so there is the same number of fields $\A{a}{\mu}$ 
as \KV/s $\kv{\mu}{a}$.
In terms of such \KV/s, 
the deformation to lowest order is given by 
\EQ
\nthXvar{1}\A{a}{\mu} 
= \c{G} 2\X{b} \kv{\nu}{b} \F{a}{\nu\mu}
\doneEQ
and 
\EQ
\nthsL{3} 
= \c{G} \invmetric{\mu\nu} \nthcurrent{2}{}{\mu}(\kv{}{a}) \A{a}{\nu}
\label{gravitycubicL}
\doneEQ
with coupling constant $\c{G}$.
Due to the explicit dependence on $\kv{\mu}{a}$,
the deformation terms break Lorentz covariance. 

An equivalent, more geometrical form for the deformed gauge symmetry 
is obtained if the gauge parameter is redefined by
$\vector{\sX'}{a} = \X{a} + \c{G}\A{a}{\mu} \kv{\mu}{b} \X{b}$,
yielding
\EQ
\nthXvar{1}\A{a}{\mu} 
= \c{G} \Lie{\sX} \A{a}{\mu} ,\quad
\X{\nu} = \X{a} \kv{\nu}{a}
\label{gravityXvarlieder}
\doneEQ 
through the Lie derivative identity 
$\Lie{\sX} \covector{f}{\mu} 
= \X{\nu} \der{[\nu} \covector{f}{\mu]} 
+\der{\mu}( \X{\nu} \covector{f}{\nu} )$
holding for any 1-form $f$ and vector field $\sX$ on $M$. 
The gravity-like nature of the deformation is seen from two observations.
First, 
this gauge symmetry \eqref{gravityXvarlieder}
expresses an infinitesimal diffeomorphism invariance
on $\A{a}{\mu}$ with respect to the vector field $\X{\nu}$
as the gauge parameter.
Second, 
the Lagrangian \eqref{gravitycubicL}
is given by a stress-energy self-coupling
\EQ
\nthsL{3} = \frac{1}{2} \T{\mu\nu}{} \nthgdeform{1}{\mu\nu} ,\quad
\nthgdeform{1}{\mu\nu} 
= \metric{\mu\nu} +2\A{a}{(\mu} \metric{\nu)\alpha} \kv{\alpha}{a} +\cdots
\label{gravityLTg}
\doneEQ
involving a deformation of the spacetime metric $\metric{\mu\nu}$ 
to $\gdeform{\mu\nu}$.
These deformation terms \eqrefs{gravityXvarlieder}{gravityLTg}
are analogous to the lowest-order nonlinear terms
in Einstein gravity theory for a vielbein. 
Notice, however, 
the field equation for $\A{a}{\mu}$ is not of a gravity-like form
as it dynamically describes a set of spin-one electromagnetic field equations
on $\F{a}{\mu\nu}$, with current sources proportional to
$\nthcurrent{2}{}{\mu}(\kv{}{a})$. 
In particular, there is no auxiliary gauge freedom on $\A{a}{\mu}$,
in contrast with the nature of the spin-two graviton field equation.
(Indeed, as is well known, in three spacetime dimensions
the graviton field equation has no dynamical degrees of freedom,
once the linearized local Lorentz rotation freedom 
on the vielbein is taken into account.)

This type of deformation \sysref{gravityXvarlieder}{gravityLTg}
can be generalized to use non-commuting \KV/s
by including a \YM/ interaction 
whose Lie algebra structure constants are tied to the \KV/ commutators,
which is explained in \Ref{Bra2}
where the deformation is completed to all orders. 
A further generalization to include a \FT/ interaction can also be considered,
which will be pursued elsewhere.
Here the objective instead will be derive a torsion generalization of
the pure gravity-like deformation of the linear abelian theory, 
extended to include a \CS/ term.

For incorporating torsion, as well as a \CS/ term, 
it is essential to keep track of how the spacetime metric $\metric{\mu\nu}$
and volume form $\vol{\mu\nu\alpha}$ appear in the linear theory 
and the deformation terms. 
To proceed, write the abelian field strength as
\EQ
\stF{\mu}{a} = \k{ab} \invvol{\mu\nu\alpha} \der{\nu} \A{b}{\alpha} ,
\label{vectorfieldstr}
\doneEQ
so the Lagrangian of the linear theory with an abelian \CS/ term 
takes the form 
\EQ
\nthsL{2} +\nthsL{2}_{\rm CS} 
= \sfrac{1}{2} \invk{ab} \metric{\mu\nu} \stF{\mu}{a} \stF{\nu}{b}
+ \sfrac{1}{2} m \A{a}{\mu} \stF{\mu}{a}
\doneEQ
where the only dependence on the spacetime metric 
occurs through $\metric{\mu\nu}$ in $\nthsL{2}$.
Torsion will now be added in the simplest manner 
by replacing the joint metric $\invk{ab} \metric{\mu\nu}$
(on vector fields with values in the dual space of $\vs$ on $M$)
with 
\EQ
\torsmetric{ab}{\mu\nu} 
= \metric{\mu\nu} \invk{ab} + \v{}{\mu\nu} \p{ab}{} ,\quad
\v{}{\mu\nu} = \vol{\mu\nu\alpha} \v{\alpha}{}
\doneEQ
where, similarly to the torsion introduced previously, 
$\v{\mu}{}$ is a constant vector on $M$
and $\p{ab}{}$ is a constant skew tensor on $\vs$. 
This yields the torsion Lagrangian 
\EQ
\nthsL{2}_{\rm tors.} 
= \sfrac{1}{2} \torsmetric{ab}{\mu\nu} \stF{\mu}{a} \stF{\nu}{b}
+\sfrac{1}{2} m \A{a}{\mu} \stF{\mu}{a}
\label{gravitytorsquadrL}
\doneEQ
with the following field-theoretic content. 
On solutions of the linear field equation 
\EQ
\linE{}{\mu a}{} +\linE{CS}{\mu a}{}
= \invvol{\mu\nu\alpha} \torsmetric{ab}{\alpha\beta} \der{\nu} \stF{\beta}{b}
+ m \invk{ab} \stF{\mu}{b} 
=0 , 
\doneEQ
the curl of 
$\torsmetric{ab}{\alpha\beta} \stF{\beta}{b} +\A{a}{\alpha}$ 
vanishes, hence
\EQ
\stF{\nu}{b} = ( \der{\mu} \wavemap{a} -m\A{a}{\mu} )
\doneEQ
with $\wavemap{a}$ representing a chiral field, as before.
Then $\nthsL{2}_{\rm tors.}$ has an equivalent dual formulation 
consisting of an abelian chiral Lagrangian theory for $\wavemap{a}$
minimally coupled to an abelian \CS/ theory for $\A{a}{\mu}$,
in which the torsion appears through the inverse metric 
$\invtorsmetric{\mu\nu}{ab}$
in contrast to the construction of the torsion Lagrangian \eqref{torsquadrL}. 

\subsection{ Torsion deformation }

Compatibility of the gravity-like deformation at lowest order
with torsion in the linear abelian theory 
\sysref{vectorfieldstr}{gravitytorsquadrL}
is easily ascertained from an analysis of the determining equations
for the deformation terms. 
The closure equation at $0$th order for the linear deformation terms 
in the gauge symmetry remains satisfied, 
because the addition of a torsion term and a \CS/ term 
in the free Lagrangian does not affect 
the abelian gauge symmetry on $\A{a}{\mu}$.
At $1$st order, the determining equation for the quadratic terms 
in the field equation is found to yield
\EQ
\nthsL{3} 
= \c{G} \stF{\mu}{a} \stF{\nu}{b} ( 
\A{c}{\mu} \torsmetric{ab}{\alpha\nu} 
-\sfrac{1}{2} \A{c}{\alpha} \torsmetric{ab}{\mu\nu} ) \kv{\alpha}{c} 
\doneEQ
under the algebraic condition
\EQ
2\metric{\alpha\beta} \v{\beta[\nu}{} \coder{\mu]} \kv{\alpha}{a} 
= \cross{\alpha}{\mu\nu} \v{}{\beta} \coder{[\alpha} \kv{\beta]}{a} 
=0 .
\label{gravitytorskv}
\doneEQ
This is the only obstruction to the deformation arising to this order.
Remarkably, no other obstructions occur to all higher orders. 

This obstruction has a simple content. 
Recall, for a rotation or boost \KV/ $\kv{\nu}{}$ 
in a three-dimensional spacetime, 
$\coder{[\mu} \kv{\nu]}{}$ is a constant antisymmetric tensor 
whose dual vector $\cross{\mu\nu}{\alpha} \coder{\mu} \kv{\nu}{}$
gives the rotation/boost axis in spacetime;
for a translation \KV/, $\coder{[\mu} \kv{\nu]}{}$ vanishes.

{\bf Theorem~5.1}:
For any flat spacetime metric $\metric{\mu\nu}$ in three dimensions,
there are two classes of commuting \KV/s $\kv{\nu}{a}$:
(i) translations, 
comprising a two or three dimensional planar isometry subgroup 
of the metric;
(ii) a rotation or boost plus an orthogonal translation,
comprising a two dimensional axisymmetric isometry subgroup 
of the metric.
The algebraic obstruction \eqref{gravitytorskv}
is satisfied for class (i) by any vector $\v{\mu}{}$,
and by a vector $\v{\mu}{}$ parallel to the rotation/boost axis
for class (ii).
Stated geometrically, 
$\Lie{\v{}{}} \kv{\mu}{a} =0$ for both classes. 

Construction of the gravity-like deformation based on this structure
now proceeds as follows. 
Geometrically, the gauge symmetry is given by 
\EQ
\Xvar\A{a}{\mu} 
= \der{\mu} \X{a} +\Lie{\X{c}\kv{}{c}} \A{a}{\mu}
\label{gravityKVthgaugesymm}
\doneEQ
where the coupling constant has been absorbed into the \KV/s $\kv{\mu}{a}$.
The Lagrangian is constructed using a deformed metric 
$\gtorsdeform{ab}{\mu\nu} 
= \invk{ab} \gdeform{\mu\nu} + \p{ab}{} \vdeform{}{\mu\nu}$
given by 
\EQs
&&
\gdeform{\mu\nu} 
= \metric{\mu\nu} + 2\A{c}{(\mu} \metric{\nu)\alpha} \kv{\alpha}{c}
+ \A{c}{\mu} \A{d}{\nu} \metric{\alpha\beta} \kv{\alpha}{c} \kv{\beta}{d} ,
\label{gdeform}\\
&&
\vdeform{}{\mu\nu}
= \v{}{\mu\nu} + 2\A{c}{[\mu} \v{}{\nu]\alpha} \kv{\alpha}{c}
+ \A{c}{\mu} \A{d}{\nu} \v{}{\alpha\beta} \kv{\alpha}{c} \kv{\beta}{d} .
\doneEQs
Under the gauge symmetry \eqref{gravityKVthgaugesymm}, 
this metric has the crucial property 
\EQ
\Xvar\gtorsdeform{ab}{\mu\nu}
= \Lie{\X{c}\kv{}{c}} \gtorsdeform{ab}{\mu\nu}
\doneEQ
by means of the algebraic relation \eqref{gravitytorskv}. 
Then a gauge-invariant Lagrangian to within a divergence is obtained by 
\EQ
L = 
\sfrac{1}{2} (\det\gdeform{}{})^{-1/2} 
\gtorsdeform{ab}{\mu\nu} \stF{\mu}{a} \stF{\nu}{b}
+\sfrac{1}{2} m \A{a}{\mu} \stF{\mu}{a}
\doneEQ
where $\det\gdeform{}$ denotes the standard determinant of 
the metric \eqref{gdeform} in spacetime coordinates.
Gauge invariance is verified through Lie derivative formulas
combined with the properties
\EQs
&&
\Xvar \ln(\det\gdeform{})
= \invgdeform{\mu\nu} \Xvar\gdeform{\mu\nu}
= \invgdeform{\mu\nu} \gdeform{\mu\nu}
= \X{c}\kv{\alpha}{c} \der{\alpha} \det\gdeform{}
+ 2\det\gdeform{} \der{\alpha}( \X{c}\kv{\alpha}{c} ) ,
\\&&
\Xvar\stF{\mu}{a}
=\k{ab} \invvol{\mu\nu\alpha} \der{\nu} \Xvar\A{b}{\alpha}
= \Lie{\X{c}\kv{}{c}} \stF{\mu}{a} 
+\stF{\mu}{a} \der{\alpha}( \X{c}\kv{\alpha}{c} ) .
\doneEQs
In particular, it is found that 
\EQ
\Xvar\sL = \der{\alpha}( \X{c}\kv{\alpha}{c} \sL )
\doneEQ
which clearly exhibits the gravity-like nature of the gauge invariance. 

Note the determinant factor in the Lagrangian can be written 
in a more familiar form if a deformed abelian field strength 
is introduced 
\EQ
\deformstF{\mu}{a} 
= \k{ab}\voldeform{\mu\nu\alpha}{} \F{b}{\nu\alpha}
= (\det\gdeform{})^{-1/2} \stF{\mu}{a} ,
\doneEQ
through a deformation of the volume tensor
$\voldeform{\mu\nu\alpha}{} = (\det\gdeform{})^{-1/2} \invvol{\mu\nu\alpha}$
corresponding to the deformed metric \eqref{gdeform}.
This yields
\EQ
\sL = \sfrac{1}{2} (\det\gdeform{}{})^{1/2} 
( (\gtorsdeform{ab}{\mu\nu} \deformstF{\mu}{a} +m \A{b}{\nu} ) 
\deformstF{\nu}{b} .
\doneEQ

An interesting geometrical feature of this theory 
comes from writing the gauge symmetry as a covariant derivative
\EQ
\Xvar\A{a}{\mu} 
= \der{\mu}\vector{\zeta'}{a} 
+ 2\F{a}{\nu\mu} \kv{\nu}{b} \kdeform{b}{c} \vector{\zeta'}{c} ,\quad
\vector{\zeta'}{a} = \kdeform{a}{b} \X{b} , 
\doneEQ
using a redefinition of the gauge parameter, where
$\kdeform{a}{b} = \id{b}{a} +\kv{\alpha}{b} \A{a}{\alpha}$.
The covariant curl of this gauge symmetry yields
the curvature of the connection 1-form 
$\conx{a}{\mu b}=  
2\F{a}{\nu\mu} \kv{\nu}{b} \kdeform{b}{c} \vector{\zeta'}{c}$,
\EQ
\D{[\mu}\Xvar\A{a}{\nu]} 
= \R{a}{\mu\nu b} \vector{\zeta'}{b} .
\doneEQ
In agreement with the geometrical deformation theorem~2.8,
this curvature is related to the deformed abelian field strength
through the action of the gauge symmetry
\EQ
\R{a}{\mu\nu b} \vector{\zeta'}{b}
= \voldeform{}{\mu\nu\rho} \invY{\ \rho}{\alpha} \invk{ab}
\Xvar( \Y{\ \alpha}{\beta} \deformstF{\beta}{b}) 
\doneEQ
where 
$\Y{\ \alpha}{\beta} = \id{\beta}{\alpha} +\kv{\alpha}{c} \A{c}{\beta}$.
Furthermore, the linear map $\Y{}{}$ here 
has a vielbein-like relation to the deformed metric 
$\gtorsdeform{ab}{\mu\nu} = 
\torsmetric{ab}{\alpha\beta} \Y{\ \alpha}{\mu} \Y{\ \beta}{\nu}$.

\section{ Concluding Remarks }
\label{conclude}

The novel nonlinear generalization of \YM/ \CS/ gauge theory with torsion 
in three spacetime dimensions 
constructed in \secref{torsionYMth}
has an extension to four dimensions when the 1-form potentials $\A{a}{\mu}$ 
are coupled to 2-form potentials $\B{a}{\mu\nu}$,
similarly to the case without torsion derived in \Ref{JMPpaper}. 
Indeed the construction extends more generally to 
a tower of coupled $p$-form potentials for $1\leq p\leq d-2$
in $d\geq 3$ dimensions,
where torsion is naturally motivated starting from 
a dual chiral field formulation of the \FT/ field strength of
the $d-2$-form potential $\B{a}{\mu_1\cdots \mu_{d-2}}$. 
A \YM/ self-interaction of the 1-form potentials $\A{a}{\mu}$, 
combined with a \YM/ Higgs interaction between the potentials
$\B{a}{\mu_1\cdots \mu_{d-2}}$ and $\A{a}{\mu}$, 
is expected to be compatible with torsion if
a $d$-dimensional \CS/ type term is included in the theory,
analogously to the case in $d=3$ dimensions. 
It would be of interest to explore whether similar considerations
apply to the gravity-like interaction of 1-form potentials
in $d\geq 3$ dimensions.

In $d=3$ dimensions, there is a special type of torsion that yields
an integrable chiral field equation due to Ward \cite{War}. 
This torsion field theory arises by a dimensional reduction of
the chiral formulation of self-dual \YM/ equations in $2+2$ dimensions
and exhibits soliton-like solutions and other characteristic features of
integrability such as a hierarchy of symmetries and conservation laws. 
Geometrically, the integrability involves parallelism \cite{torsion}, 
\ie/ flattening of the curvature on the Lie group for the target space of
the chiral field, 
as produced by adding torsion to the Cartan-Killing connection. 
For the torsion to precisely cancel the Riemannian part in the curvature,
it must be proportional to the Lie algebra structure tensor of the Lie group. 
If a dual formulation is considered for such a chiral field parallelism theory,
it leads to the condition
$\Q{a}{bc} = \lambda \liealg{a}{bc} 
= \sfrac{3}{2} \invk{ae} \p{}{d[e} \liealg{d}{bc]}$,
where $\Q{a}{bc}$ is the torsion, 
$\p{}{ab}$ is the torsion potential, 
$\liealg{a}{bc}$ denotes the Lie algebra structure constants
and $\k{ab} = -\liealg{d}{ac} \liealg{c}{bd}$ is the Cartan-Killing metric,
pulled back to the Lie algebra of the Lie group.
However, for a semisimple Lie group,
contraction of this condition with $\liealg{c}{da}$ followed by $\invk{bd}$
yields $\lambda=0$. 
Thus, Ward's integrable chiral field equation has no dual formulation 
as a gauge field theory of 1-form potentials. 

It then becomes worthwhile to ask if a self-dual gauge theory exists that
contains a \FT/ interaction of $d-2$-form potentials in $d\geq 3$ dimensions,
as this could lead to an interesting dual chiral formulation 
or perhaps to integrability properties. 
Of course, self-duality in $d$ dimensions requires pairing 
$p$-form potentials with $d-p-2$-form potentials. 
The simplest case would be 1-form potentials coupled to 
scalar (0-form) fields by an extended \FT/ type interaction 
in $d=3$ dimensions,
in analogy with three dimensional self-dual \YM/ Higgs theory. 
A natural suggestion is thus to investigate a deformation of 
self-dual \YM/ Higgs theory by including a \FT/ interaction.


\end{document}